\documentclass[usenatbib,usegraphicx]{mn2e}
\usepackage{comment}
\usepackage{amsfonts}
\usepackage{amsmath}
\usepackage{amssymb}
\usepackage{color}
\definecolor{AliceBlue}{rgb}{0.94,0.97,1.00}
\definecolor{AntiqueWhite1}{rgb}{1.00,0.94,0.86}
\definecolor{AntiqueWhite2}{rgb}{0.93,0.87,0.80}
\definecolor{AntiqueWhite3}{rgb}{0.80,0.75,0.69}
\definecolor{AntiqueWhite4}{rgb}{0.55,0.51,0.47}
\definecolor{AntiqueWhite}{rgb}{0.98,0.92,0.84}
\definecolor{BlanchedAlmond}{rgb}{1.00,0.92,0.80}
\definecolor{BlueViolet}{rgb}{0.54,0.17,0.89}
\definecolor{CadetBlue1}{rgb}{0.60,0.96,1.00}
\definecolor{CadetBlue2}{rgb}{0.56,0.90,0.93}
\definecolor{CadetBlue3}{rgb}{0.48,0.77,0.80}
\definecolor{CadetBlue4}{rgb}{0.33,0.53,0.55}
\definecolor{CadetBlue}{rgb}{0.37,0.62,0.63}
\definecolor{CornflowerBlue}{rgb}{0.39,0.58,0.93}
\definecolor{DarkBlue}{rgb}{0.00,0.00,0.55}
\definecolor{DarkCyan}{rgb}{0.00,0.55,0.55}
\definecolor{DarkGoldenrod1}{rgb}{1.00,0.73,0.06}
\definecolor{DarkGoldenrod2}{rgb}{0.93,0.68,0.05}
\definecolor{DarkGoldenrod3}{rgb}{0.80,0.58,0.05}
\definecolor{DarkGoldenrod4}{rgb}{0.55,0.40,0.03}
\definecolor{DarkGoldenrod}{rgb}{0.72,0.53,0.04}
\definecolor{DarkGray}{rgb}{0.66,0.66,0.66}
\definecolor{DarkGreen}{rgb}{0.00,0.39,0.00}
\definecolor{DarkGrey}{rgb}{0.66,0.66,0.66}
\definecolor{DarkKhaki}{rgb}{0.74,0.72,0.42}
\definecolor{DarkMagenta}{rgb}{0.55,0.00,0.55}
\definecolor{DarkOliveGreen1}{rgb}{0.79,1.00,0.44}
\definecolor{DarkOliveGreen2}{rgb}{0.74,0.93,0.41}
\definecolor{DarkOliveGreen3}{rgb}{0.64,0.80,0.35}
\definecolor{DarkOliveGreen4}{rgb}{0.43,0.55,0.24}
\definecolor{DarkOliveGreen}{rgb}{0.33,0.42,0.18}
\definecolor{DarkOrange1}{rgb}{1.00,0.50,0.00}
\definecolor{DarkOrange2}{rgb}{0.93,0.46,0.00}
\definecolor{DarkOrange3}{rgb}{0.80,0.40,0.00}
\definecolor{DarkOrange4}{rgb}{0.55,0.27,0.00}
\definecolor{DarkOrange}{rgb}{1.00,0.55,0.00}
\definecolor{DarkOrchid1}{rgb}{0.75,0.24,1.00}
\definecolor{DarkOrchid2}{rgb}{0.70,0.23,0.93}
\definecolor{DarkOrchid3}{rgb}{0.60,0.20,0.80}
\definecolor{DarkOrchid4}{rgb}{0.41,0.13,0.55}
\definecolor{DarkOrchid}{rgb}{0.60,0.20,0.80}
\definecolor{DarkRed}{rgb}{0.55,0.00,0.00}
\definecolor{DarkSalmon}{rgb}{0.91,0.59,0.48}
\definecolor{DarkSeaGreen1}{rgb}{0.76,1.00,0.76}
\definecolor{DarkSeaGreen2}{rgb}{0.71,0.93,0.71}
\definecolor{DarkSeaGreen3}{rgb}{0.61,0.80,0.61}
\definecolor{DarkSeaGreen4}{rgb}{0.41,0.55,0.41}
\definecolor{DarkSeaGreen}{rgb}{0.56,0.74,0.56}
\definecolor{DarkSlateBlue}{rgb}{0.28,0.24,0.55}
\definecolor{DarkSlateGray1}{rgb}{0.59,1.00,1.00}
\definecolor{DarkSlateGray2}{rgb}{0.55,0.93,0.93}
\definecolor{DarkSlateGray3}{rgb}{0.47,0.80,0.80}
\definecolor{DarkSlateGray4}{rgb}{0.32,0.55,0.55}
\definecolor{DarkSlateGray}{rgb}{0.18,0.31,0.31}
\definecolor{DarkSlateGrey}{rgb}{0.18,0.31,0.31}
\definecolor{DarkTurquoise}{rgb}{0.00,0.81,0.82}
\definecolor{DarkViolet}{rgb}{0.58,0.00,0.83}
\definecolor{DeepPink1}{rgb}{1.00,0.08,0.58}
\definecolor{DeepPink2}{rgb}{0.93,0.07,0.54}
\definecolor{DeepPink3}{rgb}{0.80,0.06,0.46}
\definecolor{DeepPink4}{rgb}{0.55,0.04,0.31}
\definecolor{DeepPink}{rgb}{1.00,0.08,0.58}
\definecolor{DeepSkyBlue1}{rgb}{0.00,0.75,1.00}
\definecolor{DeepSkyBlue2}{rgb}{0.00,0.70,0.93}
\definecolor{DeepSkyBlue3}{rgb}{0.00,0.60,0.80}
\definecolor{DeepSkyBlue4}{rgb}{0.00,0.41,0.55}
\definecolor{DeepSkyBlue}{rgb}{0.00,0.75,1.00}
\definecolor{DimGray}{rgb}{0.41,0.41,0.41}
\definecolor{DimGrey}{rgb}{0.41,0.41,0.41}
\definecolor{DodgerBlue1}{rgb}{0.12,0.56,1.00}
\definecolor{DodgerBlue2}{rgb}{0.11,0.53,0.93}
\definecolor{DodgerBlue3}{rgb}{0.09,0.45,0.80}
\definecolor{DodgerBlue4}{rgb}{0.06,0.31,0.55}
\definecolor{DodgerBlue}{rgb}{0.12,0.56,1.00}
\definecolor{FloralWhite}{rgb}{1.00,0.98,0.94}
\definecolor{ForestGreen}{rgb}{0.13,0.55,0.13}
\definecolor{GhostWhite}{rgb}{0.97,0.97,1.00}
\definecolor{GreenYellow}{rgb}{0.68,1.00,0.18}
\definecolor{HotPink1}{rgb}{1.00,0.43,0.71}
\definecolor{HotPink2}{rgb}{0.93,0.42,0.65}
\definecolor{HotPink3}{rgb}{0.80,0.38,0.56}
\definecolor{HotPink4}{rgb}{0.55,0.23,0.38}
\definecolor{HotPink}{rgb}{1.00,0.41,0.71}
\definecolor{IndianRed1}{rgb}{1.00,0.42,0.42}
\definecolor{IndianRed2}{rgb}{0.93,0.39,0.39}
\definecolor{IndianRed3}{rgb}{0.80,0.33,0.33}
\definecolor{IndianRed4}{rgb}{0.55,0.23,0.23}
\definecolor{IndianRed}{rgb}{0.80,0.36,0.36}
\definecolor{LavenderBlush1}{rgb}{1.00,0.94,0.96}
\definecolor{LavenderBlush2}{rgb}{0.93,0.88,0.90}
\definecolor{LavenderBlush3}{rgb}{0.80,0.76,0.77}
\definecolor{LavenderBlush4}{rgb}{0.55,0.51,0.53}
\definecolor{LavenderBlush}{rgb}{1.00,0.94,0.96}
\definecolor{LawnGreen}{rgb}{0.49,0.99,0.00}
\definecolor{LemonChiffon1}{rgb}{1.00,0.98,0.80}
\definecolor{LemonChiffon2}{rgb}{0.93,0.91,0.75}
\definecolor{LemonChiffon3}{rgb}{0.80,0.79,0.65}
\definecolor{LemonChiffon4}{rgb}{0.55,0.54,0.44}
\definecolor{LemonChiffon}{rgb}{1.00,0.98,0.80}
\definecolor{LightBlue1}{rgb}{0.75,0.94,1.00}
\definecolor{LightBlue2}{rgb}{0.70,0.87,0.93}
\definecolor{LightBlue3}{rgb}{0.60,0.75,0.80}
\definecolor{LightBlue4}{rgb}{0.41,0.51,0.55}
\definecolor{LightBlue}{rgb}{0.68,0.85,0.90}
\definecolor{LightCoral}{rgb}{0.94,0.50,0.50}
\definecolor{LightCyan1}{rgb}{0.88,1.00,1.00}
\definecolor{LightCyan2}{rgb}{0.82,0.93,0.93}
\definecolor{LightCyan3}{rgb}{0.71,0.80,0.80}
\definecolor{LightCyan4}{rgb}{0.48,0.55,0.55}
\definecolor{LightCyan}{rgb}{0.88,1.00,1.00}
\definecolor{LightGoldenrod1}{rgb}{1.00,0.93,0.55}
\definecolor{LightGoldenrod2}{rgb}{0.93,0.86,0.51}
\definecolor{LightGoldenrod3}{rgb}{0.80,0.75,0.44}
\definecolor{LightGoldenrod4}{rgb}{0.55,0.51,0.30}
\definecolor{LightGoldenrodYellow}{rgb}{0.98,0.98,0.82}
\definecolor{LightGoldenrod}{rgb}{0.93,0.87,0.51}
\definecolor{LightGray}{rgb}{0.83,0.83,0.83}
\definecolor{LightGreen}{rgb}{0.56,0.93,0.56}
\definecolor{LightGrey}{rgb}{0.83,0.83,0.83}
\definecolor{LightPink1}{rgb}{1.00,0.68,0.73}
\definecolor{LightPink2}{rgb}{0.93,0.64,0.68}
\definecolor{LightPink3}{rgb}{0.80,0.55,0.58}
\definecolor{LightPink4}{rgb}{0.55,0.37,0.40}
\definecolor{LightPink}{rgb}{1.00,0.71,0.76}
\definecolor{LightSalmon1}{rgb}{1.00,0.63,0.48}
\definecolor{LightSalmon2}{rgb}{0.93,0.58,0.45}
\definecolor{LightSalmon3}{rgb}{0.80,0.51,0.38}
\definecolor{LightSalmon4}{rgb}{0.55,0.34,0.26}
\definecolor{LightSalmon}{rgb}{1.00,0.63,0.48}
\definecolor{LightSeaGreen}{rgb}{0.13,0.70,0.67}
\definecolor{LightSkyBlue1}{rgb}{0.69,0.89,1.00}
\definecolor{LightSkyBlue2}{rgb}{0.64,0.83,0.93}
\definecolor{LightSkyBlue3}{rgb}{0.55,0.71,0.80}
\definecolor{LightSkyBlue4}{rgb}{0.38,0.48,0.55}
\definecolor{LightSkyBlue}{rgb}{0.53,0.81,0.98}
\definecolor{LightSlateBlue}{rgb}{0.52,0.44,1.00}
\definecolor{LightSlateGray}{rgb}{0.47,0.53,0.60}
\definecolor{LightSlateGrey}{rgb}{0.47,0.53,0.60}
\definecolor{LightSteelBlue1}{rgb}{0.79,0.88,1.00}
\definecolor{LightSteelBlue2}{rgb}{0.74,0.82,0.93}
\definecolor{LightSteelBlue3}{rgb}{0.64,0.71,0.80}
\definecolor{LightSteelBlue4}{rgb}{0.43,0.48,0.55}
\definecolor{LightSteelBlue}{rgb}{0.69,0.77,0.87}
\definecolor{LightYellow1}{rgb}{1.00,1.00,0.88}
\definecolor{LightYellow2}{rgb}{0.93,0.93,0.82}
\definecolor{LightYellow3}{rgb}{0.80,0.80,0.71}
\definecolor{LightYellow4}{rgb}{0.55,0.55,0.48}
\definecolor{LightYellow}{rgb}{1.00,1.00,0.88}
\definecolor{LimeGreen}{rgb}{0.20,0.80,0.20}
\definecolor{MediumAquamarine}{rgb}{0.40,0.80,0.67}
\definecolor{MediumBlue}{rgb}{0.00,0.00,0.80}
\definecolor{MediumOrchid1}{rgb}{0.88,0.40,1.00}
\definecolor{MediumOrchid2}{rgb}{0.82,0.37,0.93}
\definecolor{MediumOrchid3}{rgb}{0.71,0.32,0.80}
\definecolor{MediumOrchid4}{rgb}{0.48,0.22,0.55}
\definecolor{MediumOrchid}{rgb}{0.73,0.33,0.83}
\definecolor{MediumPurple1}{rgb}{0.67,0.51,1.00}
\definecolor{MediumPurple2}{rgb}{0.62,0.47,0.93}
\definecolor{MediumPurple3}{rgb}{0.54,0.41,0.80}
\definecolor{MediumPurple4}{rgb}{0.36,0.28,0.55}
\definecolor{MediumPurple}{rgb}{0.58,0.44,0.86}
\definecolor{MediumSeaGreen}{rgb}{0.24,0.70,0.44}
\definecolor{MediumSlateBlue}{rgb}{0.48,0.41,0.93}
\definecolor{MediumSpringGreen}{rgb}{0.00,0.98,0.60}
\definecolor{MediumTurquoise}{rgb}{0.28,0.82,0.80}
\definecolor{MediumVioletRed}{rgb}{0.78,0.08,0.52}
\definecolor{MidnightBlue}{rgb}{0.10,0.10,0.44}
\definecolor{MintCream}{rgb}{0.96,1.00,0.98}
\definecolor{MistyRose1}{rgb}{1.00,0.89,0.88}
\definecolor{MistyRose2}{rgb}{0.93,0.84,0.82}
\definecolor{MistyRose3}{rgb}{0.80,0.72,0.71}
\definecolor{MistyRose4}{rgb}{0.55,0.49,0.48}
\definecolor{MistyRose}{rgb}{1.00,0.89,0.88}
\definecolor{NavajoWhite1}{rgb}{1.00,0.87,0.68}
\definecolor{NavajoWhite2}{rgb}{0.93,0.81,0.63}
\definecolor{NavajoWhite3}{rgb}{0.80,0.70,0.55}
\definecolor{NavajoWhite4}{rgb}{0.55,0.47,0.37}
\definecolor{NavajoWhite}{rgb}{1.00,0.87,0.68}
\definecolor{NavyBlue}{rgb}{0.00,0.00,0.50}
\definecolor{OldLace}{rgb}{0.99,0.96,0.90}
\definecolor{OliveDrab1}{rgb}{0.75,1.00,0.24}
\definecolor{OliveDrab2}{rgb}{0.70,0.93,0.23}
\definecolor{OliveDrab3}{rgb}{0.60,0.80,0.20}
\definecolor{OliveDrab4}{rgb}{0.41,0.55,0.13}
\definecolor{OliveDrab}{rgb}{0.42,0.56,0.14}
\definecolor{OrangeRed1}{rgb}{1.00,0.27,0.00}
\definecolor{OrangeRed2}{rgb}{0.93,0.25,0.00}
\definecolor{OrangeRed3}{rgb}{0.80,0.22,0.00}
\definecolor{OrangeRed4}{rgb}{0.55,0.15,0.00}
\definecolor{OrangeRed}{rgb}{1.00,0.27,0.00}
\definecolor{PaleGoldenrod}{rgb}{0.93,0.91,0.67}
\definecolor{PaleGreen1}{rgb}{0.60,1.00,0.60}
\definecolor{PaleGreen2}{rgb}{0.56,0.93,0.56}
\definecolor{PaleGreen3}{rgb}{0.49,0.80,0.49}
\definecolor{PaleGreen4}{rgb}{0.33,0.55,0.33}
\definecolor{PaleGreen}{rgb}{0.60,0.98,0.60}
\definecolor{PaleTurquoise1}{rgb}{0.73,1.00,1.00}
\definecolor{PaleTurquoise2}{rgb}{0.68,0.93,0.93}
\definecolor{PaleTurquoise3}{rgb}{0.59,0.80,0.80}
\definecolor{PaleTurquoise4}{rgb}{0.40,0.55,0.55}
\definecolor{PaleTurquoise}{rgb}{0.69,0.93,0.93}
\definecolor{PaleVioletRed1}{rgb}{1.00,0.51,0.67}
\definecolor{PaleVioletRed2}{rgb}{0.93,0.47,0.62}
\definecolor{PaleVioletRed3}{rgb}{0.80,0.41,0.54}
\definecolor{PaleVioletRed4}{rgb}{0.55,0.28,0.36}
\definecolor{PaleVioletRed}{rgb}{0.86,0.44,0.58}
\definecolor{PapayaWhip}{rgb}{1.00,0.94,0.84}
\definecolor{PeachPuff1}{rgb}{1.00,0.85,0.73}
\definecolor{PeachPuff2}{rgb}{0.93,0.80,0.68}
\definecolor{PeachPuff3}{rgb}{0.80,0.69,0.58}
\definecolor{PeachPuff4}{rgb}{0.55,0.47,0.40}
\definecolor{PeachPuff}{rgb}{1.00,0.85,0.73}
\definecolor{PowderBlue}{rgb}{0.69,0.88,0.90}
\definecolor{RosyBrown1}{rgb}{1.00,0.76,0.76}
\definecolor{RosyBrown2}{rgb}{0.93,0.71,0.71}
\definecolor{RosyBrown3}{rgb}{0.80,0.61,0.61}
\definecolor{RosyBrown4}{rgb}{0.55,0.41,0.41}
\definecolor{RosyBrown}{rgb}{0.74,0.56,0.56}
\definecolor{RoyalBlue1}{rgb}{0.28,0.46,1.00}
\definecolor{RoyalBlue2}{rgb}{0.26,0.43,0.93}
\definecolor{RoyalBlue3}{rgb}{0.23,0.37,0.80}
\definecolor{RoyalBlue4}{rgb}{0.15,0.25,0.55}
\definecolor{RoyalBlue}{rgb}{0.25,0.41,0.88}
\definecolor{SaddleBrown}{rgb}{0.55,0.27,0.07}
\definecolor{SandyBrown}{rgb}{0.96,0.64,0.38}
\definecolor{SeaGreen1}{rgb}{0.33,1.00,0.62}
\definecolor{SeaGreen2}{rgb}{0.31,0.93,0.58}
\definecolor{SeaGreen3}{rgb}{0.26,0.80,0.50}
\definecolor{SeaGreen4}{rgb}{0.18,0.55,0.34}
\definecolor{SeaGreen}{rgb}{0.18,0.55,0.34}
\definecolor{SkyBlue1}{rgb}{0.53,0.81,1.00}
\definecolor{SkyBlue2}{rgb}{0.49,0.75,0.93}
\definecolor{SkyBlue3}{rgb}{0.42,0.65,0.80}
\definecolor{SkyBlue4}{rgb}{0.29,0.44,0.55}
\definecolor{SkyBlue}{rgb}{0.53,0.81,0.92}
\definecolor{SlateBlue1}{rgb}{0.51,0.44,1.00}
\definecolor{SlateBlue2}{rgb}{0.48,0.40,0.93}
\definecolor{SlateBlue3}{rgb}{0.41,0.35,0.80}
\definecolor{SlateBlue4}{rgb}{0.28,0.24,0.55}
\definecolor{SlateBlue}{rgb}{0.42,0.35,0.80}
\definecolor{SlateGray1}{rgb}{0.78,0.89,1.00}
\definecolor{SlateGray2}{rgb}{0.73,0.83,0.93}
\definecolor{SlateGray3}{rgb}{0.62,0.71,0.80}
\definecolor{SlateGray4}{rgb}{0.42,0.48,0.55}
\definecolor{SlateGray}{rgb}{0.44,0.50,0.56}
\definecolor{SlateGrey}{rgb}{0.44,0.50,0.56}
\definecolor{SpringGreen1}{rgb}{0.00,1.00,0.50}
\definecolor{SpringGreen2}{rgb}{0.00,0.93,0.46}
\definecolor{SpringGreen3}{rgb}{0.00,0.80,0.40}
\definecolor{SpringGreen4}{rgb}{0.00,0.55,0.27}
\definecolor{SpringGreen}{rgb}{0.00,1.00,0.50}
\definecolor{SteelBlue1}{rgb}{0.39,0.72,1.00}
\definecolor{SteelBlue2}{rgb}{0.36,0.67,0.93}
\definecolor{SteelBlue3}{rgb}{0.31,0.58,0.80}
\definecolor{SteelBlue4}{rgb}{0.21,0.39,0.55}
\definecolor{SteelBlue}{rgb}{0.27,0.51,0.71}
\definecolor{VioletRed1}{rgb}{1.00,0.24,0.59}
\definecolor{VioletRed2}{rgb}{0.93,0.23,0.55}
\definecolor{VioletRed3}{rgb}{0.80,0.20,0.47}
\definecolor{VioletRed4}{rgb}{0.55,0.13,0.32}
\definecolor{VioletRed}{rgb}{0.82,0.13,0.56}
\definecolor{WhiteSmoke}{rgb}{0.96,0.96,0.96}
\definecolor{YellowGreen}{rgb}{0.60,0.80,0.20}
\definecolor{aliceblue}{rgb}{0.94,0.97,1.00}
\definecolor{antiquewhite}{rgb}{0.98,0.92,0.84}
\definecolor{aquamarine1}{rgb}{0.50,1.00,0.83}
\definecolor{aquamarine2}{rgb}{0.46,0.93,0.78}
\definecolor{aquamarine3}{rgb}{0.40,0.80,0.67}
\definecolor{aquamarine4}{rgb}{0.27,0.55,0.45}
\definecolor{aquamarine}{rgb}{0.50,1.00,0.83}
\definecolor{azure1}{rgb}{0.94,1.00,1.00}
\definecolor{azure2}{rgb}{0.88,0.93,0.93}
\definecolor{azure3}{rgb}{0.76,0.80,0.80}
\definecolor{azure4}{rgb}{0.51,0.55,0.55}
\definecolor{azure}{rgb}{0.94,1.00,1.00}
\definecolor{beige}{rgb}{0.96,0.96,0.86}
\definecolor{bisque1}{rgb}{1.00,0.89,0.77}
\definecolor{bisque2}{rgb}{0.93,0.84,0.72}
\definecolor{bisque3}{rgb}{0.80,0.72,0.62}
\definecolor{bisque4}{rgb}{0.55,0.49,0.42}
\definecolor{bisque}{rgb}{1.00,0.89,0.77}
\definecolor{black}{rgb}{0.00,0.00,0.00}
\definecolor{blanchedalmond}{rgb}{1.00,0.92,0.80}
\definecolor{blue1}{rgb}{0.00,0.00,1.00}
\definecolor{blue2}{rgb}{0.00,0.00,0.93}
\definecolor{blue3}{rgb}{0.00,0.00,0.80}
\definecolor{blue4}{rgb}{0.00,0.00,0.55}
\definecolor{blueviolet}{rgb}{0.54,0.17,0.89}
\definecolor{blue}{rgb}{0.00,0.00,1.00}
\definecolor{brown1}{rgb}{1.00,0.25,0.25}
\definecolor{brown2}{rgb}{0.93,0.23,0.23}
\definecolor{brown3}{rgb}{0.80,0.20,0.20}
\definecolor{brown4}{rgb}{0.55,0.14,0.14}
\definecolor{brown}{rgb}{0.65,0.16,0.16}
\definecolor{burlywood1}{rgb}{1.00,0.83,0.61}
\definecolor{burlywood2}{rgb}{0.93,0.77,0.57}
\definecolor{burlywood3}{rgb}{0.80,0.67,0.49}
\definecolor{burlywood4}{rgb}{0.55,0.45,0.33}
\definecolor{burlywood}{rgb}{0.87,0.72,0.53}
\definecolor{cadetblue}{rgb}{0.37,0.62,0.63}
\definecolor{chartreuse1}{rgb}{0.50,1.00,0.00}
\definecolor{chartreuse2}{rgb}{0.46,0.93,0.00}
\definecolor{chartreuse3}{rgb}{0.40,0.80,0.00}
\definecolor{chartreuse4}{rgb}{0.27,0.55,0.00}
\definecolor{chartreuse}{rgb}{0.50,1.00,0.00}
\definecolor{chocolate1}{rgb}{1.00,0.50,0.14}
\definecolor{chocolate2}{rgb}{0.93,0.46,0.13}
\definecolor{chocolate3}{rgb}{0.80,0.40,0.11}
\definecolor{chocolate4}{rgb}{0.55,0.27,0.07}
\definecolor{chocolate}{rgb}{0.82,0.41,0.12}
\definecolor{coral1}{rgb}{1.00,0.45,0.34}
\definecolor{coral2}{rgb}{0.93,0.42,0.31}
\definecolor{coral3}{rgb}{0.80,0.36,0.27}
\definecolor{coral4}{rgb}{0.55,0.24,0.18}
\definecolor{coral}{rgb}{1.00,0.50,0.31}
\definecolor{cornflowerblue}{rgb}{0.39,0.58,0.93}
\definecolor{cornsilk1}{rgb}{1.00,0.97,0.86}
\definecolor{cornsilk2}{rgb}{0.93,0.91,0.80}
\definecolor{cornsilk3}{rgb}{0.80,0.78,0.69}
\definecolor{cornsilk4}{rgb}{0.55,0.53,0.47}
\definecolor{cornsilk}{rgb}{1.00,0.97,0.86}
\definecolor{cyan1}{rgb}{0.00,1.00,1.00}
\definecolor{cyan2}{rgb}{0.00,0.93,0.93}
\definecolor{cyan3}{rgb}{0.00,0.80,0.80}
\definecolor{cyan4}{rgb}{0.00,0.55,0.55}
\definecolor{cyan}{rgb}{0.00,1.00,1.00}
\definecolor{darkblue}{rgb}{0.00,0.00,0.55}
\definecolor{darkcyan}{rgb}{0.00,0.55,0.55}
\definecolor{darkgoldenrod}{rgb}{0.72,0.53,0.04}
\definecolor{darkgray}{rgb}{0.66,0.66,0.66}
\definecolor{darkgreen}{rgb}{0.00,0.39,0.00}
\definecolor{darkgrey}{rgb}{0.66,0.66,0.66}
\definecolor{darkkhaki}{rgb}{0.74,0.72,0.42}
\definecolor{darkmagenta}{rgb}{0.55,0.00,0.55}
\definecolor{darkolive}{rgb}{0.33,0.42,0.18}
\definecolor{darkorange}{rgb}{1.00,0.55,0.00}
\definecolor{darkorchid}{rgb}{0.60,0.20,0.80}
\definecolor{darkred}{rgb}{0.55,0.00,0.00}
\definecolor{darksalmon}{rgb}{0.91,0.59,0.48}
\definecolor{darksea}{rgb}{0.56,0.74,0.56}
\definecolor{darkslate}{rgb}{0.18,0.31,0.31}
\definecolor{darkslate}{rgb}{0.18,0.31,0.31}
\definecolor{darkslate}{rgb}{0.28,0.24,0.55}
\definecolor{darkturquoise}{rgb}{0.00,0.81,0.82}
\definecolor{darkviolet}{rgb}{0.58,0.00,0.83}
\definecolor{deeppink}{rgb}{1.00,0.08,0.58}
\definecolor{deepsky}{rgb}{0.00,0.75,1.00}
\definecolor{dimgray}{rgb}{0.41,0.41,0.41}
\definecolor{dimgrey}{rgb}{0.41,0.41,0.41}
\definecolor{dodgerblue}{rgb}{0.12,0.56,1.00}
\definecolor{firebrick1}{rgb}{1.00,0.19,0.19}
\definecolor{firebrick2}{rgb}{0.93,0.17,0.17}
\definecolor{firebrick3}{rgb}{0.80,0.15,0.15}
\definecolor{firebrick4}{rgb}{0.55,0.10,0.10}
\definecolor{firebrick}{rgb}{0.70,0.13,0.13}
\definecolor{floralwhite}{rgb}{1.00,0.98,0.94}
\definecolor{forestgreen}{rgb}{0.13,0.55,0.13}
\definecolor{gainsboro}{rgb}{0.86,0.86,0.86}
\definecolor{ghostwhite}{rgb}{0.97,0.97,1.00}
\definecolor{gold1}{rgb}{1.00,0.84,0.00}
\definecolor{gold2}{rgb}{0.93,0.79,0.00}
\definecolor{gold3}{rgb}{0.80,0.68,0.00}
\definecolor{gold4}{rgb}{0.55,0.46,0.00}
\definecolor{goldenrod1}{rgb}{1.00,0.76,0.15}
\definecolor{goldenrod2}{rgb}{0.93,0.71,0.13}
\definecolor{goldenrod3}{rgb}{0.80,0.61,0.11}
\definecolor{goldenrod4}{rgb}{0.55,0.41,0.08}
\definecolor{goldenrod}{rgb}{0.85,0.65,0.13}
\definecolor{gold}{rgb}{1.00,0.84,0.00}
\definecolor{gray0}{rgb}{0.00,0.00,0.00}
\definecolor{gray100}{rgb}{1.00,1.00,1.00}
\definecolor{gray10}{rgb}{0.10,0.10,0.10}
\definecolor{gray11}{rgb}{0.11,0.11,0.11}
\definecolor{gray12}{rgb}{0.12,0.12,0.12}
\definecolor{gray13}{rgb}{0.13,0.13,0.13}
\definecolor{gray14}{rgb}{0.14,0.14,0.14}
\definecolor{gray15}{rgb}{0.15,0.15,0.15}
\definecolor{gray16}{rgb}{0.16,0.16,0.16}
\definecolor{gray17}{rgb}{0.17,0.17,0.17}
\definecolor{gray18}{rgb}{0.18,0.18,0.18}
\definecolor{gray19}{rgb}{0.19,0.19,0.19}
\definecolor{gray1}{rgb}{0.01,0.01,0.01}
\definecolor{gray20}{rgb}{0.20,0.20,0.20}
\definecolor{gray21}{rgb}{0.21,0.21,0.21}
\definecolor{gray22}{rgb}{0.22,0.22,0.22}
\definecolor{gray23}{rgb}{0.23,0.23,0.23}
\definecolor{gray24}{rgb}{0.24,0.24,0.24}
\definecolor{gray25}{rgb}{0.25,0.25,0.25}
\definecolor{gray26}{rgb}{0.26,0.26,0.26}
\definecolor{gray27}{rgb}{0.27,0.27,0.27}
\definecolor{gray28}{rgb}{0.28,0.28,0.28}
\definecolor{gray29}{rgb}{0.29,0.29,0.29}
\definecolor{gray2}{rgb}{0.02,0.02,0.02}
\definecolor{gray30}{rgb}{0.30,0.30,0.30}
\definecolor{gray31}{rgb}{0.31,0.31,0.31}
\definecolor{gray32}{rgb}{0.32,0.32,0.32}
\definecolor{gray33}{rgb}{0.33,0.33,0.33}
\definecolor{gray34}{rgb}{0.34,0.34,0.34}
\definecolor{gray35}{rgb}{0.35,0.35,0.35}
\definecolor{gray36}{rgb}{0.36,0.36,0.36}
\definecolor{gray37}{rgb}{0.37,0.37,0.37}
\definecolor{gray38}{rgb}{0.38,0.38,0.38}
\definecolor{gray39}{rgb}{0.39,0.39,0.39}
\definecolor{gray3}{rgb}{0.03,0.03,0.03}
\definecolor{gray40}{rgb}{0.40,0.40,0.40}
\definecolor{gray41}{rgb}{0.41,0.41,0.41}
\definecolor{gray42}{rgb}{0.42,0.42,0.42}
\definecolor{gray43}{rgb}{0.43,0.43,0.43}
\definecolor{gray44}{rgb}{0.44,0.44,0.44}
\definecolor{gray45}{rgb}{0.45,0.45,0.45}
\definecolor{gray46}{rgb}{0.46,0.46,0.46}
\definecolor{gray47}{rgb}{0.47,0.47,0.47}
\definecolor{gray48}{rgb}{0.48,0.48,0.48}
\definecolor{gray49}{rgb}{0.49,0.49,0.49}
\definecolor{gray4}{rgb}{0.04,0.04,0.04}
\definecolor{gray50}{rgb}{0.50,0.50,0.50}
\definecolor{gray51}{rgb}{0.51,0.51,0.51}
\definecolor{gray52}{rgb}{0.52,0.52,0.52}
\definecolor{gray53}{rgb}{0.53,0.53,0.53}
\definecolor{gray54}{rgb}{0.54,0.54,0.54}
\definecolor{gray55}{rgb}{0.55,0.55,0.55}
\definecolor{gray56}{rgb}{0.56,0.56,0.56}
\definecolor{gray57}{rgb}{0.57,0.57,0.57}
\definecolor{gray58}{rgb}{0.58,0.58,0.58}
\definecolor{gray59}{rgb}{0.59,0.59,0.59}
\definecolor{gray5}{rgb}{0.05,0.05,0.05}
\definecolor{gray60}{rgb}{0.60,0.60,0.60}
\definecolor{gray61}{rgb}{0.61,0.61,0.61}
\definecolor{gray62}{rgb}{0.62,0.62,0.62}
\definecolor{gray63}{rgb}{0.63,0.63,0.63}
\definecolor{gray64}{rgb}{0.64,0.64,0.64}
\definecolor{gray65}{rgb}{0.65,0.65,0.65}
\definecolor{gray66}{rgb}{0.66,0.66,0.66}
\definecolor{gray67}{rgb}{0.67,0.67,0.67}
\definecolor{gray68}{rgb}{0.68,0.68,0.68}
\definecolor{gray69}{rgb}{0.69,0.69,0.69}
\definecolor{gray6}{rgb}{0.06,0.06,0.06}
\definecolor{gray70}{rgb}{0.70,0.70,0.70}
\definecolor{gray71}{rgb}{0.71,0.71,0.71}
\definecolor{gray72}{rgb}{0.72,0.72,0.72}
\definecolor{gray73}{rgb}{0.73,0.73,0.73}
\definecolor{gray74}{rgb}{0.74,0.74,0.74}
\definecolor{gray75}{rgb}{0.75,0.75,0.75}
\definecolor{gray76}{rgb}{0.76,0.76,0.76}
\definecolor{gray77}{rgb}{0.77,0.77,0.77}
\definecolor{gray78}{rgb}{0.78,0.78,0.78}
\definecolor{gray79}{rgb}{0.79,0.79,0.79}
\definecolor{gray7}{rgb}{0.07,0.07,0.07}
\definecolor{gray80}{rgb}{0.80,0.80,0.80}
\definecolor{gray81}{rgb}{0.81,0.81,0.81}
\definecolor{gray82}{rgb}{0.82,0.82,0.82}
\definecolor{gray83}{rgb}{0.83,0.83,0.83}
\definecolor{gray84}{rgb}{0.84,0.84,0.84}
\definecolor{gray85}{rgb}{0.85,0.85,0.85}
\definecolor{gray86}{rgb}{0.86,0.86,0.86}
\definecolor{gray87}{rgb}{0.87,0.87,0.87}
\definecolor{gray88}{rgb}{0.88,0.88,0.88}
\definecolor{gray89}{rgb}{0.89,0.89,0.89}
\definecolor{gray8}{rgb}{0.08,0.08,0.08}
\definecolor{gray90}{rgb}{0.90,0.90,0.90}
\definecolor{gray91}{rgb}{0.91,0.91,0.91}
\definecolor{gray92}{rgb}{0.92,0.92,0.92}
\definecolor{gray93}{rgb}{0.93,0.93,0.93}
\definecolor{gray94}{rgb}{0.94,0.94,0.94}
\definecolor{gray95}{rgb}{0.95,0.95,0.95}
\definecolor{gray96}{rgb}{0.96,0.96,0.96}
\definecolor{gray97}{rgb}{0.97,0.97,0.97}
\definecolor{gray98}{rgb}{0.98,0.98,0.98}
\definecolor{gray99}{rgb}{0.99,0.99,0.99}
\definecolor{gray9}{rgb}{0.09,0.09,0.09}
\definecolor{gray}{rgb}{0.75,0.75,0.75}
\definecolor{green1}{rgb}{0.00,1.00,0.00}
\definecolor{green2}{rgb}{0.00,0.93,0.00}
\definecolor{green3}{rgb}{0.00,0.80,0.00}
\definecolor{green4}{rgb}{0.00,0.55,0.00}
\definecolor{greenyellow}{rgb}{0.68,1.00,0.18}
\definecolor{green}{rgb}{0.00,1.00,0.00}
\definecolor{grey0}{rgb}{0.00,0.00,0.00}
\definecolor{grey100}{rgb}{1.00,1.00,1.00}
\definecolor{grey10}{rgb}{0.10,0.10,0.10}
\definecolor{grey11}{rgb}{0.11,0.11,0.11}
\definecolor{grey12}{rgb}{0.12,0.12,0.12}
\definecolor{grey13}{rgb}{0.13,0.13,0.13}
\definecolor{grey14}{rgb}{0.14,0.14,0.14}
\definecolor{grey15}{rgb}{0.15,0.15,0.15}
\definecolor{grey16}{rgb}{0.16,0.16,0.16}
\definecolor{grey17}{rgb}{0.17,0.17,0.17}
\definecolor{grey18}{rgb}{0.18,0.18,0.18}
\definecolor{grey19}{rgb}{0.19,0.19,0.19}
\definecolor{grey1}{rgb}{0.01,0.01,0.01}
\definecolor{grey20}{rgb}{0.20,0.20,0.20}
\definecolor{grey21}{rgb}{0.21,0.21,0.21}
\definecolor{grey22}{rgb}{0.22,0.22,0.22}
\definecolor{grey23}{rgb}{0.23,0.23,0.23}
\definecolor{grey24}{rgb}{0.24,0.24,0.24}
\definecolor{grey25}{rgb}{0.25,0.25,0.25}
\definecolor{grey26}{rgb}{0.26,0.26,0.26}
\definecolor{grey27}{rgb}{0.27,0.27,0.27}
\definecolor{grey28}{rgb}{0.28,0.28,0.28}
\definecolor{grey29}{rgb}{0.29,0.29,0.29}
\definecolor{grey2}{rgb}{0.02,0.02,0.02}
\definecolor{grey30}{rgb}{0.30,0.30,0.30}
\definecolor{grey31}{rgb}{0.31,0.31,0.31}
\definecolor{grey32}{rgb}{0.32,0.32,0.32}
\definecolor{grey33}{rgb}{0.33,0.33,0.33}
\definecolor{grey34}{rgb}{0.34,0.34,0.34}
\definecolor{grey35}{rgb}{0.35,0.35,0.35}
\definecolor{grey36}{rgb}{0.36,0.36,0.36}
\definecolor{grey37}{rgb}{0.37,0.37,0.37}
\definecolor{grey38}{rgb}{0.38,0.38,0.38}
\definecolor{grey39}{rgb}{0.39,0.39,0.39}
\definecolor{grey3}{rgb}{0.03,0.03,0.03}
\definecolor{grey40}{rgb}{0.40,0.40,0.40}
\definecolor{grey41}{rgb}{0.41,0.41,0.41}
\definecolor{grey42}{rgb}{0.42,0.42,0.42}
\definecolor{grey43}{rgb}{0.43,0.43,0.43}
\definecolor{grey44}{rgb}{0.44,0.44,0.44}
\definecolor{grey45}{rgb}{0.45,0.45,0.45}
\definecolor{grey46}{rgb}{0.46,0.46,0.46}
\definecolor{grey47}{rgb}{0.47,0.47,0.47}
\definecolor{grey48}{rgb}{0.48,0.48,0.48}
\definecolor{grey49}{rgb}{0.49,0.49,0.49}
\definecolor{grey4}{rgb}{0.04,0.04,0.04}
\definecolor{grey50}{rgb}{0.50,0.50,0.50}
\definecolor{grey51}{rgb}{0.51,0.51,0.51}
\definecolor{grey52}{rgb}{0.52,0.52,0.52}
\definecolor{grey53}{rgb}{0.53,0.53,0.53}
\definecolor{grey54}{rgb}{0.54,0.54,0.54}
\definecolor{grey55}{rgb}{0.55,0.55,0.55}
\definecolor{grey56}{rgb}{0.56,0.56,0.56}
\definecolor{grey57}{rgb}{0.57,0.57,0.57}
\definecolor{grey58}{rgb}{0.58,0.58,0.58}
\definecolor{grey59}{rgb}{0.59,0.59,0.59}
\definecolor{grey5}{rgb}{0.05,0.05,0.05}
\definecolor{grey60}{rgb}{0.60,0.60,0.60}
\definecolor{grey61}{rgb}{0.61,0.61,0.61}
\definecolor{grey62}{rgb}{0.62,0.62,0.62}
\definecolor{grey63}{rgb}{0.63,0.63,0.63}
\definecolor{grey64}{rgb}{0.64,0.64,0.64}
\definecolor{grey65}{rgb}{0.65,0.65,0.65}
\definecolor{grey66}{rgb}{0.66,0.66,0.66}
\definecolor{grey67}{rgb}{0.67,0.67,0.67}
\definecolor{grey68}{rgb}{0.68,0.68,0.68}
\definecolor{grey69}{rgb}{0.69,0.69,0.69}
\definecolor{grey6}{rgb}{0.06,0.06,0.06}
\definecolor{grey70}{rgb}{0.70,0.70,0.70}
\definecolor{grey71}{rgb}{0.71,0.71,0.71}
\definecolor{grey72}{rgb}{0.72,0.72,0.72}
\definecolor{grey73}{rgb}{0.73,0.73,0.73}
\definecolor{grey74}{rgb}{0.74,0.74,0.74}
\definecolor{grey75}{rgb}{0.75,0.75,0.75}
\definecolor{grey76}{rgb}{0.76,0.76,0.76}
\definecolor{grey77}{rgb}{0.77,0.77,0.77}
\definecolor{grey78}{rgb}{0.78,0.78,0.78}
\definecolor{grey79}{rgb}{0.79,0.79,0.79}
\definecolor{grey7}{rgb}{0.07,0.07,0.07}
\definecolor{grey80}{rgb}{0.80,0.80,0.80}
\definecolor{grey81}{rgb}{0.81,0.81,0.81}
\definecolor{grey82}{rgb}{0.82,0.82,0.82}
\definecolor{grey83}{rgb}{0.83,0.83,0.83}
\definecolor{grey84}{rgb}{0.84,0.84,0.84}
\definecolor{grey85}{rgb}{0.85,0.85,0.85}
\definecolor{grey86}{rgb}{0.86,0.86,0.86}
\definecolor{grey87}{rgb}{0.87,0.87,0.87}
\definecolor{grey88}{rgb}{0.88,0.88,0.88}
\definecolor{grey89}{rgb}{0.89,0.89,0.89}
\definecolor{grey8}{rgb}{0.08,0.08,0.08}
\definecolor{grey90}{rgb}{0.90,0.90,0.90}
\definecolor{grey91}{rgb}{0.91,0.91,0.91}
\definecolor{grey92}{rgb}{0.92,0.92,0.92}
\definecolor{grey93}{rgb}{0.93,0.93,0.93}
\definecolor{grey94}{rgb}{0.94,0.94,0.94}
\definecolor{grey95}{rgb}{0.95,0.95,0.95}
\definecolor{grey96}{rgb}{0.96,0.96,0.96}
\definecolor{grey97}{rgb}{0.97,0.97,0.97}
\definecolor{grey98}{rgb}{0.98,0.98,0.98}
\definecolor{grey99}{rgb}{0.99,0.99,0.99}
\definecolor{grey9}{rgb}{0.09,0.09,0.09}
\definecolor{grey}{rgb}{0.75,0.75,0.75}
\definecolor{honeydew1}{rgb}{0.94,1.00,0.94}
\definecolor{honeydew2}{rgb}{0.88,0.93,0.88}
\definecolor{honeydew3}{rgb}{0.76,0.80,0.76}
\definecolor{honeydew4}{rgb}{0.51,0.55,0.51}
\definecolor{honeydew}{rgb}{0.94,1.00,0.94}
\definecolor{hotpink}{rgb}{1.00,0.41,0.71}
\definecolor{indianred}{rgb}{0.80,0.36,0.36}
\definecolor{ivory1}{rgb}{1.00,1.00,0.94}
\definecolor{ivory2}{rgb}{0.93,0.93,0.88}
\definecolor{ivory3}{rgb}{0.80,0.80,0.76}
\definecolor{ivory4}{rgb}{0.55,0.55,0.51}
\definecolor{ivory}{rgb}{1.00,1.00,0.94}
\definecolor{khaki1}{rgb}{1.00,0.96,0.56}
\definecolor{khaki2}{rgb}{0.93,0.90,0.52}
\definecolor{khaki3}{rgb}{0.80,0.78,0.45}
\definecolor{khaki4}{rgb}{0.55,0.53,0.31}
\definecolor{khaki}{rgb}{0.94,0.90,0.55}
\definecolor{lavenderblush}{rgb}{1.00,0.94,0.96}
\definecolor{lavender}{rgb}{0.90,0.90,0.98}
\definecolor{lawngreen}{rgb}{0.49,0.99,0.00}
\definecolor{lemonchiffon}{rgb}{1.00,0.98,0.80}
\definecolor{lightblue}{rgb}{0.68,0.85,0.90}
\definecolor{lightcoral}{rgb}{0.94,0.50,0.50}
\definecolor{lightcyan}{rgb}{0.88,1.00,1.00}
\definecolor{lightgoldenrod}{rgb}{0.93,0.87,0.51}
\definecolor{lightgoldenrod}{rgb}{0.98,0.98,0.82}
\definecolor{lightgray}{rgb}{0.83,0.83,0.83}
\definecolor{lightgreen}{rgb}{0.56,0.93,0.56}
\definecolor{lightgrey}{rgb}{0.83,0.83,0.83}
\definecolor{lightpink}{rgb}{1.00,0.71,0.76}
\definecolor{lightsalmon}{rgb}{1.00,0.63,0.48}
\definecolor{lightsea}{rgb}{0.13,0.70,0.67}
\definecolor{lightsky}{rgb}{0.53,0.81,0.98}
\definecolor{lightslate}{rgb}{0.47,0.53,0.60}
\definecolor{lightslate}{rgb}{0.47,0.53,0.60}
\definecolor{lightslate}{rgb}{0.52,0.44,1.00}
\definecolor{lightsteel}{rgb}{0.69,0.77,0.87}
\definecolor{lightyellow}{rgb}{1.00,1.00,0.88}
\definecolor{limegreen}{rgb}{0.20,0.80,0.20}
\definecolor{linen}{rgb}{0.98,0.94,0.90}
\definecolor{magenta1}{rgb}{1.00,0.00,1.00}
\definecolor{magenta2}{rgb}{0.93,0.00,0.93}
\definecolor{magenta3}{rgb}{0.80,0.00,0.80}
\definecolor{magenta4}{rgb}{0.55,0.00,0.55}
\definecolor{magenta}{rgb}{1.00,0.00,1.00}
\definecolor{maroon1}{rgb}{1.00,0.20,0.70}
\definecolor{maroon2}{rgb}{0.93,0.19,0.65}
\definecolor{maroon3}{rgb}{0.80,0.16,0.56}
\definecolor{maroon4}{rgb}{0.55,0.11,0.38}
\definecolor{maroon}{rgb}{0.69,0.19,0.38}
\definecolor{mediumaquamarine}{rgb}{0.40,0.80,0.67}
\definecolor{mediumblue}{rgb}{0.00,0.00,0.80}
\definecolor{mediumorchid}{rgb}{0.73,0.33,0.83}
\definecolor{mediumpurple}{rgb}{0.58,0.44,0.86}
\definecolor{mediumsea}{rgb}{0.24,0.70,0.44}
\definecolor{mediumslate}{rgb}{0.48,0.41,0.93}
\definecolor{mediumspring}{rgb}{0.00,0.98,0.60}
\definecolor{mediumturquoise}{rgb}{0.28,0.82,0.80}
\definecolor{mediumviolet}{rgb}{0.78,0.08,0.52}
\definecolor{midnightblue}{rgb}{0.10,0.10,0.44}
\definecolor{mintcream}{rgb}{0.96,1.00,0.98}
\definecolor{mistyrose}{rgb}{1.00,0.89,0.88}
\definecolor{moccasin}{rgb}{1.00,0.89,0.71}
\definecolor{navajowhite}{rgb}{1.00,0.87,0.68}
\definecolor{navyblue}{rgb}{0.00,0.00,0.50}
\definecolor{navy}{rgb}{0.00,0.00,0.50}
\definecolor{oldlace}{rgb}{0.99,0.96,0.90}
\definecolor{olivedrab}{rgb}{0.42,0.56,0.14}
\definecolor{orange1}{rgb}{1.00,0.65,0.00}
\definecolor{orange2}{rgb}{0.93,0.60,0.00}
\definecolor{orange3}{rgb}{0.80,0.52,0.00}
\definecolor{orange4}{rgb}{0.55,0.35,0.00}
\definecolor{orangered}{rgb}{1.00,0.27,0.00}
\definecolor{orange}{rgb}{1.00,0.65,0.00}
\definecolor{orchid1}{rgb}{1.00,0.51,0.98}
\definecolor{orchid2}{rgb}{0.93,0.48,0.91}
\definecolor{orchid3}{rgb}{0.80,0.41,0.79}
\definecolor{orchid4}{rgb}{0.55,0.28,0.54}
\definecolor{orchid}{rgb}{0.85,0.44,0.84}
\definecolor{palegoldenrod}{rgb}{0.93,0.91,0.67}
\definecolor{palegreen}{rgb}{0.60,0.98,0.60}
\definecolor{paleturquoise}{rgb}{0.69,0.93,0.93}
\definecolor{paleviolet}{rgb}{0.86,0.44,0.58}
\definecolor{papayawhip}{rgb}{1.00,0.94,0.84}
\definecolor{peachpuff}{rgb}{1.00,0.85,0.73}
\definecolor{peru}{rgb}{0.80,0.52,0.25}
\definecolor{pink1}{rgb}{1.00,0.71,0.77}
\definecolor{pink2}{rgb}{0.93,0.66,0.72}
\definecolor{pink3}{rgb}{0.80,0.57,0.62}
\definecolor{pink4}{rgb}{0.55,0.39,0.42}
\definecolor{pink}{rgb}{1.00,0.75,0.80}
\definecolor{plum1}{rgb}{1.00,0.73,1.00}
\definecolor{plum2}{rgb}{0.93,0.68,0.93}
\definecolor{plum3}{rgb}{0.80,0.59,0.80}
\definecolor{plum4}{rgb}{0.55,0.40,0.55}
\definecolor{plum}{rgb}{0.87,0.63,0.87}
\definecolor{powderblue}{rgb}{0.69,0.88,0.90}
\definecolor{purple1}{rgb}{0.61,0.19,1.00}
\definecolor{purple2}{rgb}{0.57,0.17,0.93}
\definecolor{purple3}{rgb}{0.49,0.15,0.80}
\definecolor{purple4}{rgb}{0.33,0.10,0.55}
\definecolor{purple}{rgb}{0.63,0.13,0.94}
\definecolor{red1}{rgb}{1.00,0.00,0.00}
\definecolor{red2}{rgb}{0.93,0.00,0.00}
\definecolor{red3}{rgb}{0.80,0.00,0.00}
\definecolor{red4}{rgb}{0.55,0.00,0.00}
\definecolor{red}{rgb}{1.00,0.00,0.00}
\definecolor{rosybrown}{rgb}{0.74,0.56,0.56}
\definecolor{royalblue}{rgb}{0.25,0.41,0.88}
\definecolor{saddlebrown}{rgb}{0.55,0.27,0.07}
\definecolor{salmon1}{rgb}{1.00,0.55,0.41}
\definecolor{salmon2}{rgb}{0.93,0.51,0.38}
\definecolor{salmon3}{rgb}{0.80,0.44,0.33}
\definecolor{salmon4}{rgb}{0.55,0.30,0.22}
\definecolor{salmon}{rgb}{0.98,0.50,0.45}
\definecolor{sandybrown}{rgb}{0.96,0.64,0.38}
\definecolor{seagreen}{rgb}{0.18,0.55,0.34}
\definecolor{seashell1}{rgb}{1.00,0.96,0.93}
\definecolor{seashell2}{rgb}{0.93,0.90,0.87}
\definecolor{seashell3}{rgb}{0.80,0.77,0.75}
\definecolor{seashell4}{rgb}{0.55,0.53,0.51}
\definecolor{seashell}{rgb}{1.00,0.96,0.93}
\definecolor{sienna1}{rgb}{1.00,0.51,0.28}
\definecolor{sienna2}{rgb}{0.93,0.47,0.26}
\definecolor{sienna3}{rgb}{0.80,0.41,0.22}
\definecolor{sienna4}{rgb}{0.55,0.28,0.15}
\definecolor{sienna}{rgb}{0.63,0.32,0.18}
\definecolor{skyblue}{rgb}{0.53,0.81,0.92}
\definecolor{slateblue}{rgb}{0.42,0.35,0.80}
\definecolor{slategray}{rgb}{0.44,0.50,0.56}
\definecolor{slategrey}{rgb}{0.44,0.50,0.56}
\definecolor{snow1}{rgb}{1.00,0.98,0.98}
\definecolor{snow2}{rgb}{0.93,0.91,0.91}
\definecolor{snow3}{rgb}{0.80,0.79,0.79}
\definecolor{snow4}{rgb}{0.55,0.54,0.54}
\definecolor{snow}{rgb}{1.00,0.98,0.98}
\definecolor{springgreen}{rgb}{0.00,1.00,0.50}
\definecolor{steelblue}{rgb}{0.27,0.51,0.71}
\definecolor{tan1}{rgb}{1.00,0.65,0.31}
\definecolor{tan2}{rgb}{0.93,0.60,0.29}
\definecolor{tan3}{rgb}{0.80,0.52,0.25}
\definecolor{tan4}{rgb}{0.55,0.35,0.17}
\definecolor{tan}{rgb}{0.82,0.71,0.55}
\definecolor{thistle1}{rgb}{1.00,0.88,1.00}
\definecolor{thistle2}{rgb}{0.93,0.82,0.93}
\definecolor{thistle3}{rgb}{0.80,0.71,0.80}
\definecolor{thistle4}{rgb}{0.55,0.48,0.55}
\definecolor{thistle}{rgb}{0.85,0.75,0.85}
\definecolor{tomato1}{rgb}{1.00,0.39,0.28}
\definecolor{tomato2}{rgb}{0.93,0.36,0.26}
\definecolor{tomato3}{rgb}{0.80,0.31,0.22}
\definecolor{tomato4}{rgb}{0.55,0.21,0.15}
\definecolor{tomato}{rgb}{1.00,0.39,0.28}
\definecolor{turquoise1}{rgb}{0.00,0.96,1.00}
\definecolor{turquoise2}{rgb}{0.00,0.90,0.93}
\definecolor{turquoise3}{rgb}{0.00,0.77,0.80}
\definecolor{turquoise4}{rgb}{0.00,0.53,0.55}
\definecolor{turquoise}{rgb}{0.25,0.88,0.82}
\definecolor{violetred}{rgb}{0.82,0.13,0.56}
\definecolor{violet}{rgb}{0.93,0.51,0.93}
\definecolor{wheat1}{rgb}{1.00,0.91,0.73}
\definecolor{wheat2}{rgb}{0.93,0.85,0.68}
\definecolor{wheat3}{rgb}{0.80,0.73,0.59}
\definecolor{wheat4}{rgb}{0.55,0.49,0.40}
\definecolor{wheat}{rgb}{0.96,0.87,0.70}
\definecolor{whitesmoke}{rgb}{0.96,0.96,0.96}
\definecolor{white}{rgb}{1.00,1.00,1.00}
\definecolor{yellow1}{rgb}{1.00,1.00,0.00}
\definecolor{yellow2}{rgb}{0.93,0.93,0.00}
\definecolor{yellow3}{rgb}{0.80,0.80,0.00}
\definecolor{yellow4}{rgb}{0.55,0.55,0.00}
\definecolor{yellowgreen}{rgb}{0.60,0.80,0.20}
\definecolor{yellow}{rgb}{1.00,1.00,0.00}

\bibpunct[; ]{(}{)}{;}{a}{}{;}

\title[Peculiar early-type galaxies]
    {Peculiar early-type galaxies in the SDSS Stripe82}

\author[Sugata Kaviraj]
{Sugata Kaviraj\thanks{E-mail: s.kaviraj@imperial.ac.uk}$^{1,2,3}$\\
$^1$Blackett Laboratory, Imperial College London, London SW7 2AZ, UK\\
$^2$Mullard Space Science Laboratory, Holmbury St. Mary, Dorking,
Surrey RH5 6NT UK\\
$^3$Department of Physics, University of Oxford, Keble Road,
Oxford OX1 3RH, UK}

\begin{document}

\pagerange{\pageref{firstpage}--\pageref{lastpage}} \pubyear{2009}

\maketitle

\label{firstpage}


\begin{abstract}
We explore the properties of `peculiar' early-type galaxies (ETGs)
in the local Universe, that show (faint) morphological signatures
of recent interactions such as tidal tails, shells and dust lanes.
Standard-depth ($\sim51$s exposure) multi-colour galaxy images
from the Sloan Digital Sky Survey (SDSS) are combined with the
significantly ($\sim$2 mags) deeper {\color{black}monochromatic}
images from the public SDSS Stripe82 ($-50^{\circ} < \alpha <
59^{\circ}, -1.25^{\circ} < \delta < 1.25^{\circ}$) to extract,
through careful visual inspection, a robust sample of nearby
($z<0.05$), luminous ($M_r<-20.5$) ETGs, including a subset of
$\sim70$ peculiar systems. $\sim18$\% of ETGs exhibit signs of
disturbed morphologies (e.g. shells), while $\sim7$\% show
evidence of dust lanes and patches. An analysis of optical
emission-line ratios indicates that the fraction of peculiar ETGs
that are Seyferts or LINERs (19.4\%) is twice the corresponding
values in their relaxed counterparts (10.1\%). LINER-like emission
is the dominant type of nebular activity in all ETG classes,
plausibly driven by stellar photoionisation associated with recent
star formation. An analysis of UV-optical colours indicates that,
regardless of the luminosity range being considered, the fraction
of peculiar ETGs that have experienced star formation in the last
Gyr is a factor of $\sim1.5$ higher than that in their relaxed
counterparts. The spectro-photometric results strongly suggest
that the interactions that produce the morphological peculiarities
also induce low-level recent star formation which, based on the
recent literature, are likely to contribute a few percent of the
stellar mass over the last $\sim1$ Gyr. Peculiar ETGs
preferentially inhabit low-density environments (outskirts of
clusters, groups or the field), either due to high peculiar
velocities in clusters making merging unlikely or because shell
systems are disrupted through frequent interactions within a
cluster crossing time. {\color{black}The catalogue of galaxies
that forms the basis of this paper can be obtained at:
http://www.mssl.ucl.ac.uk/$\sim$ska/stripe82/skaviraj\_stripe82.dat
or on request from the author.}
\end{abstract}




\section{Introduction}
Given their dominance of the stellar mass density in the local
Universe (Bernardi et al. 2003a), understanding the formation and
subsequent evolution of early-type galaxies (ETGs) is a key
element of the modern galaxy evolution effort. Their obedience to
simple optical scaling relations, such as the Fundamental Plane
\citep[e.g.][]{Jorg1996,VD1996,Saglia1997} and the
colour-magnitude relation
\citep[e.g.][]{BLE92,Ellis1997,Stanford98,Gladders98,VD2000},
indicates that the bulk of the stellar populations in these
galaxies has formed at high redshift ($z>1$), possibly over short
timescales of $\sim1$ Gyr \citep{Thomas1999}. The mass assembly at
late epochs ($z<1$) is better probed using rest-frame UV data (Yi
et al. 2005; Kaviraj et al. 2007a, 2007b, 2008; Schawinski et al.
2007a), which is more sensitive to the small fractions of young
stars that are expected to form in early-types in the standard
model
\citep[e.g.][]{Cole2000,Hatton2003,Kaviraj2005a,deLucia2006,deLucia2007},
plausibly through repeated minor merger activity
\citep{Kaviraj2009,Kaviraj2010}.

{\color{black}In the context of the standard model, in which
mergers play an important role in the buildup of stellar mass in
present-day ETGs \citep[e.g.][]{Cole2000,Hatton2003,deLucia2006},
there is a need to study the role of galaxy interactions and to
explore the star formation that accompanies these events. The role
of mergers in the nearby Universe has recently been explored
through studies of close pairs \citep[e.g.][Darg et al.
2010a,b]{Patton2000,Lefevre2000,Patton2002,Lin2004,Bell2006,McIntosh2008,Bundy2009}
and quantitative estimates of structural morphology
\citep[e.g.][]{Conselice2003}. The tell-tale signatures of recent
interactions and their impact on the ETG population can be most
efficiently studied by exploring ETGs that exhibit morphological
peculiarities and many important studies have contributed to a
sizeable literature on peculiar ETGs. Early-type galaxies which
exhibit systems of shells \citep[see
e.g.][]{Malin1980,Malin1983a,Malin1983b,Schweizer1990,Schweizer1992,Sikkema2007}
have long been recognised as remnants of recent merging, most
probably through minor mergers with various orbital confiurations
\citep[see e.g.][]{Quinn1984,Prieur1988,Hernquist1992}. The
presence of dust lanes in spheroidal systems - NGC 5128 being the
archetype in the local Universe \citep[see e.g.][]{Graham1979} -
is thought to be the result of accretion of cold material from
gas-rich companions within the last few Gyrs \citep[see][and
references therein]{Hawarden1981}.}

With the advent of the Sloan Digital Sky Survey (SDSS; Abazajian
et al. 2009), representative samples of local ETGs have been
culled and their star formation histories (SFHs) explored using
photometric and spectroscopic diagnostics of star formation
activity (e.g. Bernardi et al. 2003a; Kaviraj et al. 2007b,
Kaviraj 2008). A key aspect of such work is the identification of
galaxy morphologies from survey images. Given the prodigious
output of modern observational surveys such as the SDSS, a useful
strategy to tackle the large volumes of data is often to use
proxies for galaxy morphology, such as colour (since early-types
dominate the red sequence) or pipeline parameters such as the
concentration index (since early-types have more strongly peaked
light profiles than spirals). Significant efforts have been made
to exploit the SDSS pipeline to separate the galaxy population
into its constituent morphological types and study the properties
of early-type systems (see e.g. Eisenstein et al. 2001; Bernardi
et al. 2003a,b,c; Park \& Choi 2005).

While pipeline classification of morphologies is reasonably
efficient, there has been a realisation that the most accurate
method of estimating galaxy morphology remains direct visual
inspection of galaxy images (e.g. Fukugita et al. 2007; Schawinski
et al. 2007; Lintott et al. 2008). While inspection of standard
SDSS images does allow us to identify early-type and late-type
systems with reasonable accuracy, at least in the redshift range
$z<0.1$ (see e.g. Lintott et al. 2008), a drawback is the
relatively low exposure time ($\sim51$ seconds) and the resultant
lack of depth. Deep images inevitably show richer morphological
detail and, in particular for early-type galaxies, often reveal
signatures of recent merging such as fans, shells and tidal debris
\citep{VD2005}.

In this study we aim to combine standard SDSS imaging with that in
the SDSS `Stripe82', a stripe along the celestial equator in the
Southern Galactic Cap $(-50^{\circ} < \alpha < 59^{\circ},
-1.25^{\circ} < \delta < 1.25^{\circ})$ that has been imaged
multiple times, both to allow the stacked data to reach fainter
magnitudes and as part of the SDSS Supernova Survey (Frieman et
al. 2008). The public Stripe82 imaging (released with the SDSS
DR7) offers an {\color{black}$r$-band} co-addition of 122 runs,
with individual areas of the $\sim300$ deg$^2$ stripe area having
been covered between 20 and 40 times (Abazajian et al. 2009). The
Stripe82 imaging reaches $\sim$2 mags deeper than single SDSS
scans {\color{black}(which have magnitude limits of $22.2$, $22.2$
and $21.3$ mags in the $g$, $r$ and $i$-bands respectively). The
$r$-band coadds used in this paper thus have a nominal depth of
$\sim24$ mags.}

The novelties of this paper are two-fold. First, we simultaneously
inspect standard multi-colour images from the SDSS DR7 and their
deeper Stripe82 counterparts to both identify ETGs and flag
members of the ETG population that show morphological signatures
of recent interactions (e.g. shells, tidal tails and dust lanes).
Second, we explore the spectro-photometric properties of the
peculiar ETG population (exploiting optical emission lines and
UV-optical colours) and study their local environments.

The catalogue presented here is expected to have higher purity in
the morphological classifications than its predecessors, with the
added advantage of identifying ETGs that exhibit morphological
signatures of recent interactions. It provides a useful reference
sample of peculiar ETGs from the SDSS, with a homogeneous set of
spectro-photometric products available from the SDSS pipeline,
which is suitable for follow up using current and future
instruments. Forthcoming companion papers will explore the
structural peculiarities such as shell systems and dust lanes in
more detail and correlate them with the star formation and
chemical enrichment histories of the individual galaxies.


\begin{figure*}
\begin{minipage}{172mm}
\begin{center}
$\begin{array}{cccc}

\includegraphics[width=1.4in]{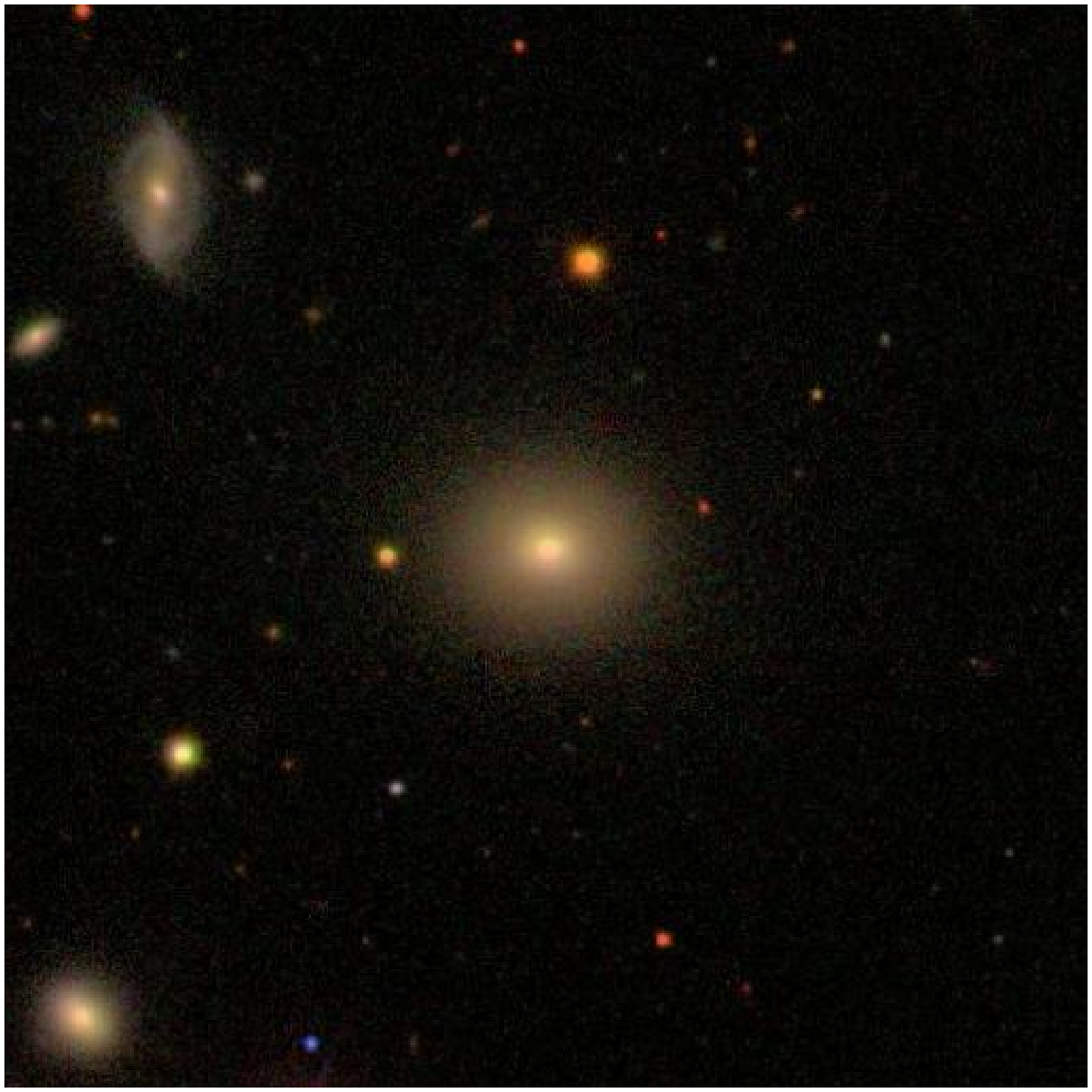}&\includegraphics[width=1.4in]{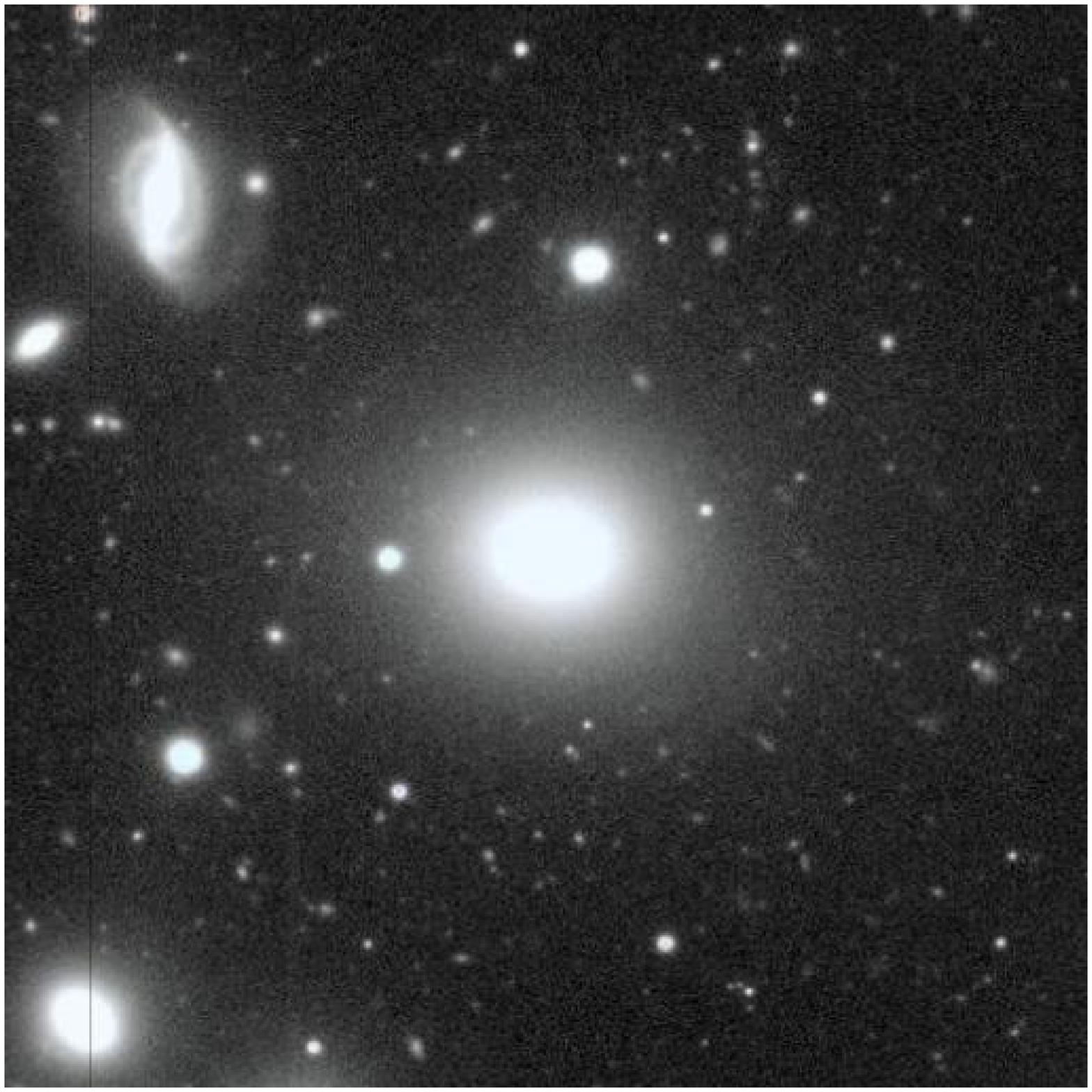}&
\includegraphics[width=1.4in]{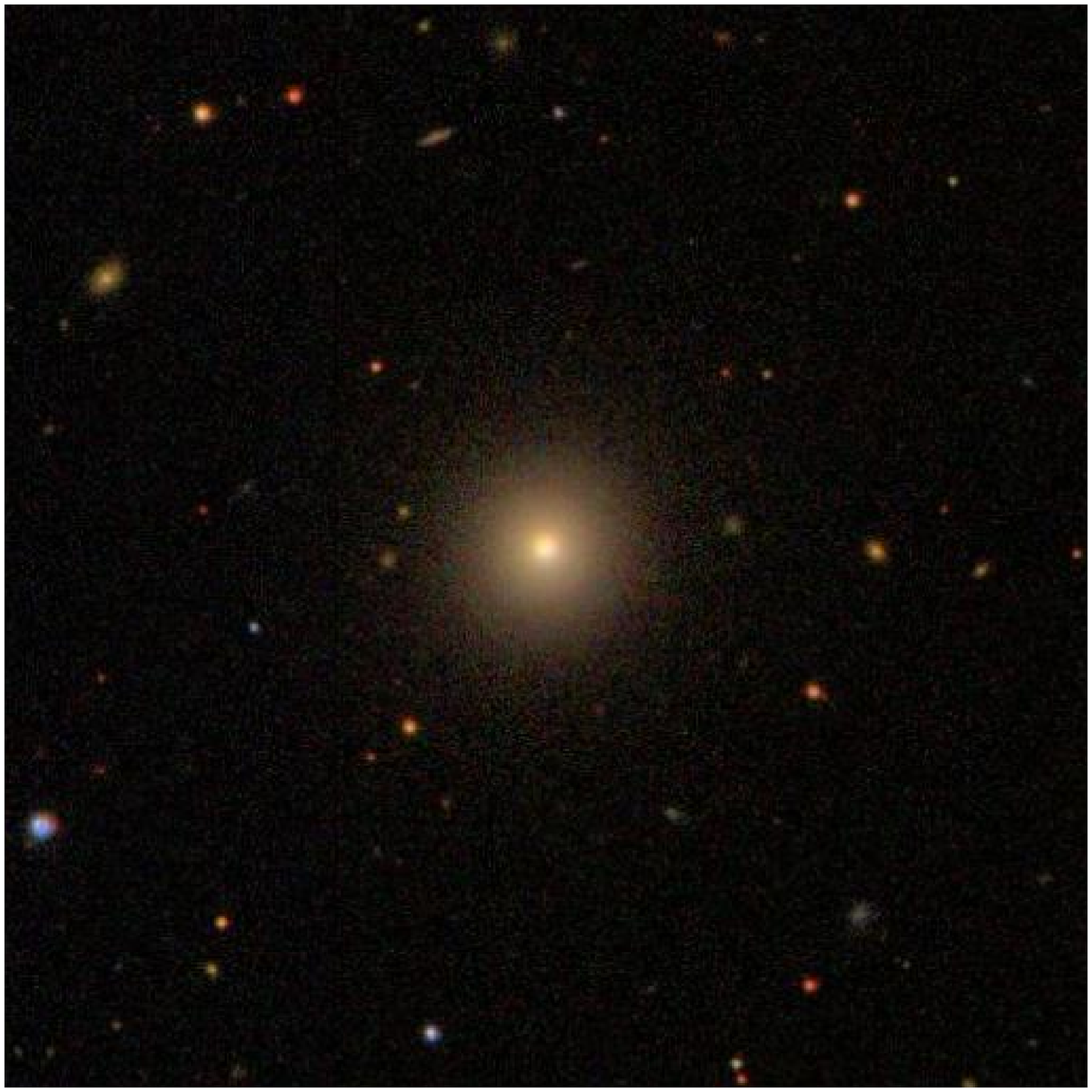}&\includegraphics[width=1.4in]{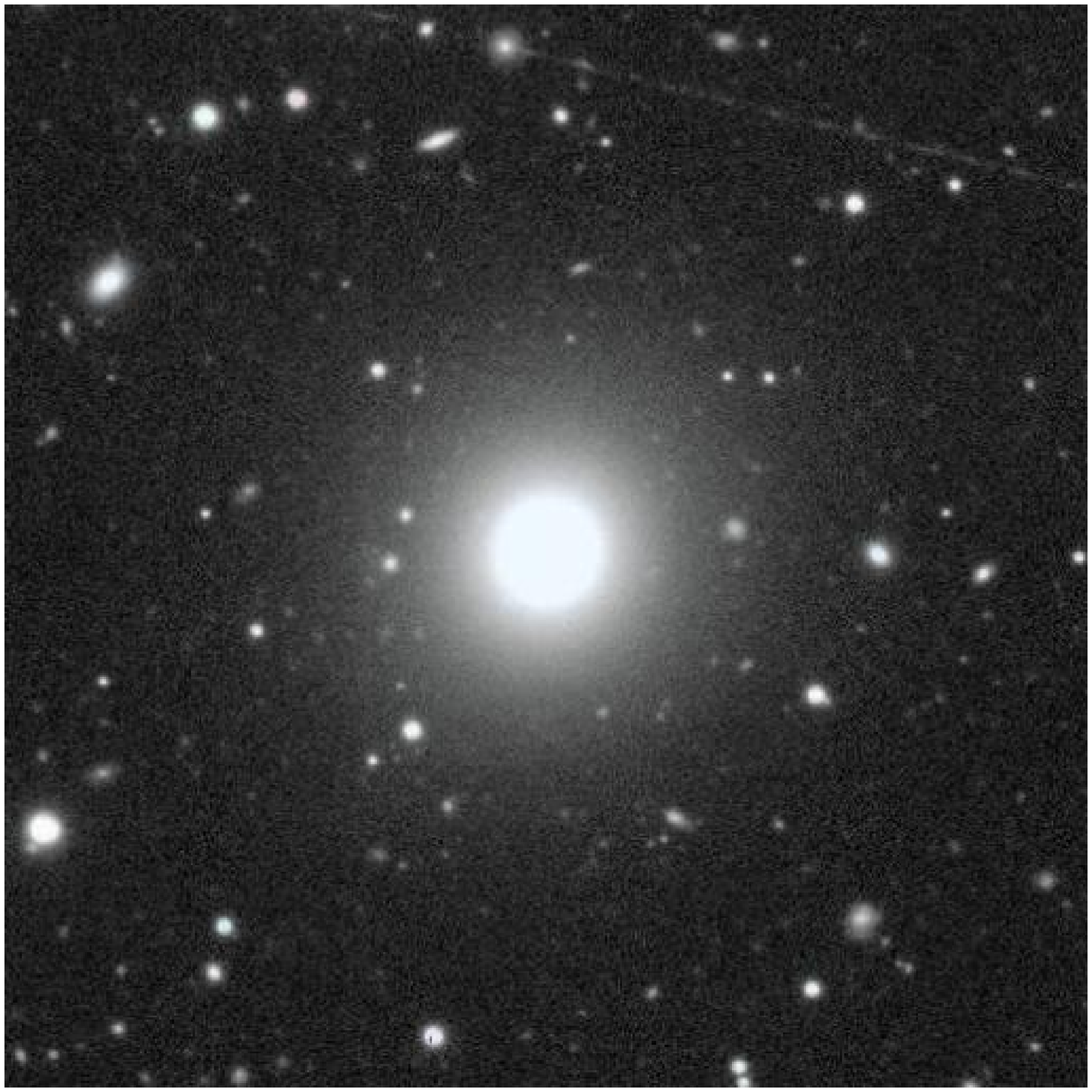}\\

\includegraphics[width=1.4in]{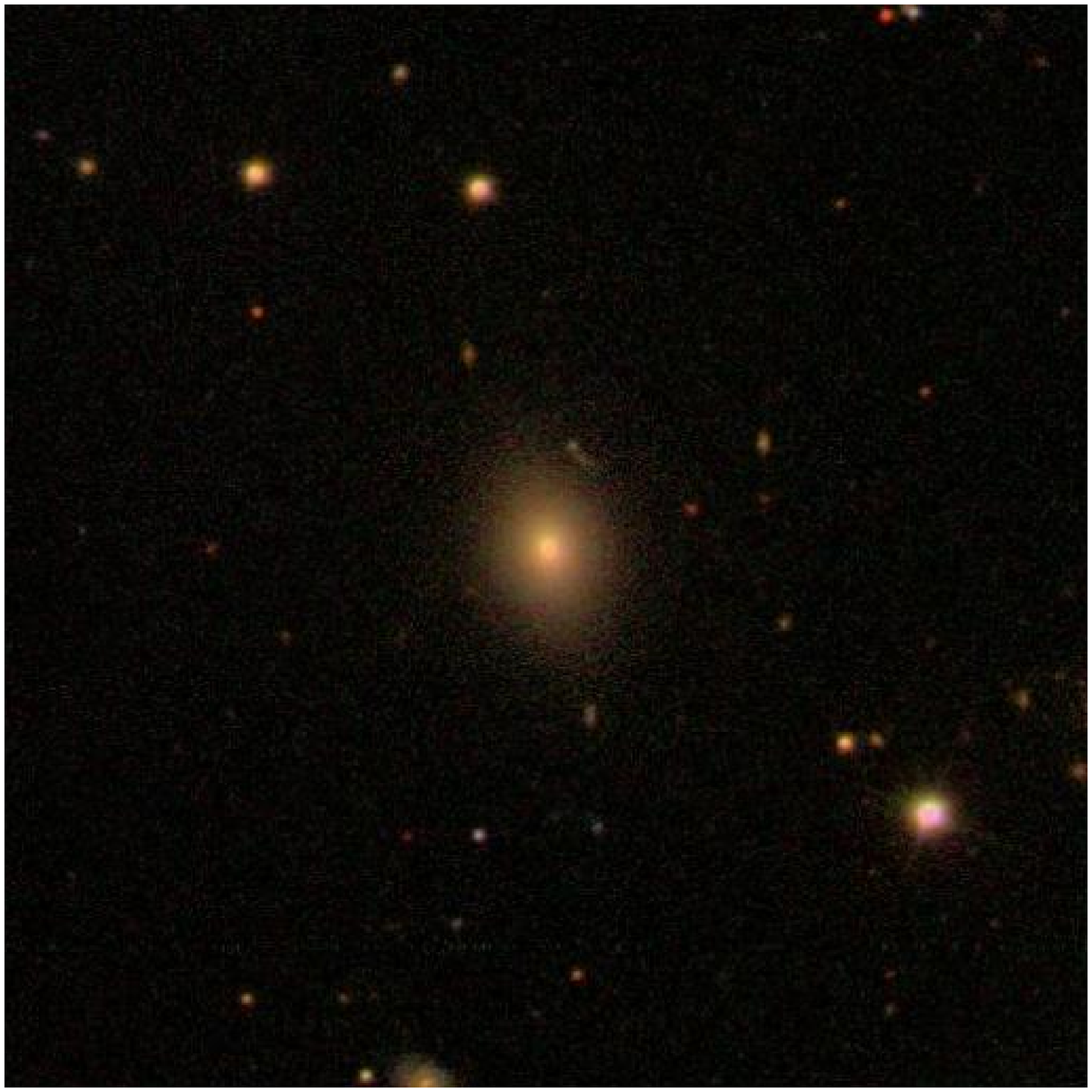}&\includegraphics[width=1.4in]{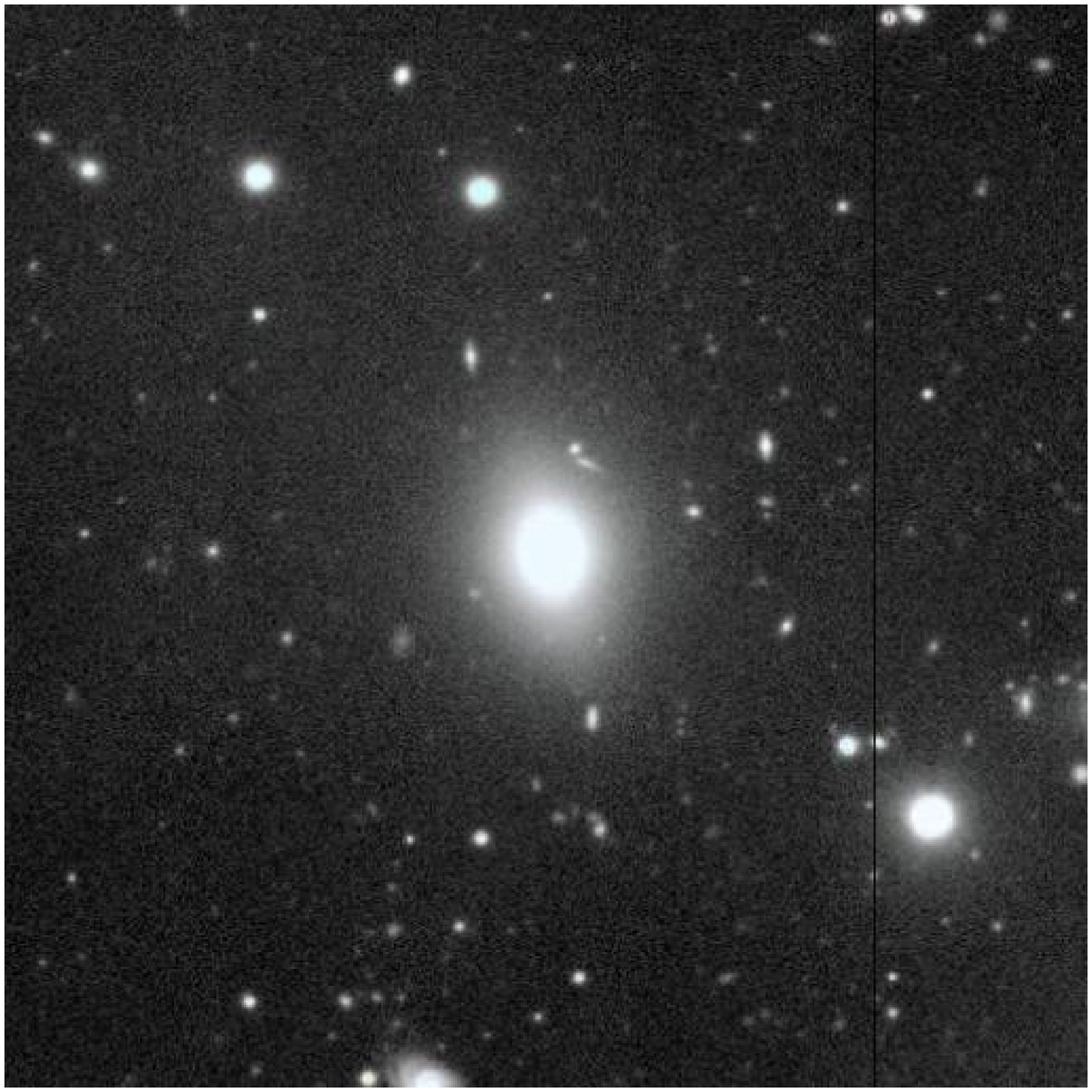}&
\includegraphics[width=1.4in]{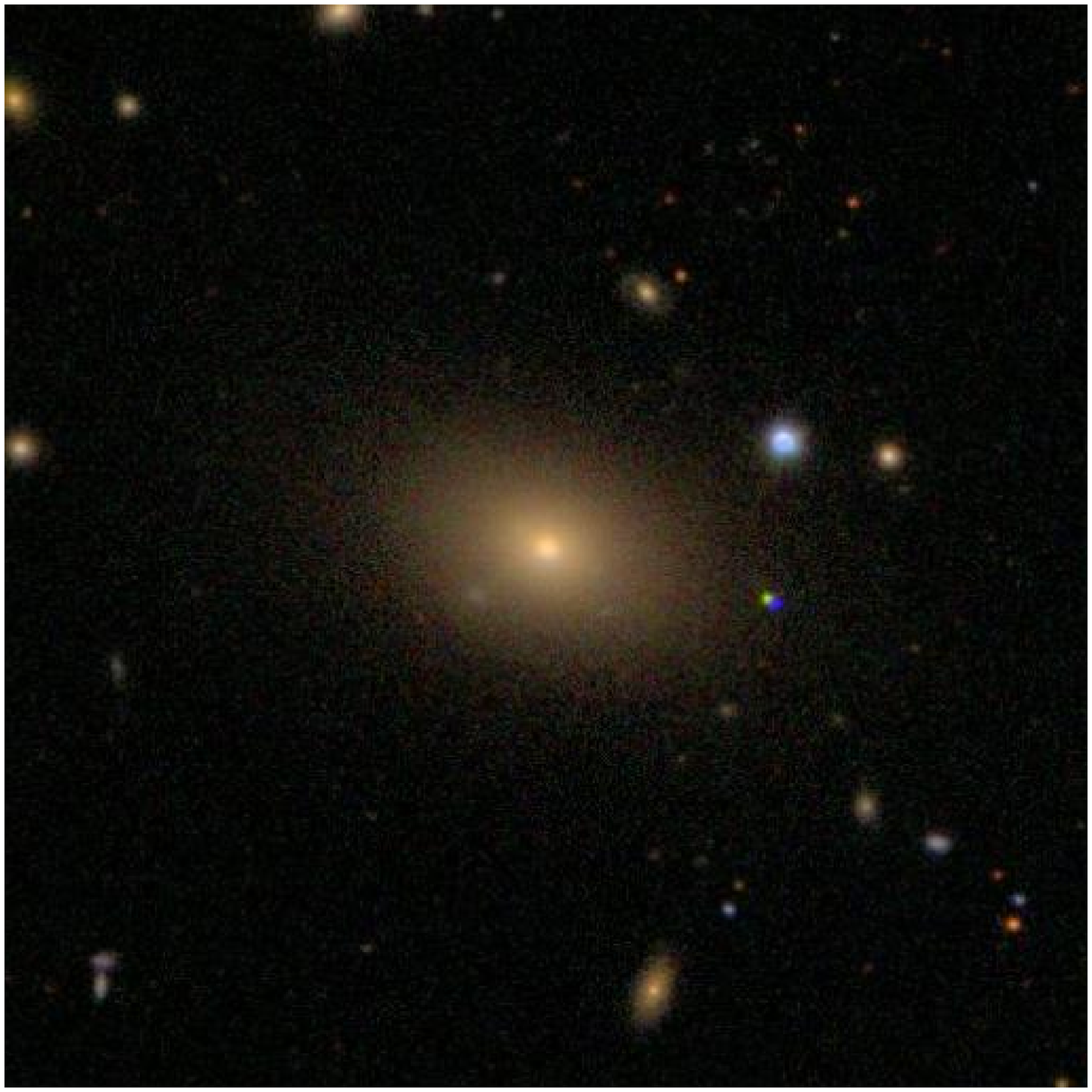}&\includegraphics[width=1.4in]{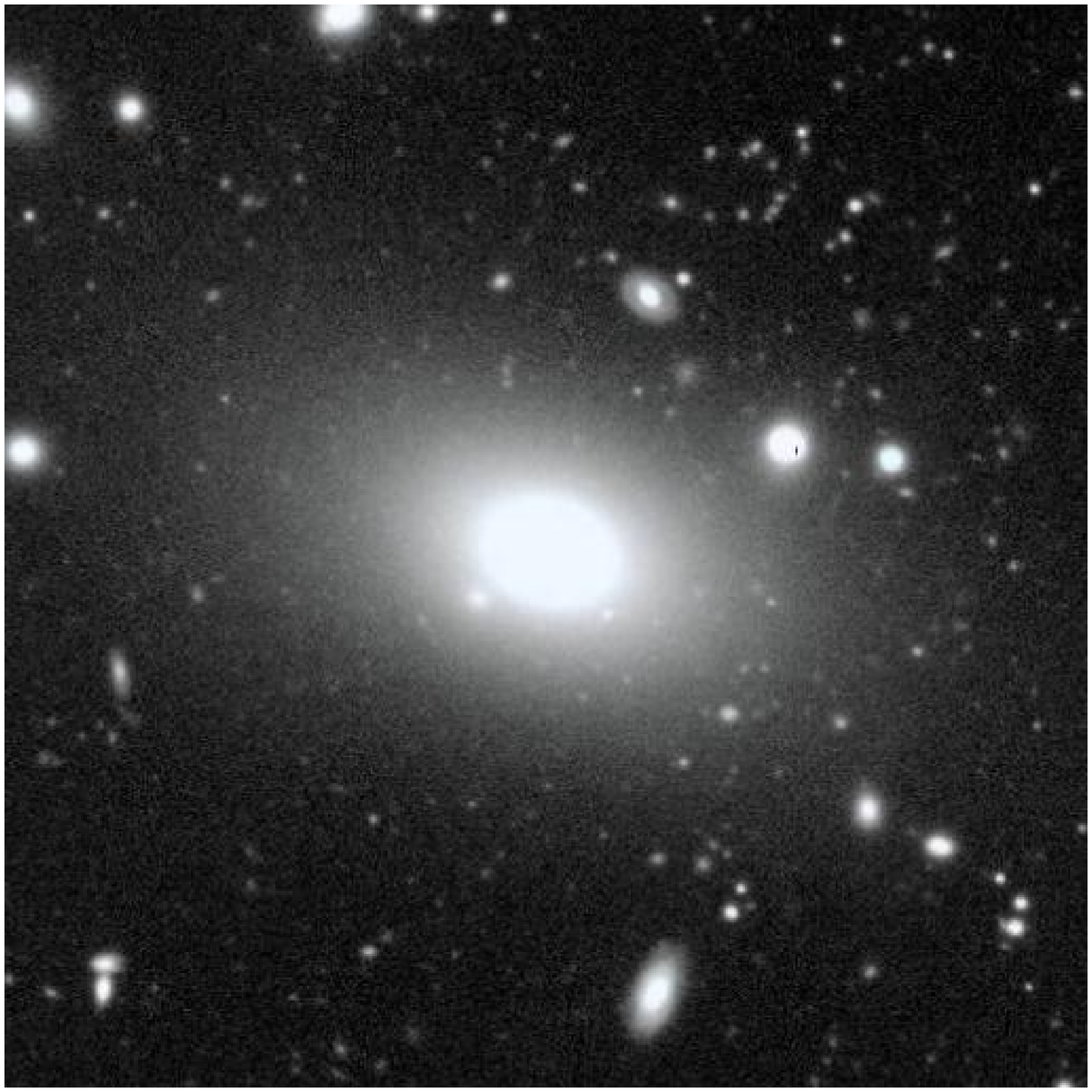}\\

\includegraphics[width=1.4in]{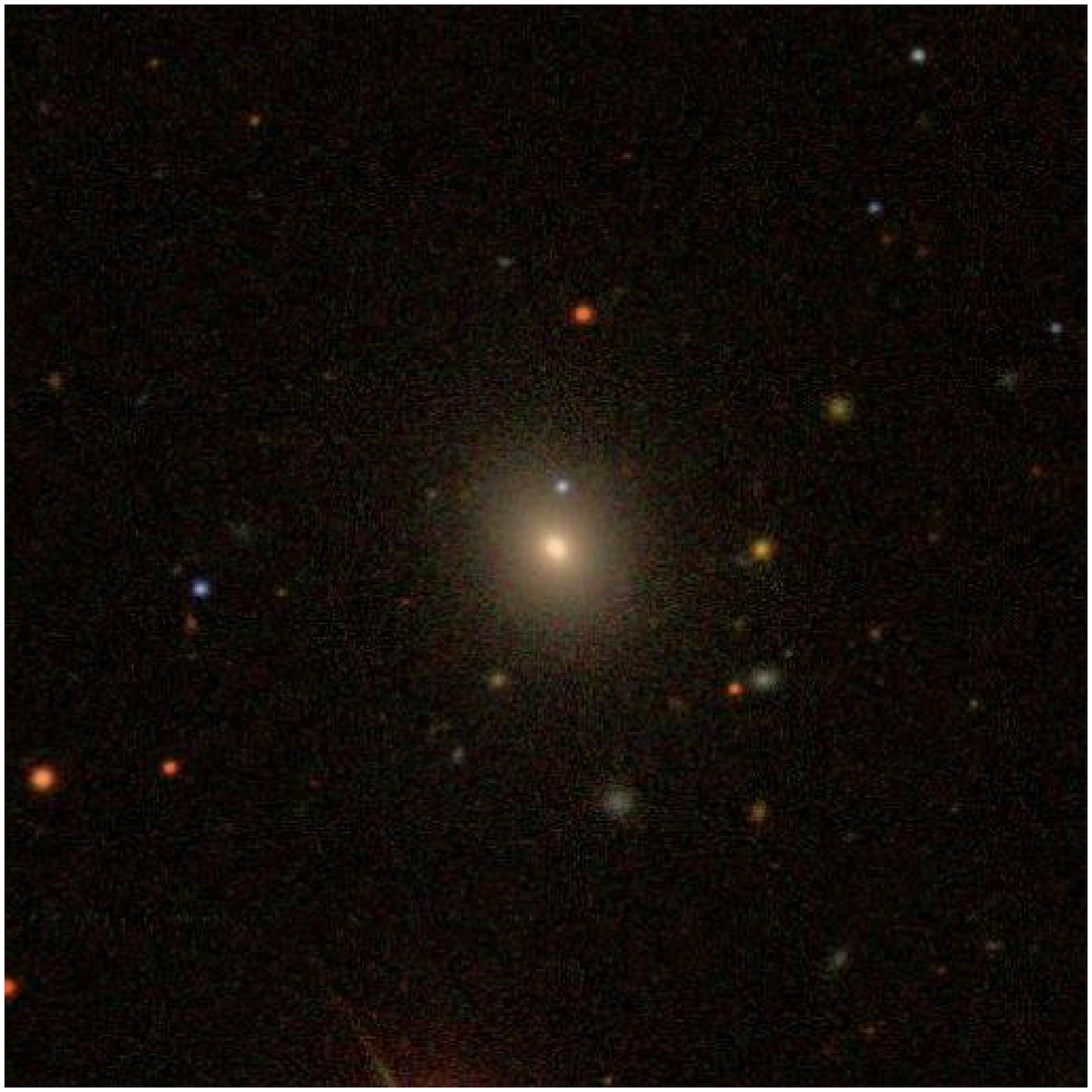}&\includegraphics[width=1.4in]{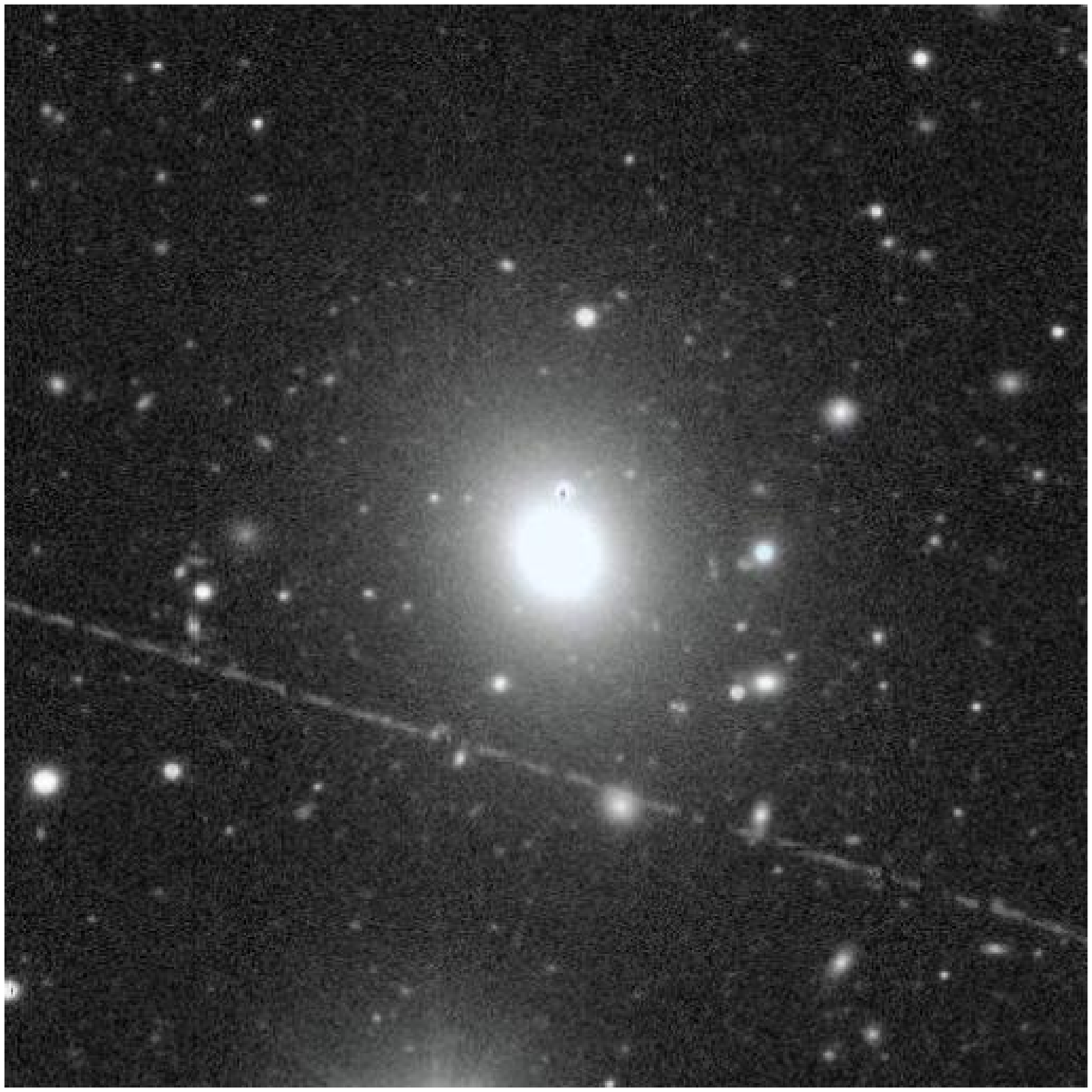}&
\includegraphics[width=1.4in]{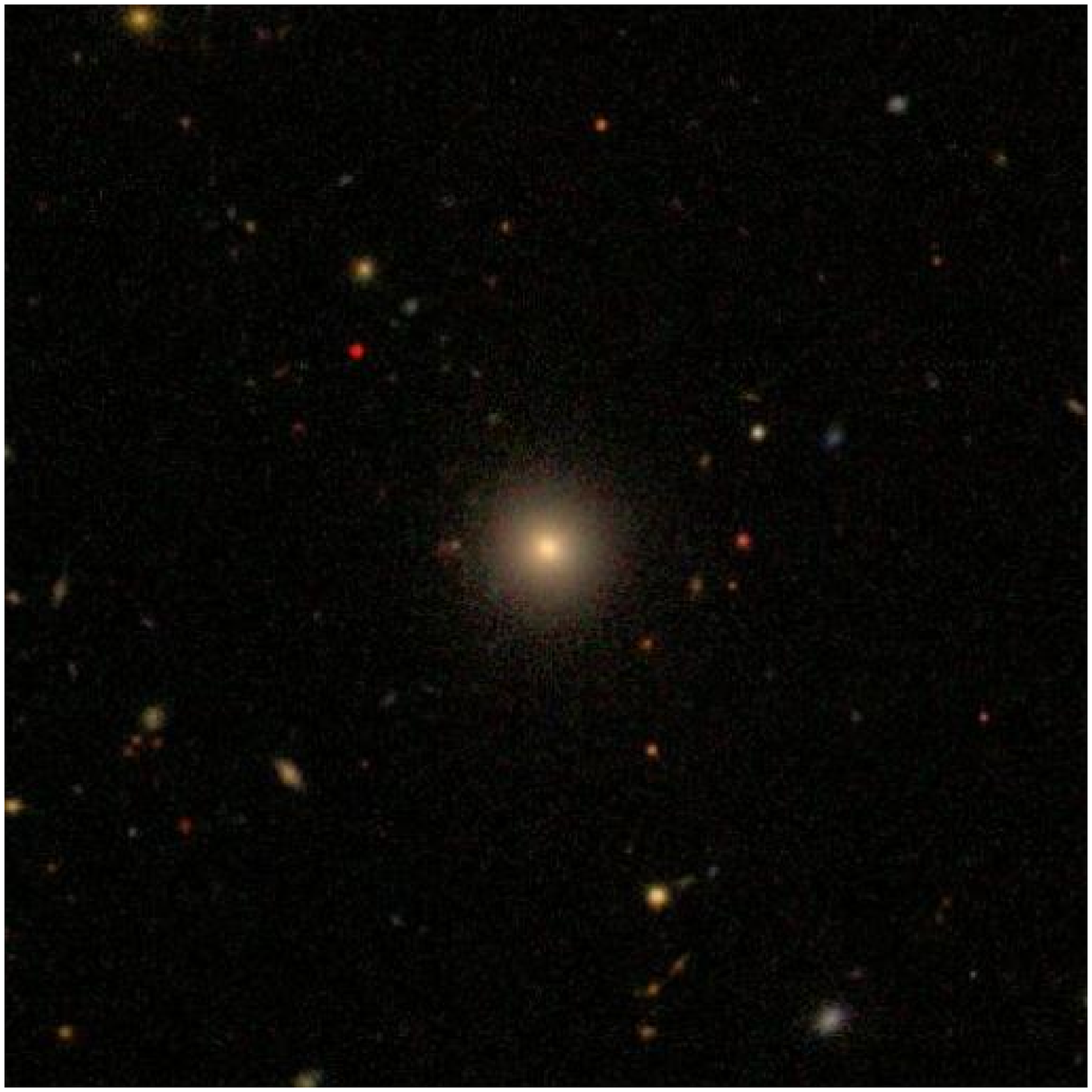}&\includegraphics[width=1.4in]{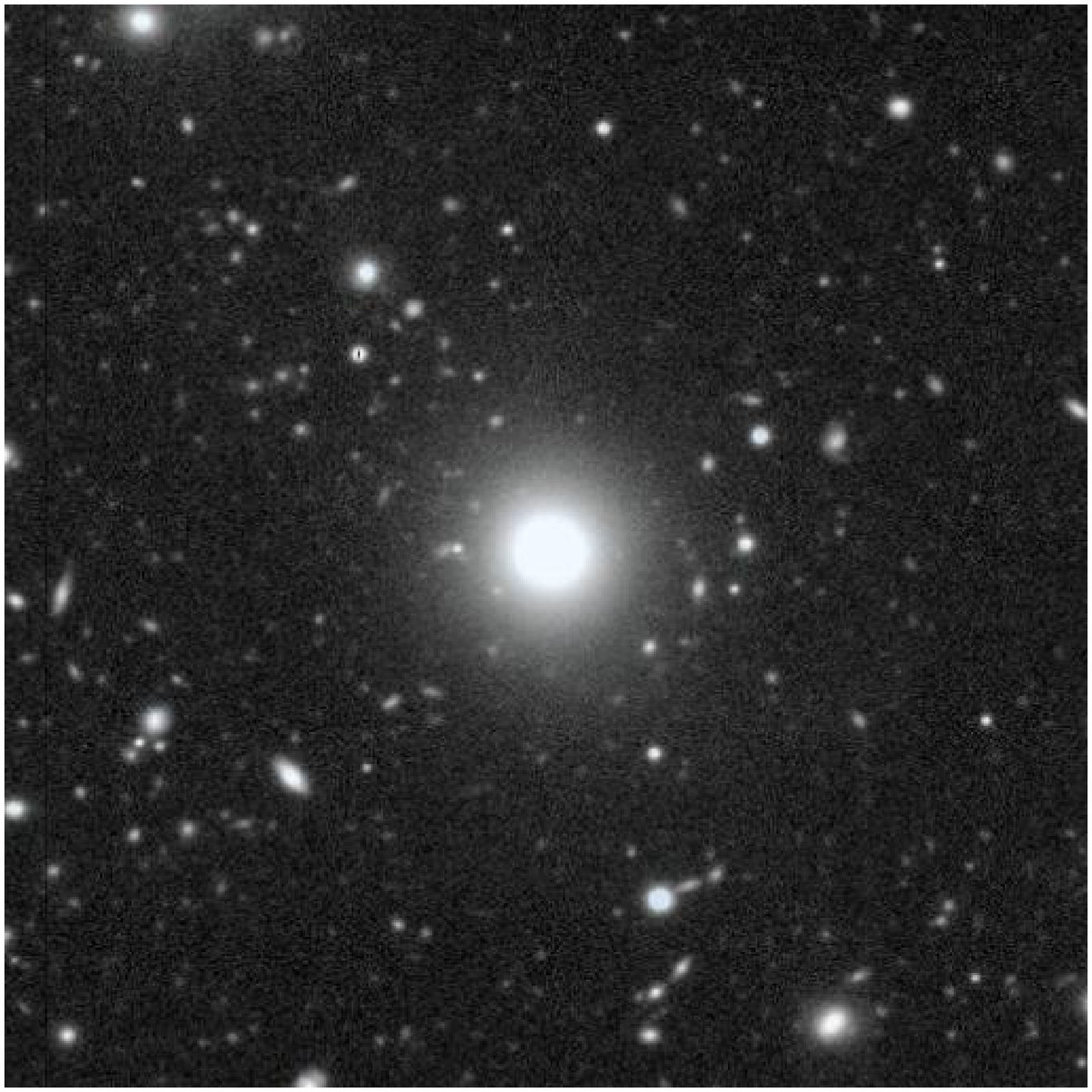}\\\\\\\\

\includegraphics[width=1.4in]{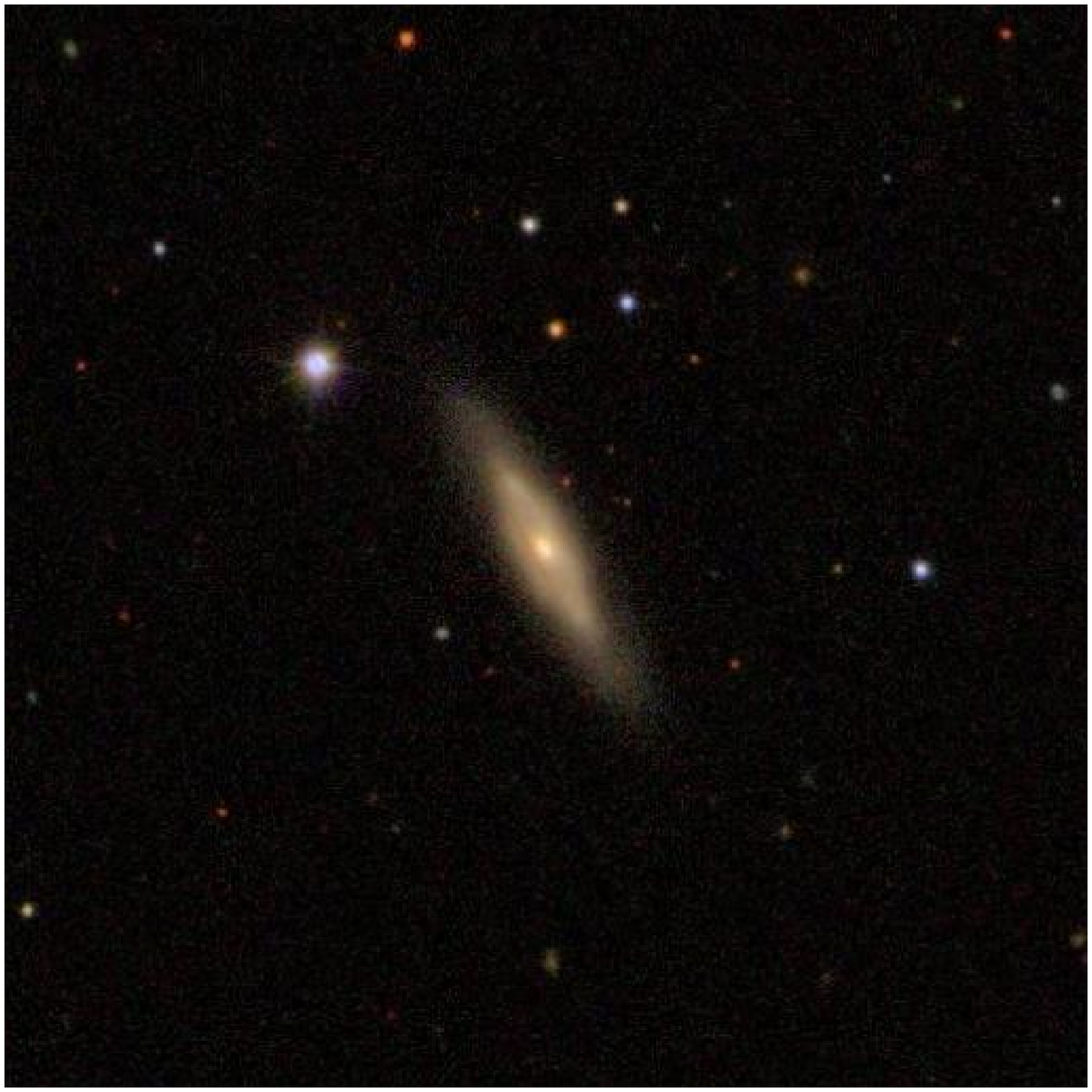}&\includegraphics[width=1.4in]{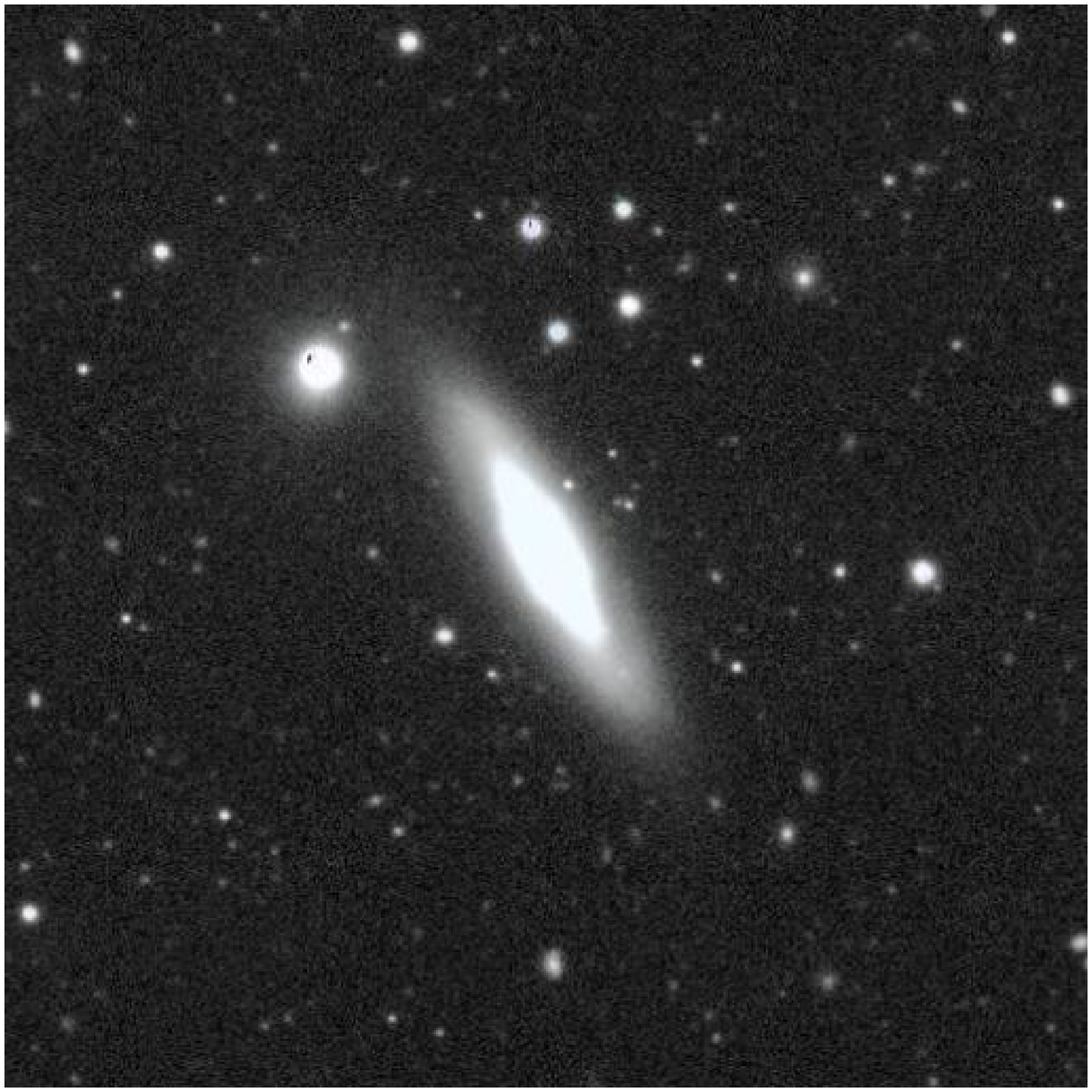}&
\includegraphics[width=1.4in]{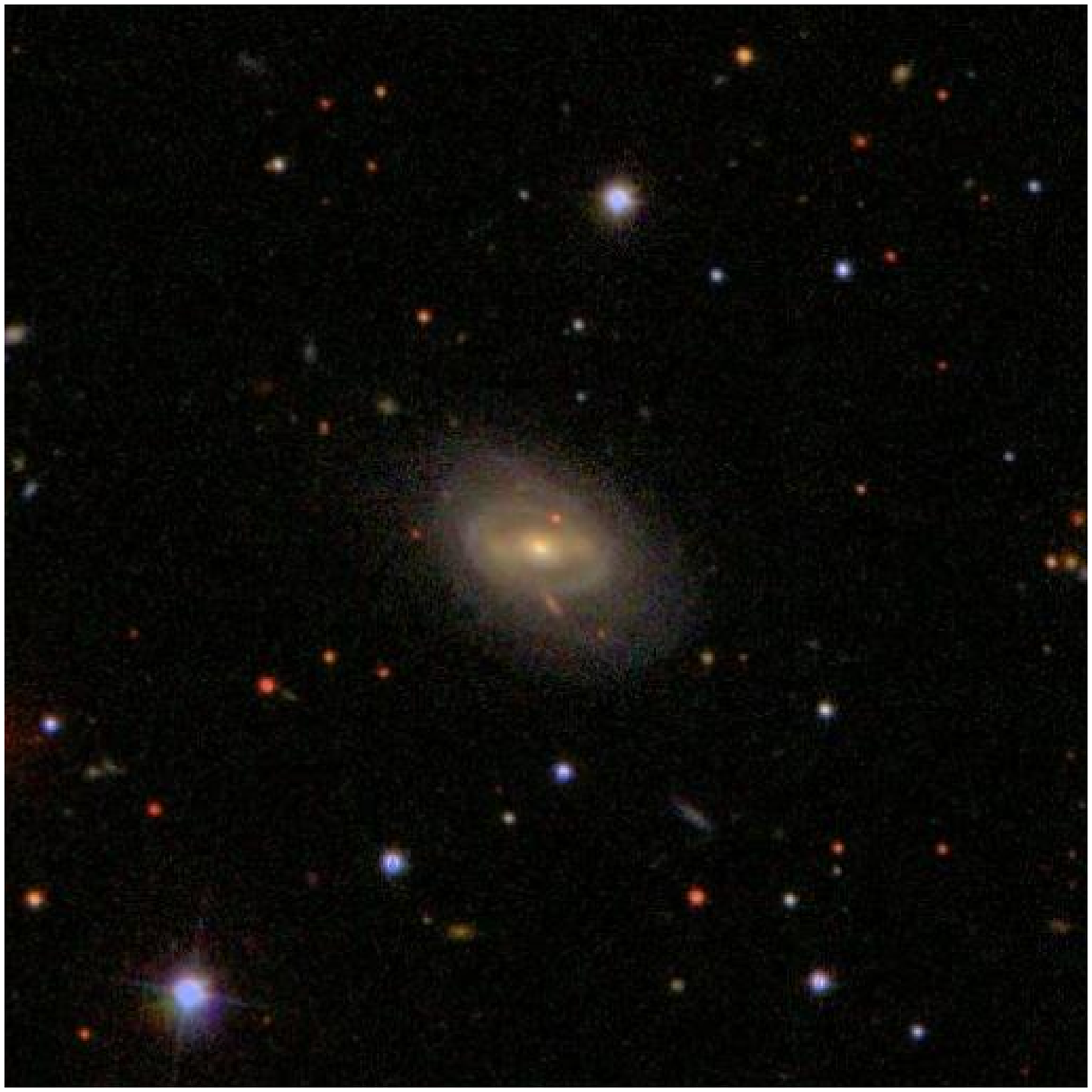}&\includegraphics[width=1.4in]{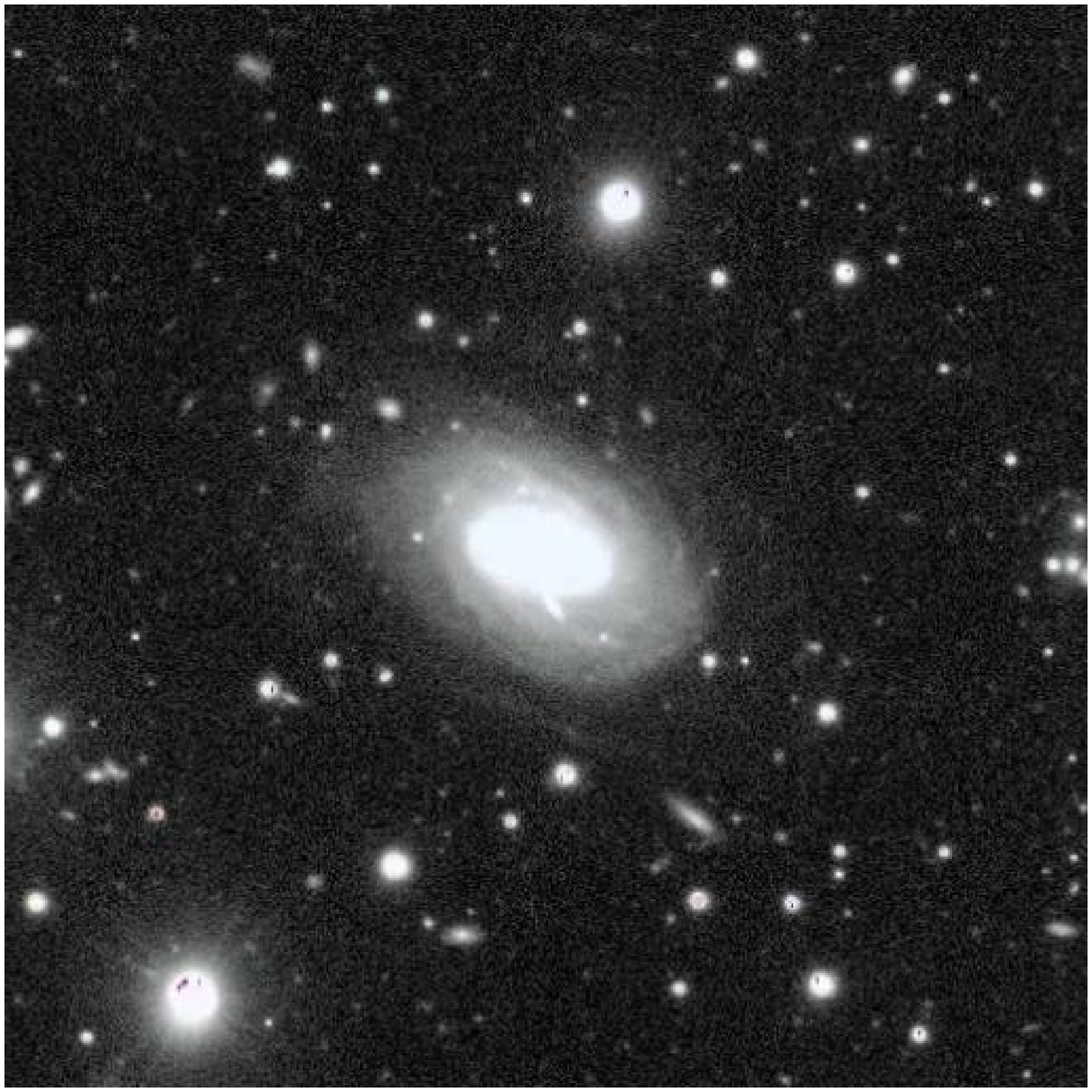}\\

\includegraphics[width=1.4in]{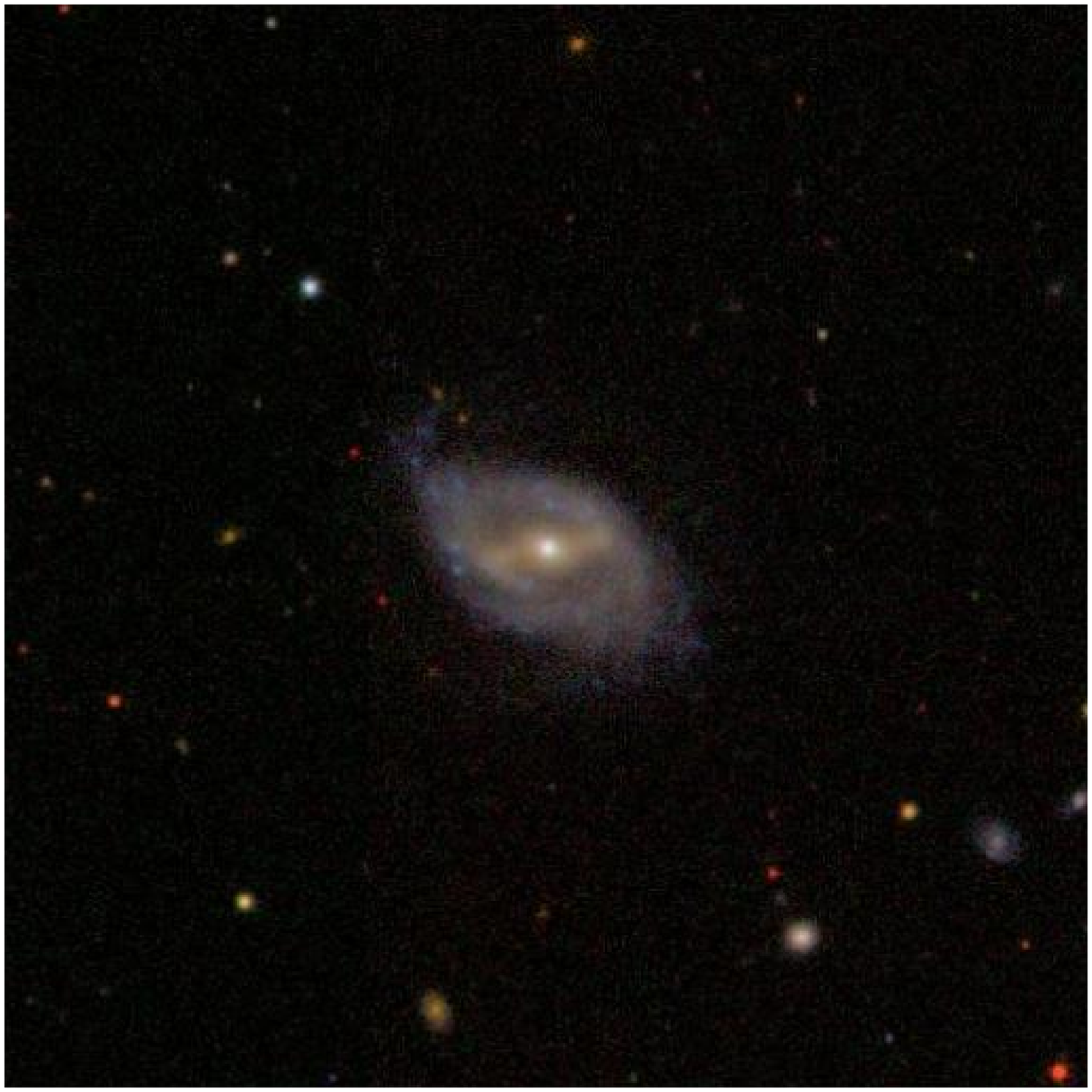}&\includegraphics[width=1.4in]{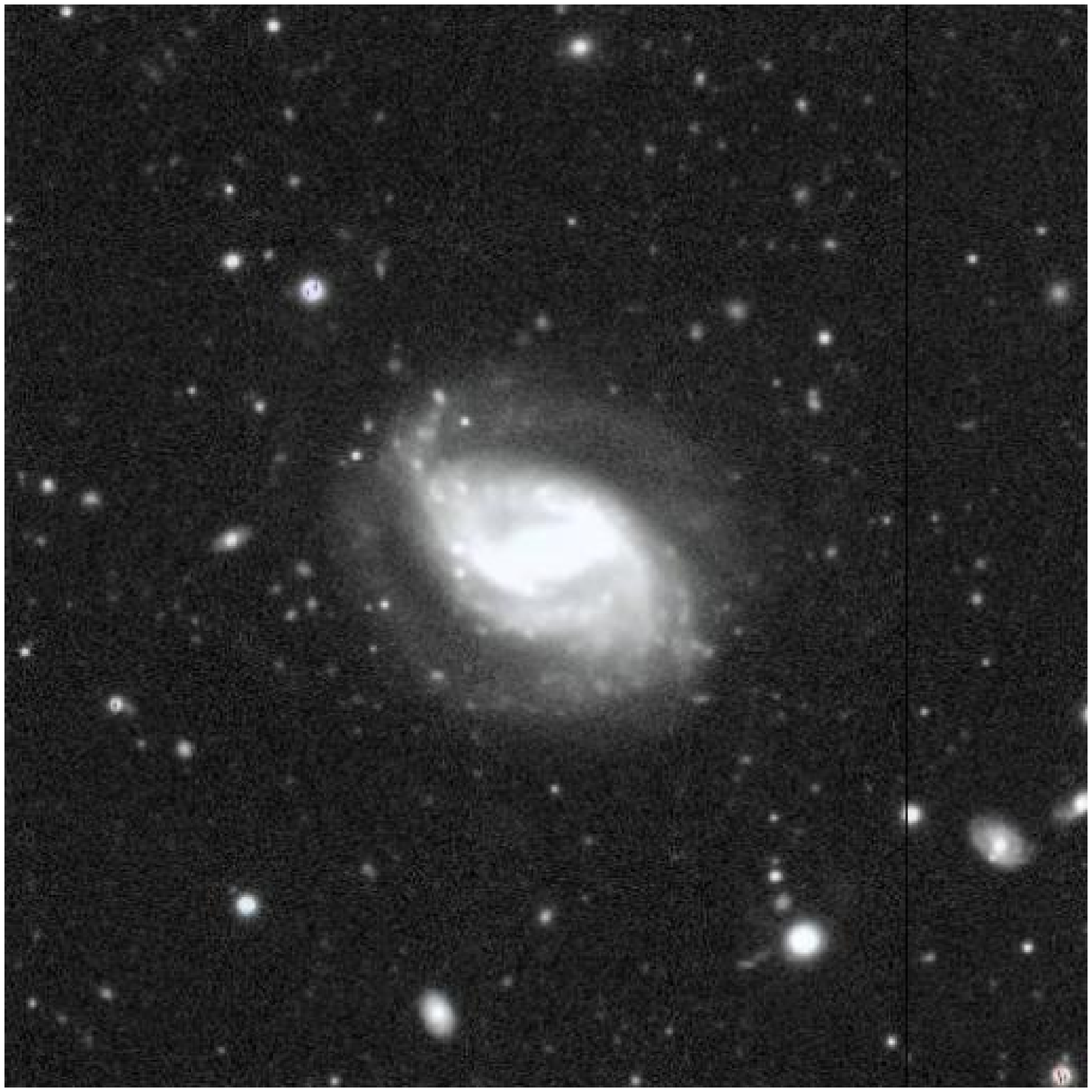}&
\includegraphics[width=1.4in]{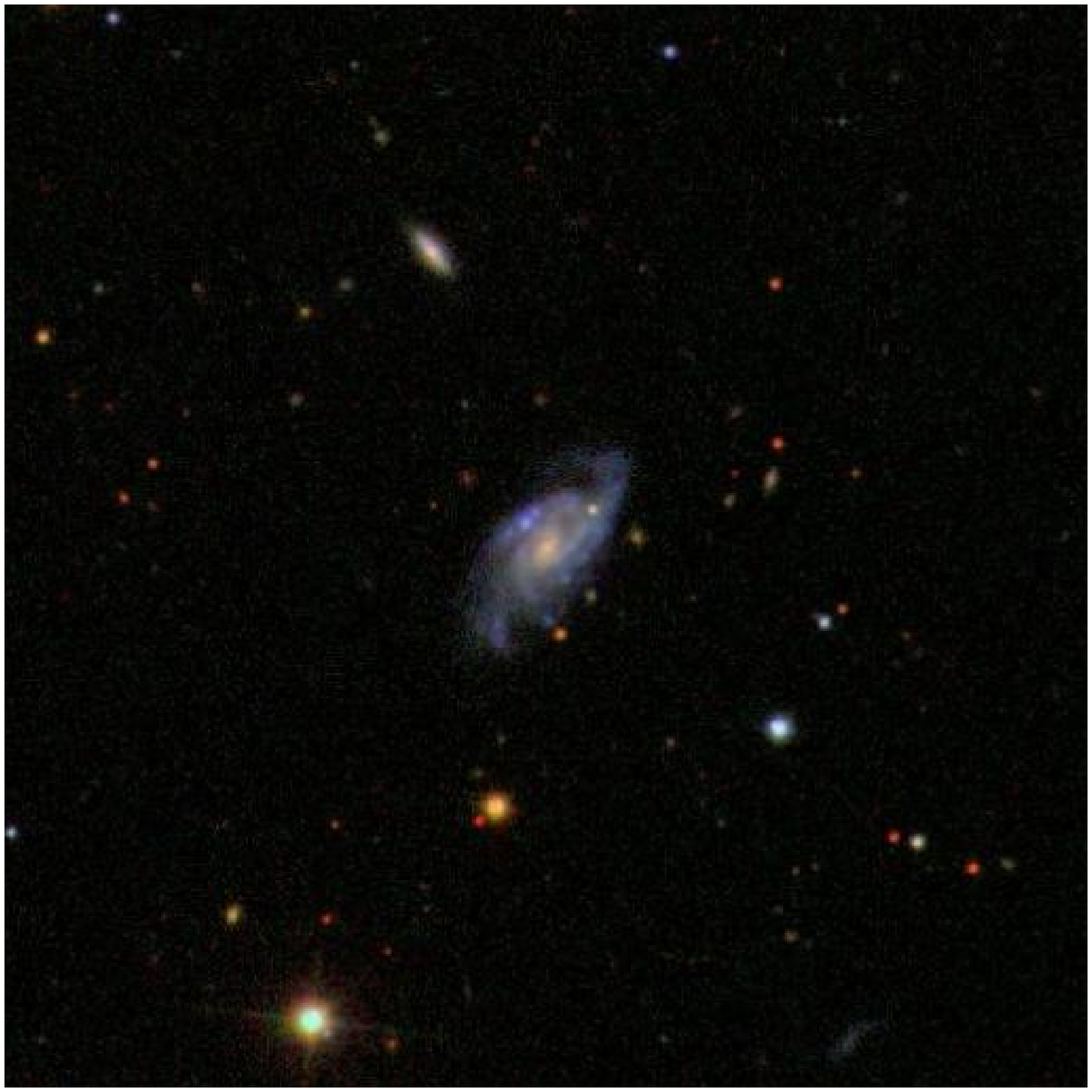}&\includegraphics[width=1.4in]{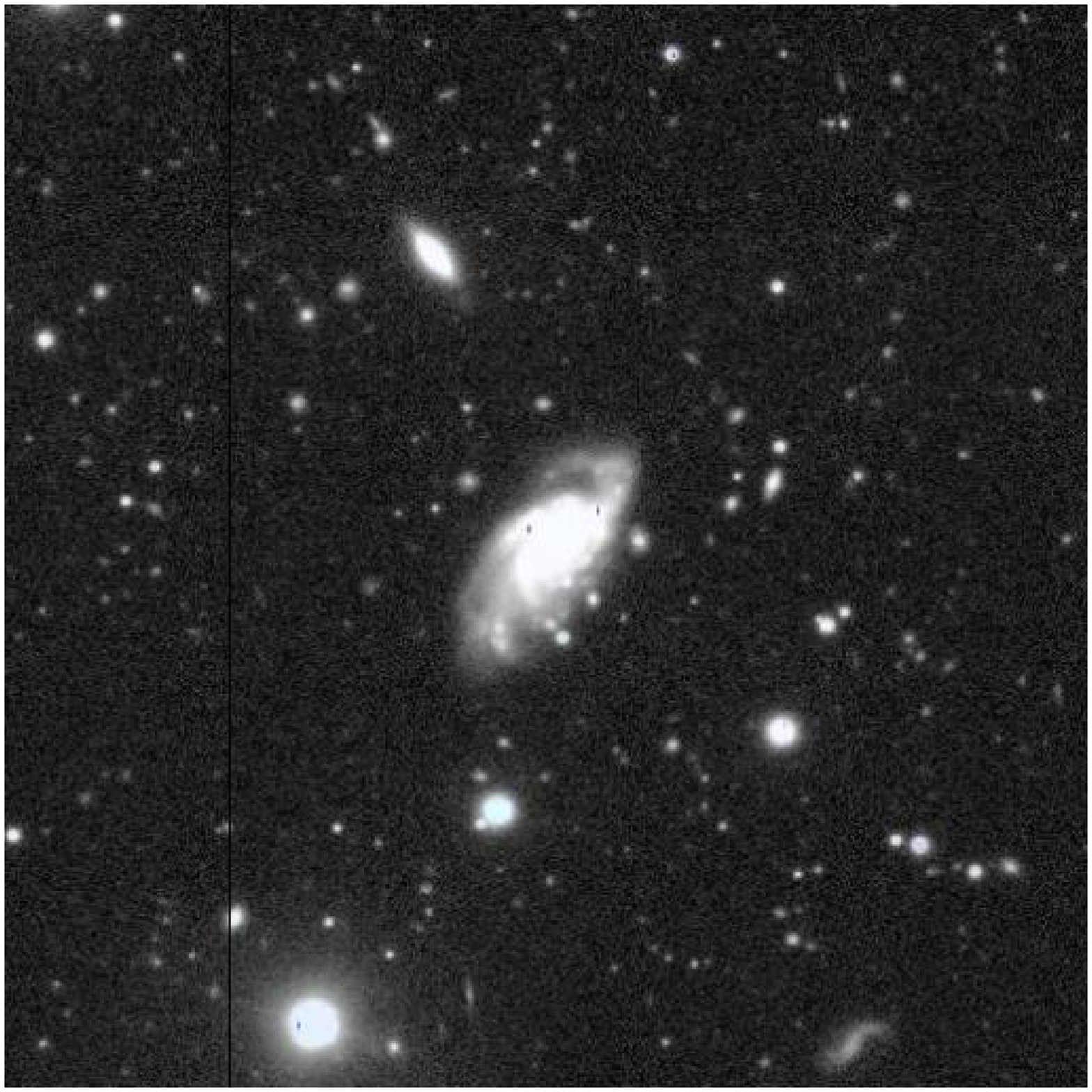}\\

\includegraphics[width=1.4in]{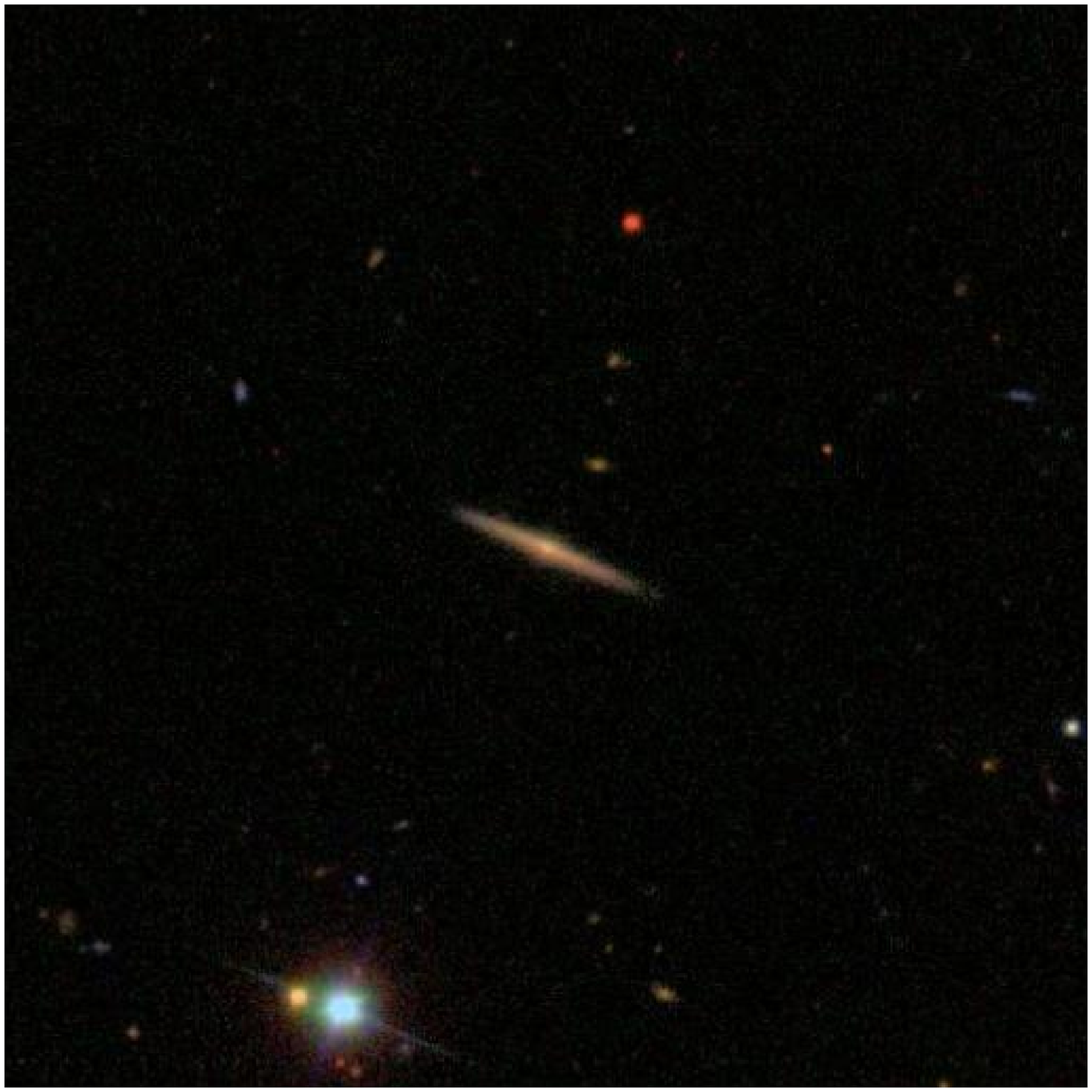}&\includegraphics[width=1.4in]{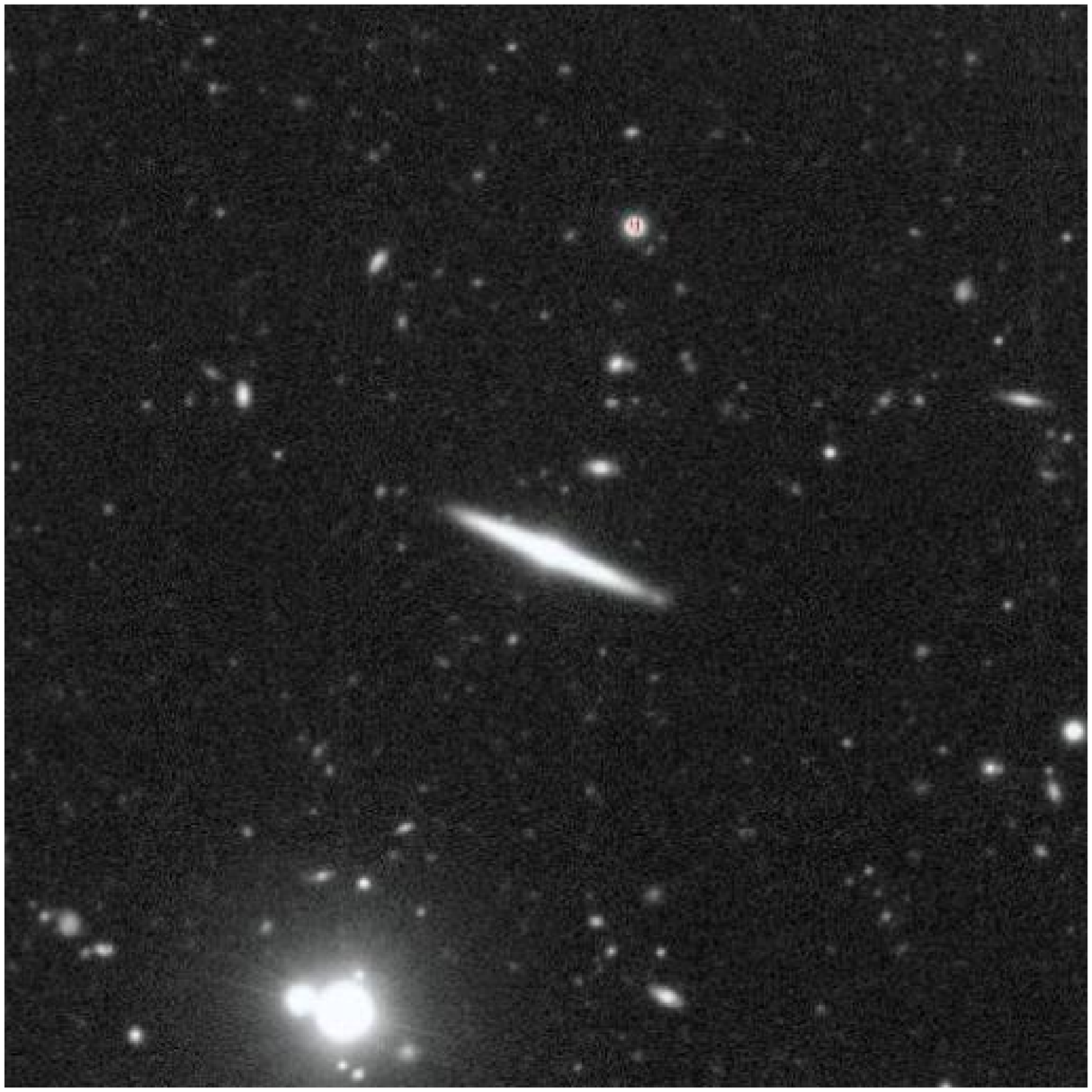}&
\includegraphics[width=1.4in]{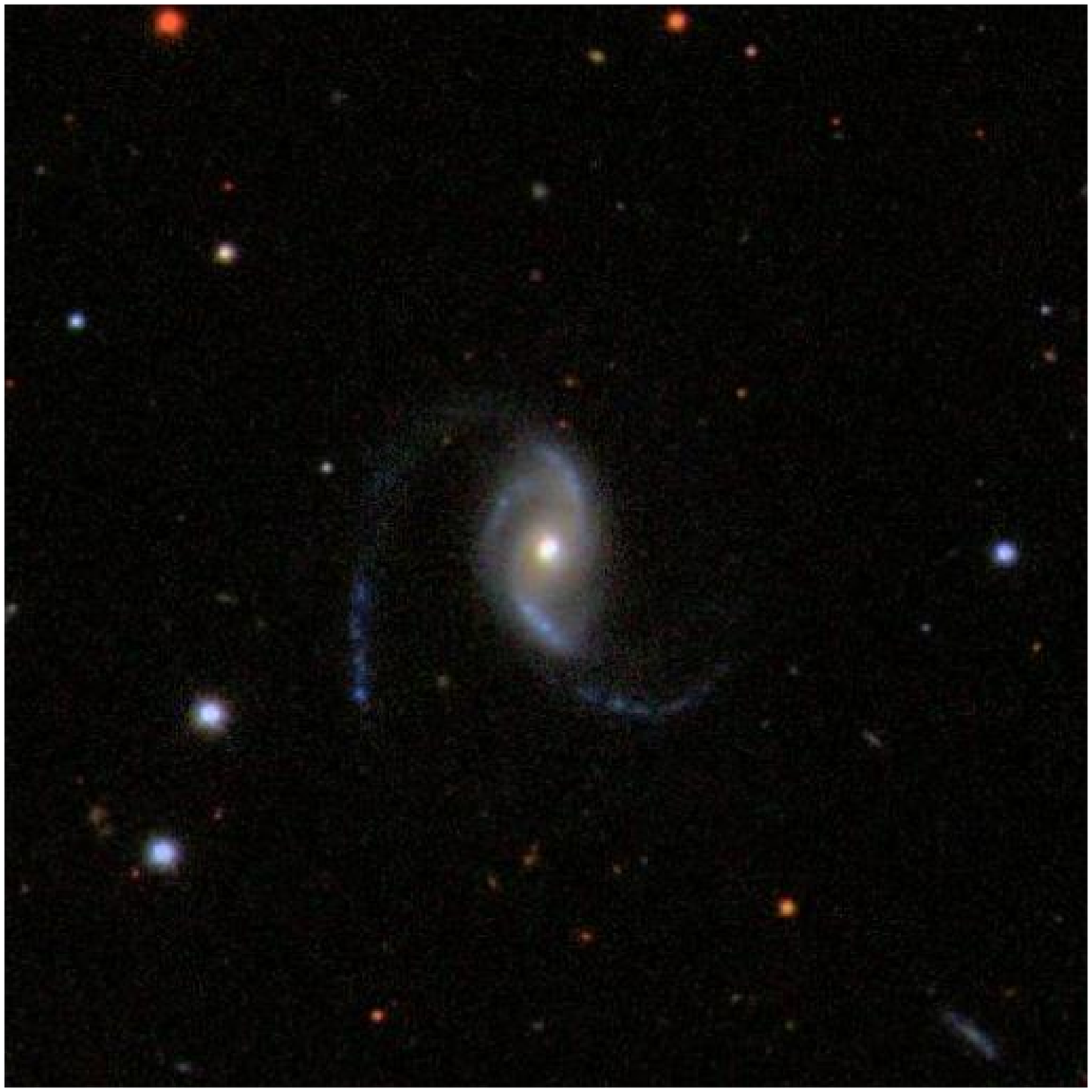}&\includegraphics[width=1.4in]{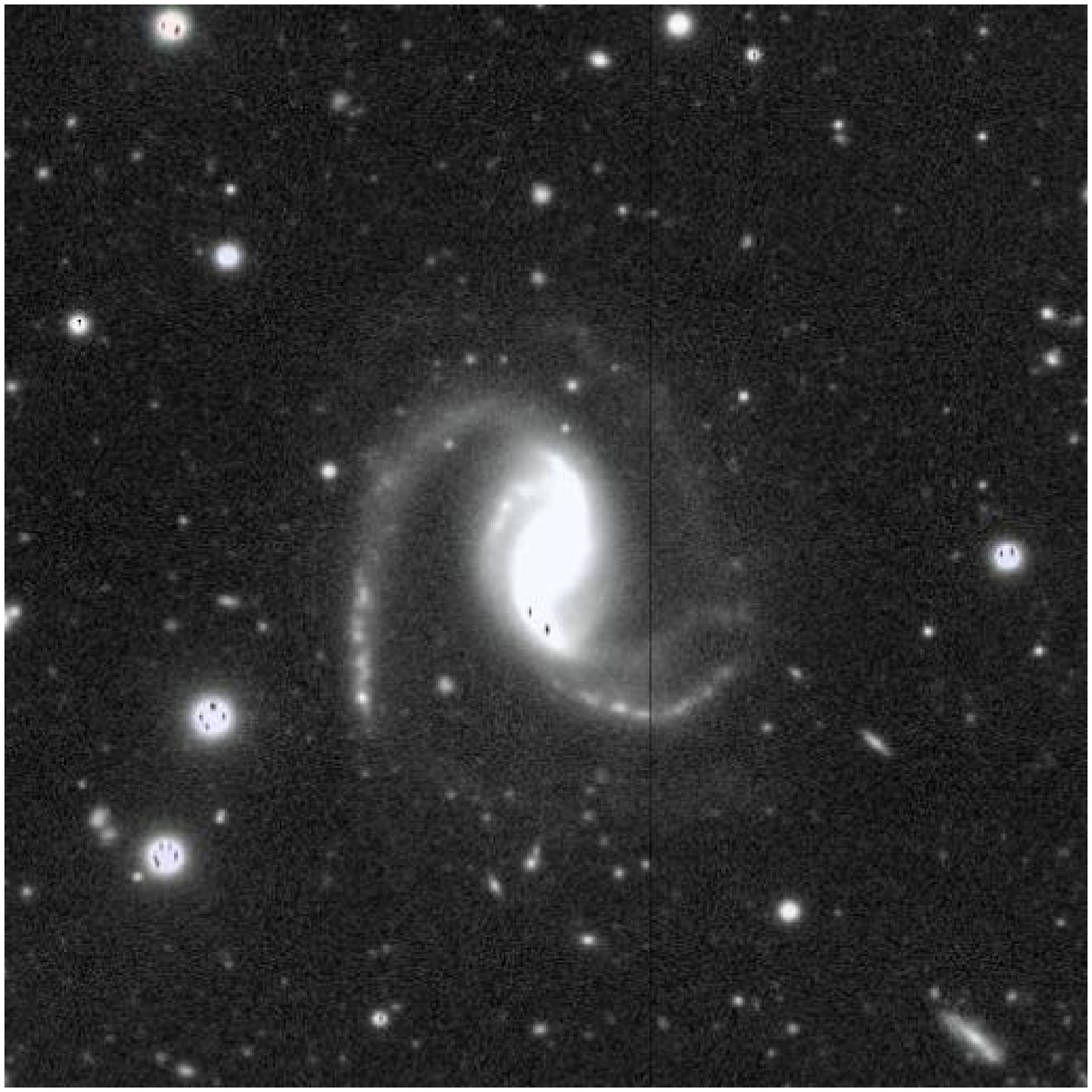}

\end{array}$
\caption{Examples of relaxed early-types (top three rows) and
galaxies classified as late-type (bottom three rows). We show both
the multi-colour standard-depth image (left-hand column) and its
deeper Stripe82 counterpart (right-hand column).}
\label{fig:etg_ltg}
\end{center}
\end{minipage}
\end{figure*}


\begin{figure*}
\begin{minipage}{172mm}
\begin{center}
$\begin{array}{cc}
\includegraphics[width=3in]{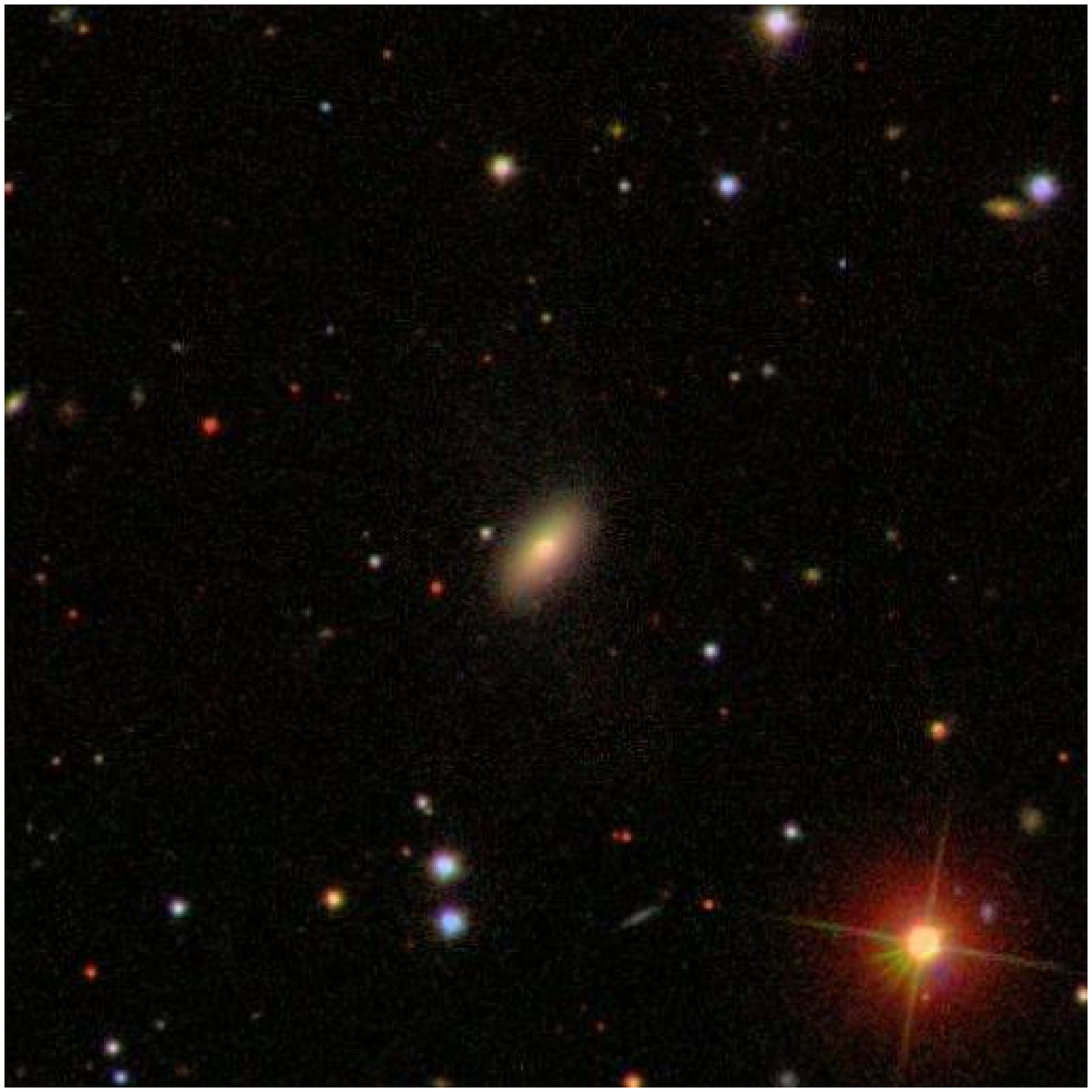}&\includegraphics[width=3in]{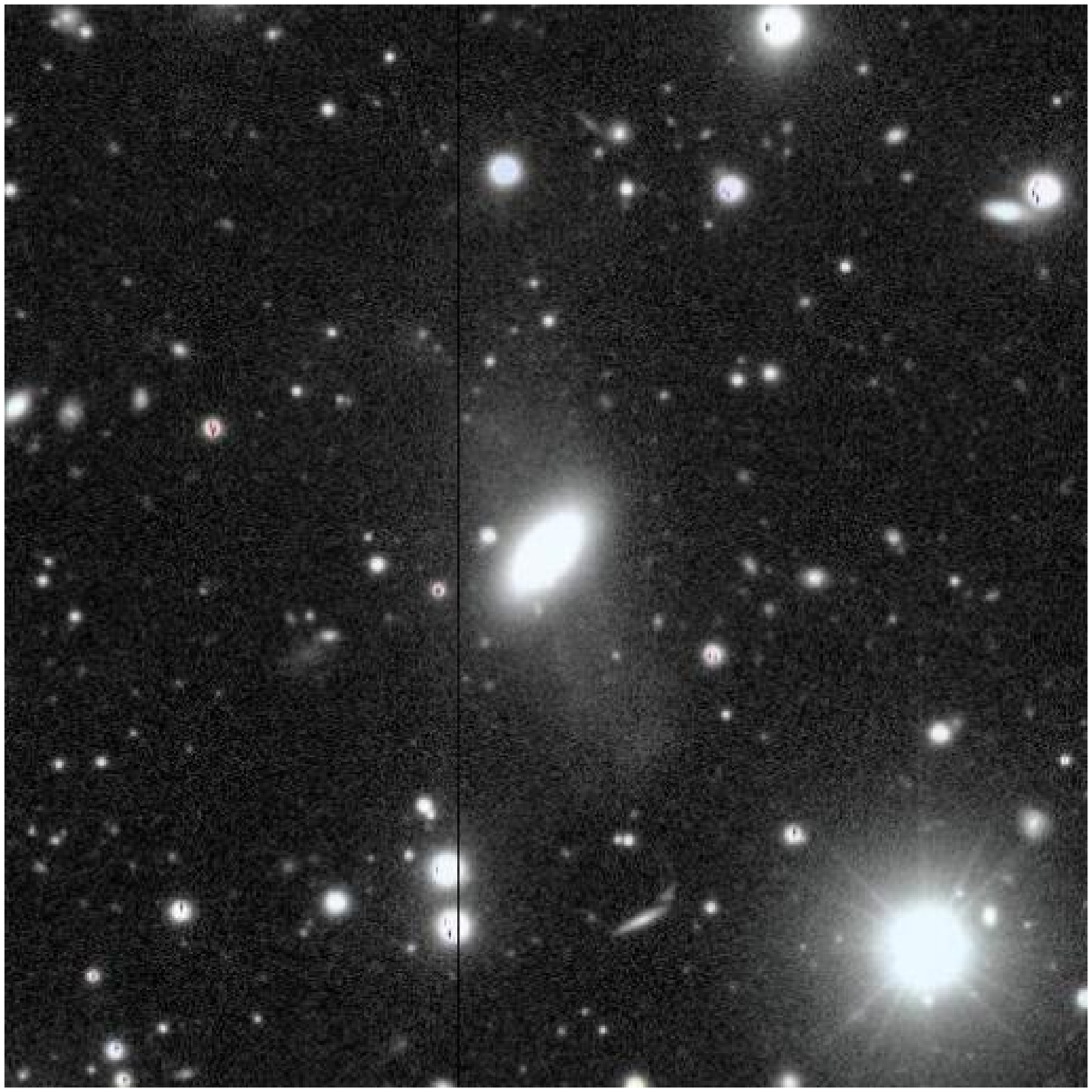}\\
\includegraphics[width=3in]{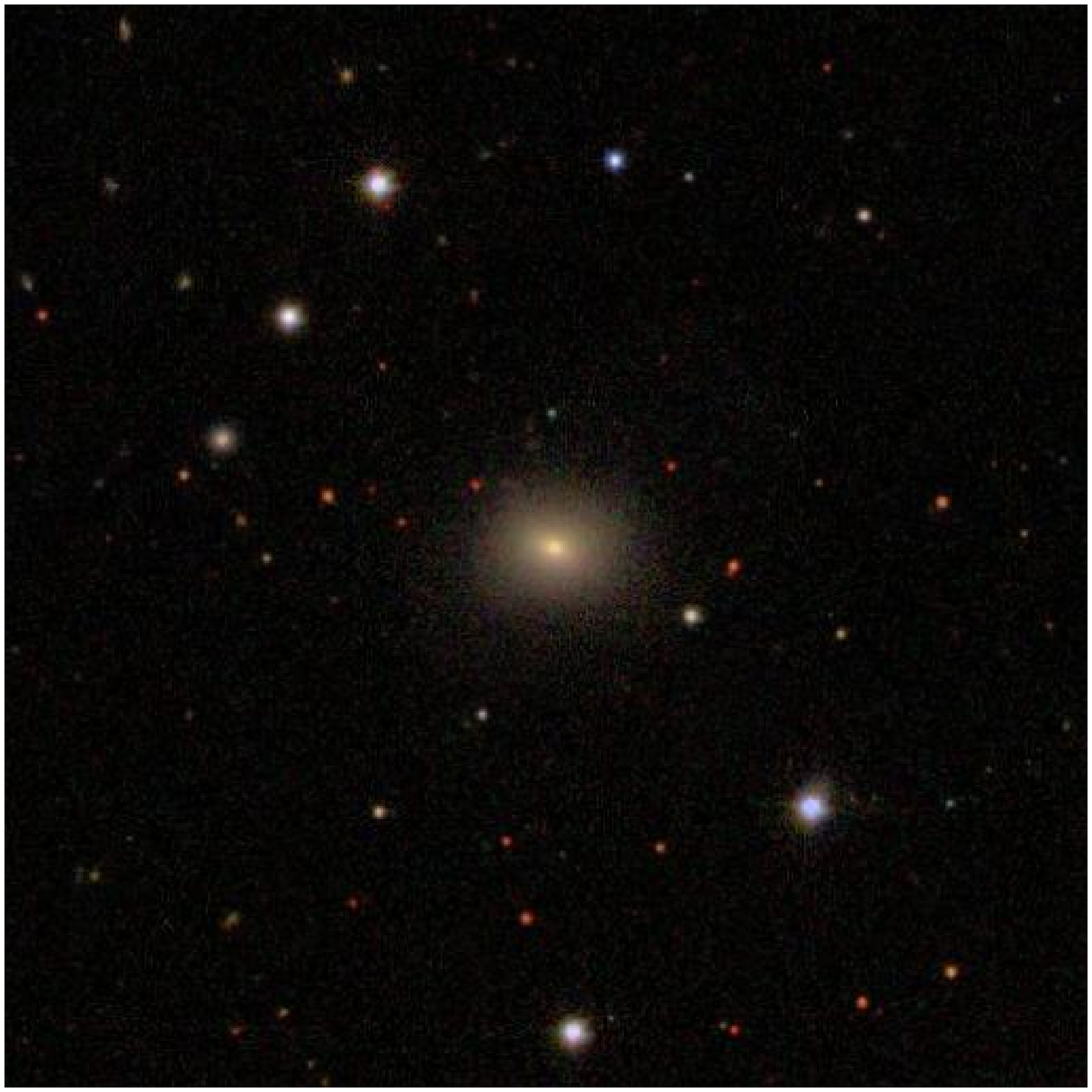}&\includegraphics[width=3in]{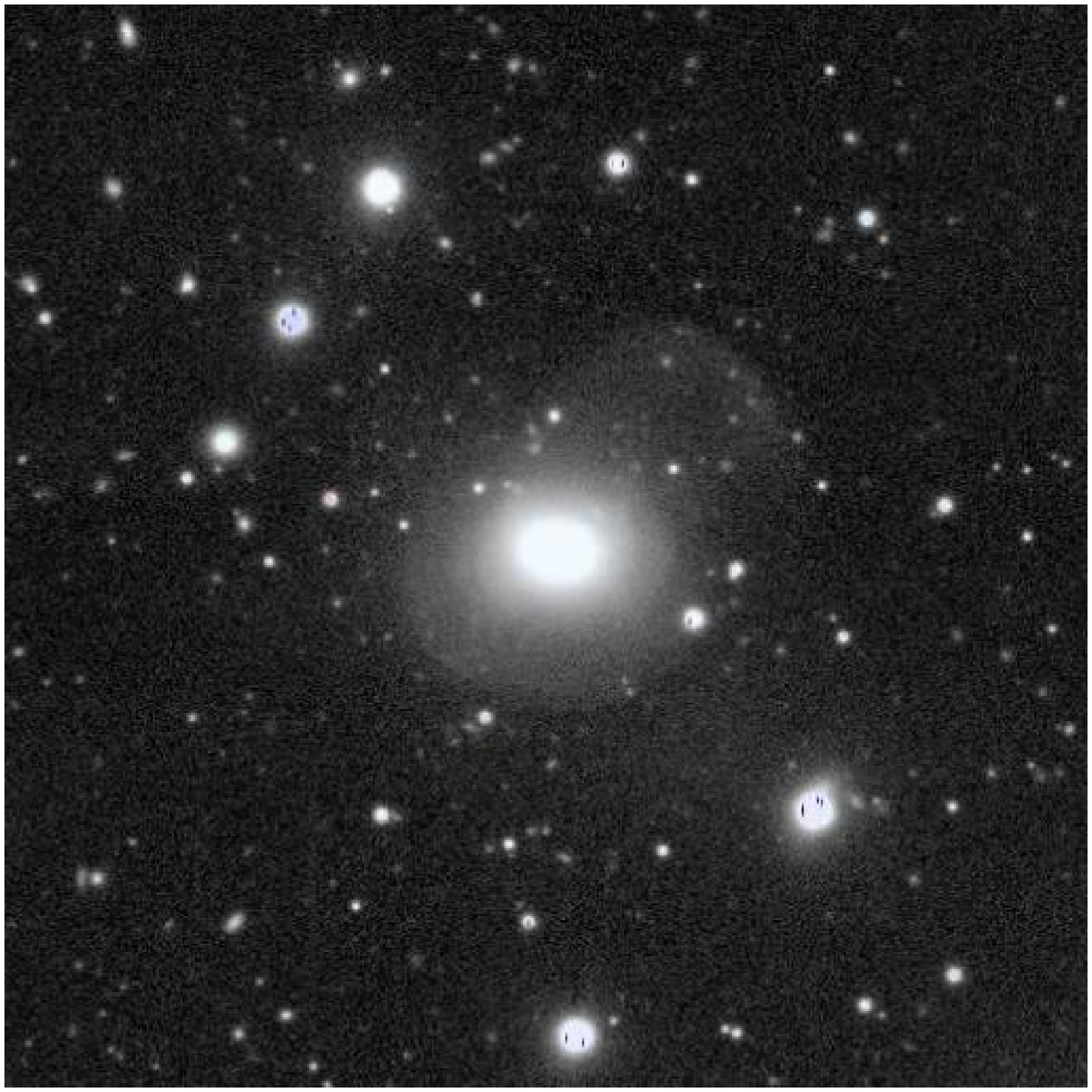}\\
\includegraphics[width=3in]{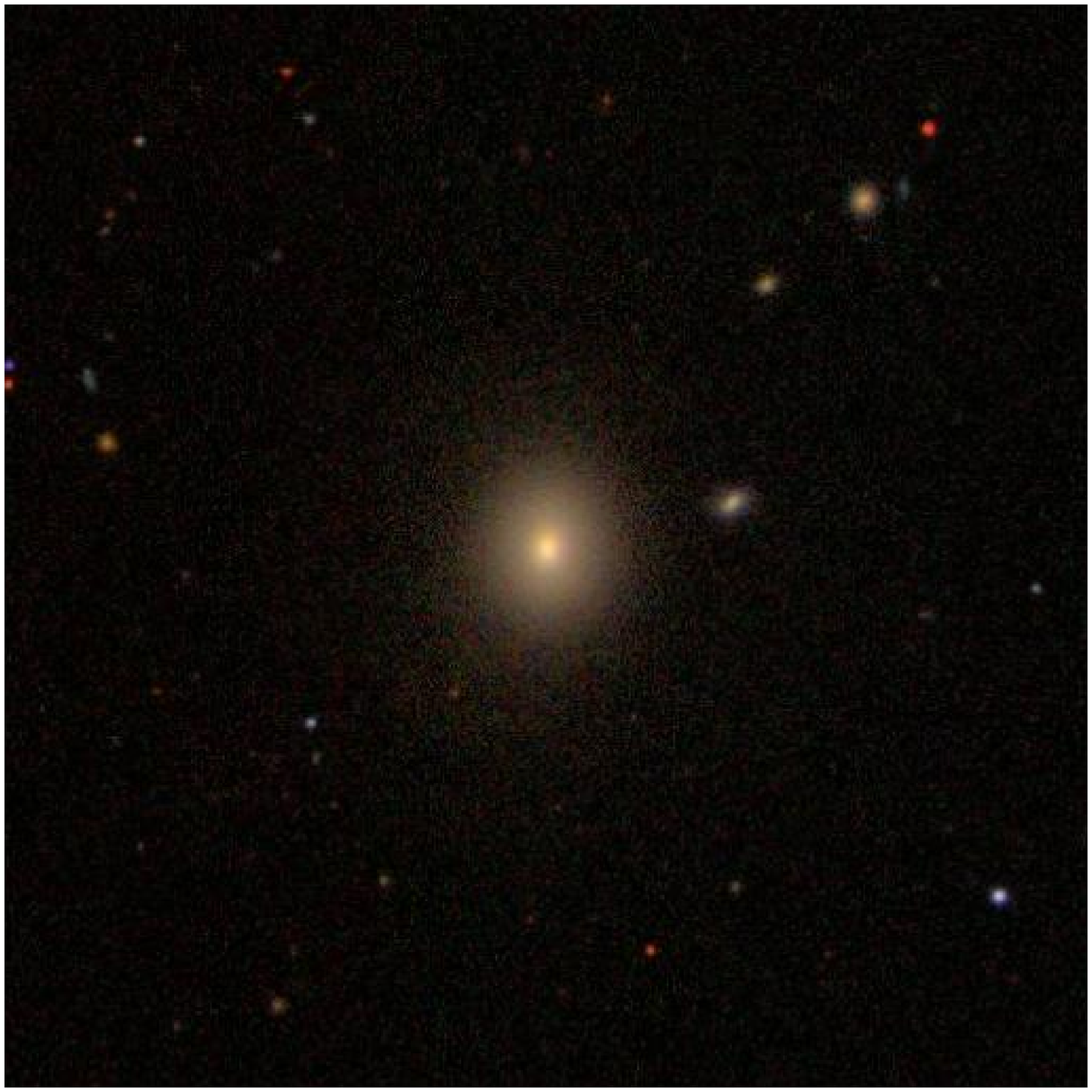}&\includegraphics[width=3in]{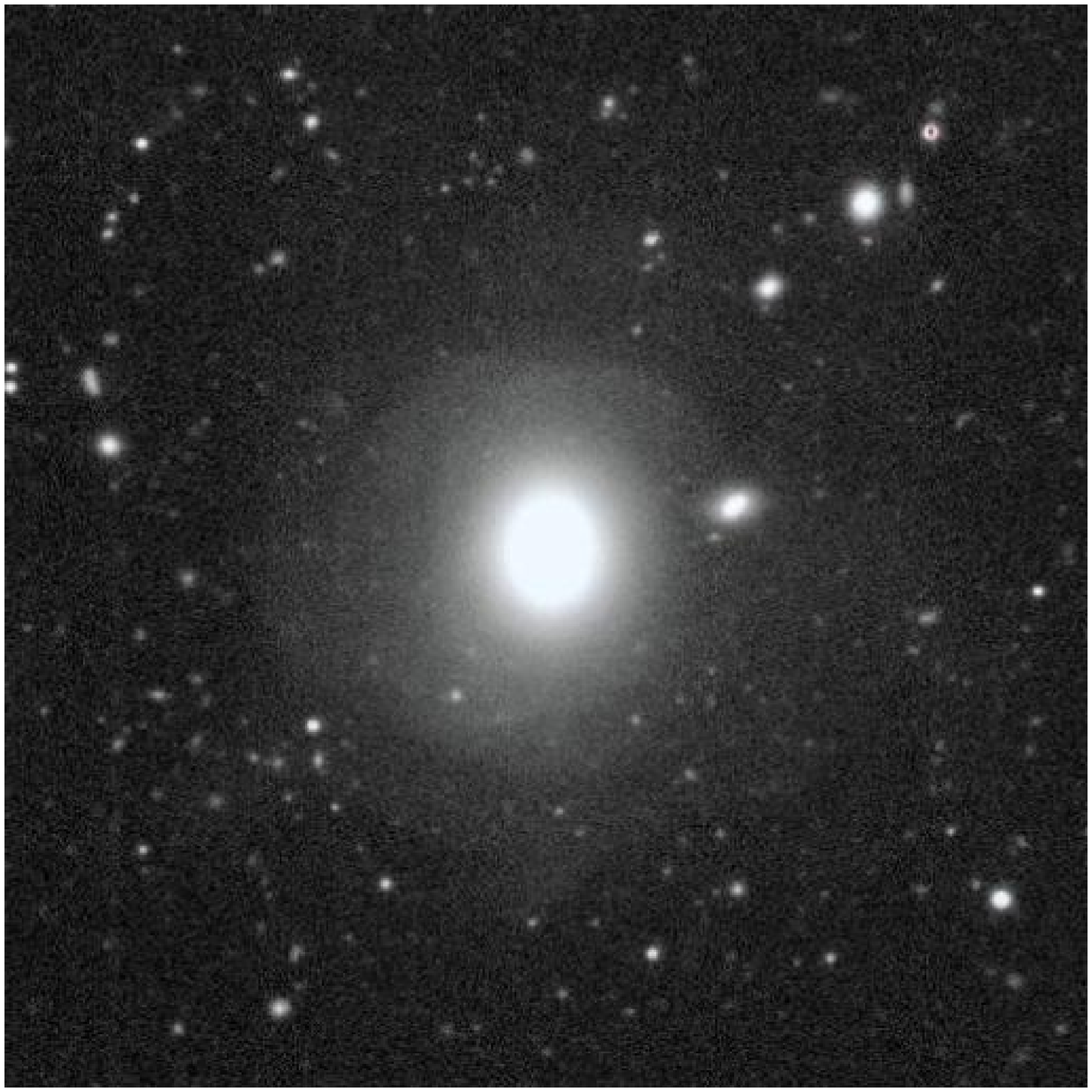}\\
\end{array}$
\caption{Examples of early-types with morphological disturbances
(shells and tidal debris). Note that the features are only visible
in the Stripe82 images (right-hand column) and are invisible in
the standard-depth images (left-hand column).}
\label{fig:tid_etg1}
\end{center}
\end{minipage}
\end{figure*}


\begin{figure*}
\begin{minipage}{172mm}
\begin{center}
$\begin{array}{cc}
\includegraphics[width=3in]{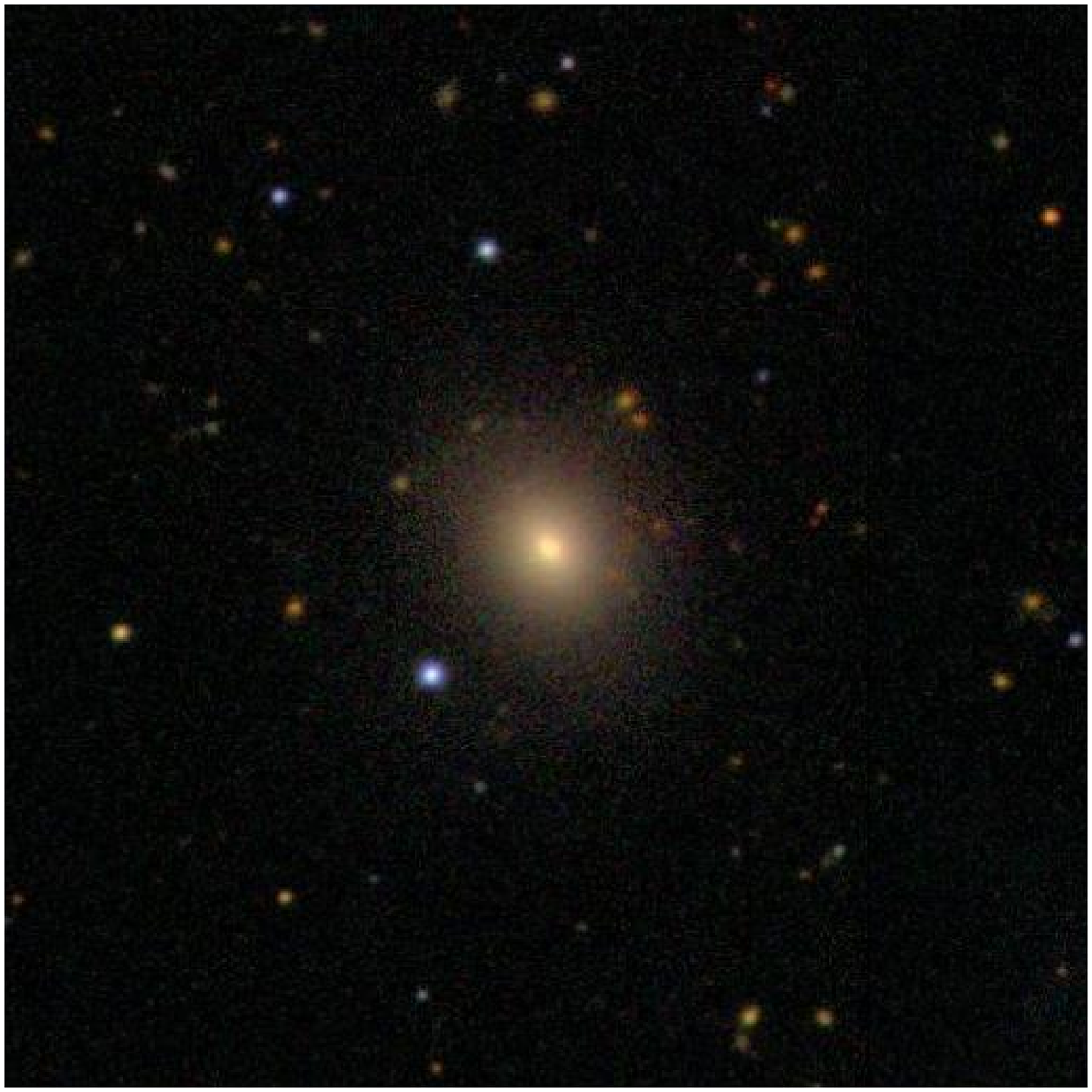}&\includegraphics[width=3in]{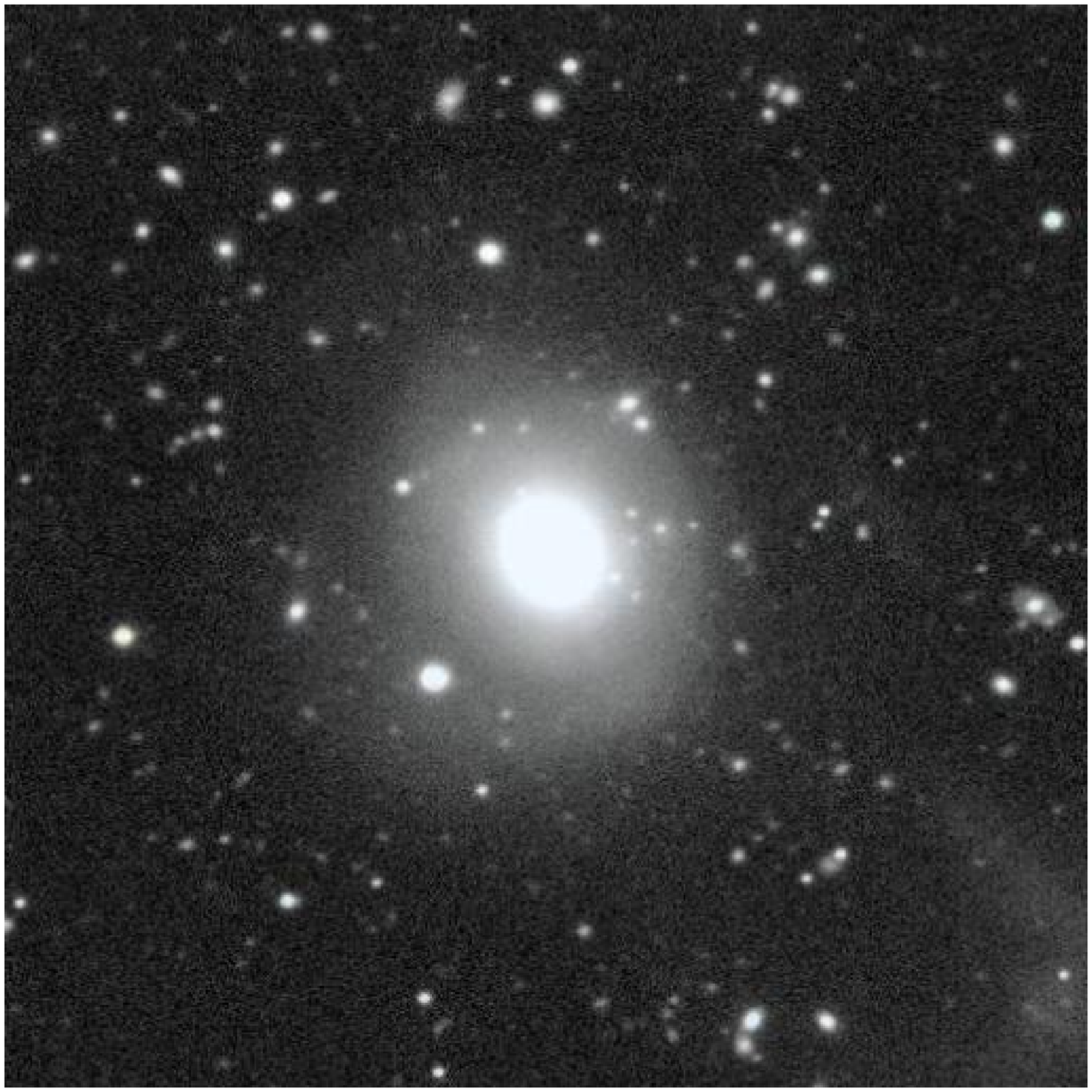}\\
\includegraphics[width=3in]{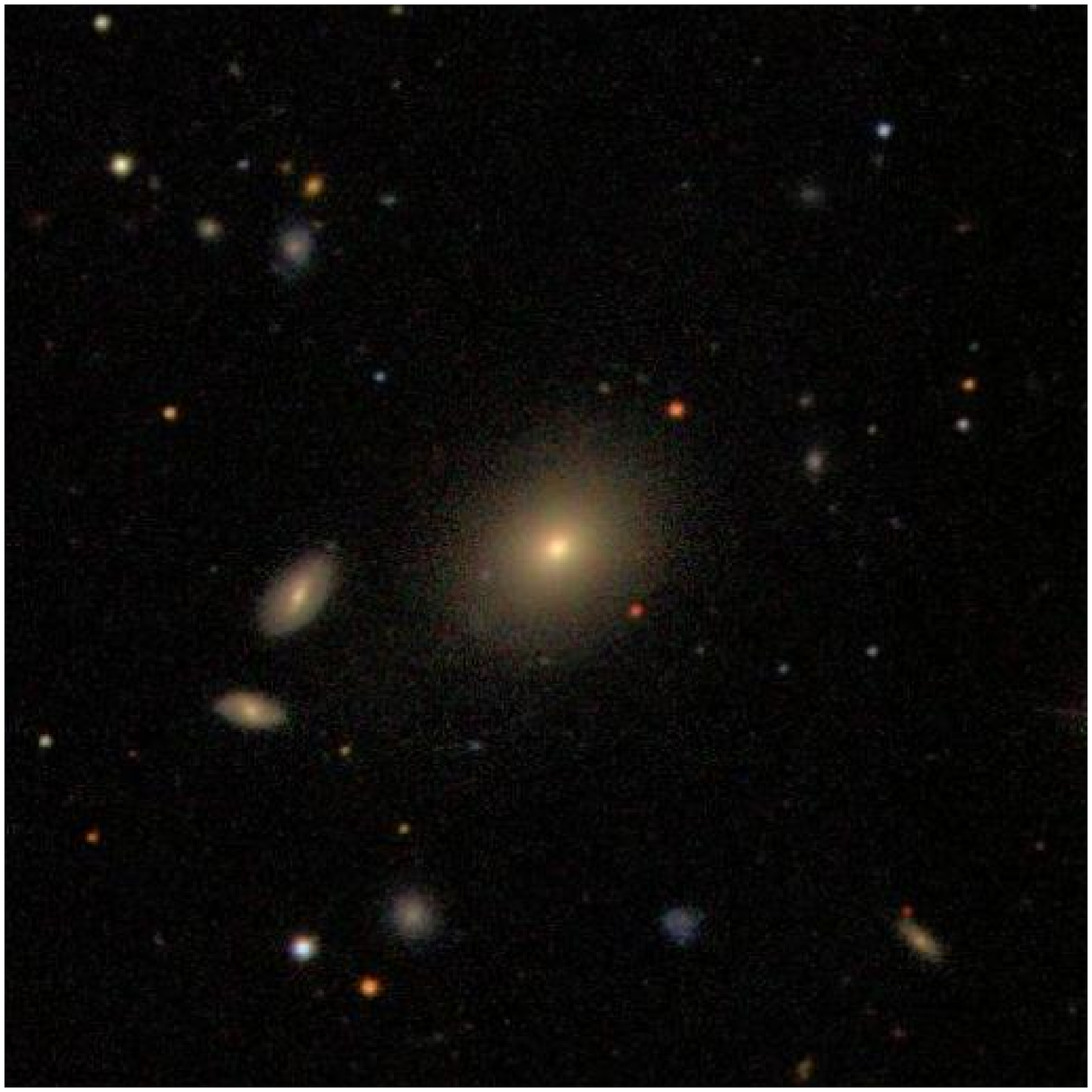}&\includegraphics[width=3in]{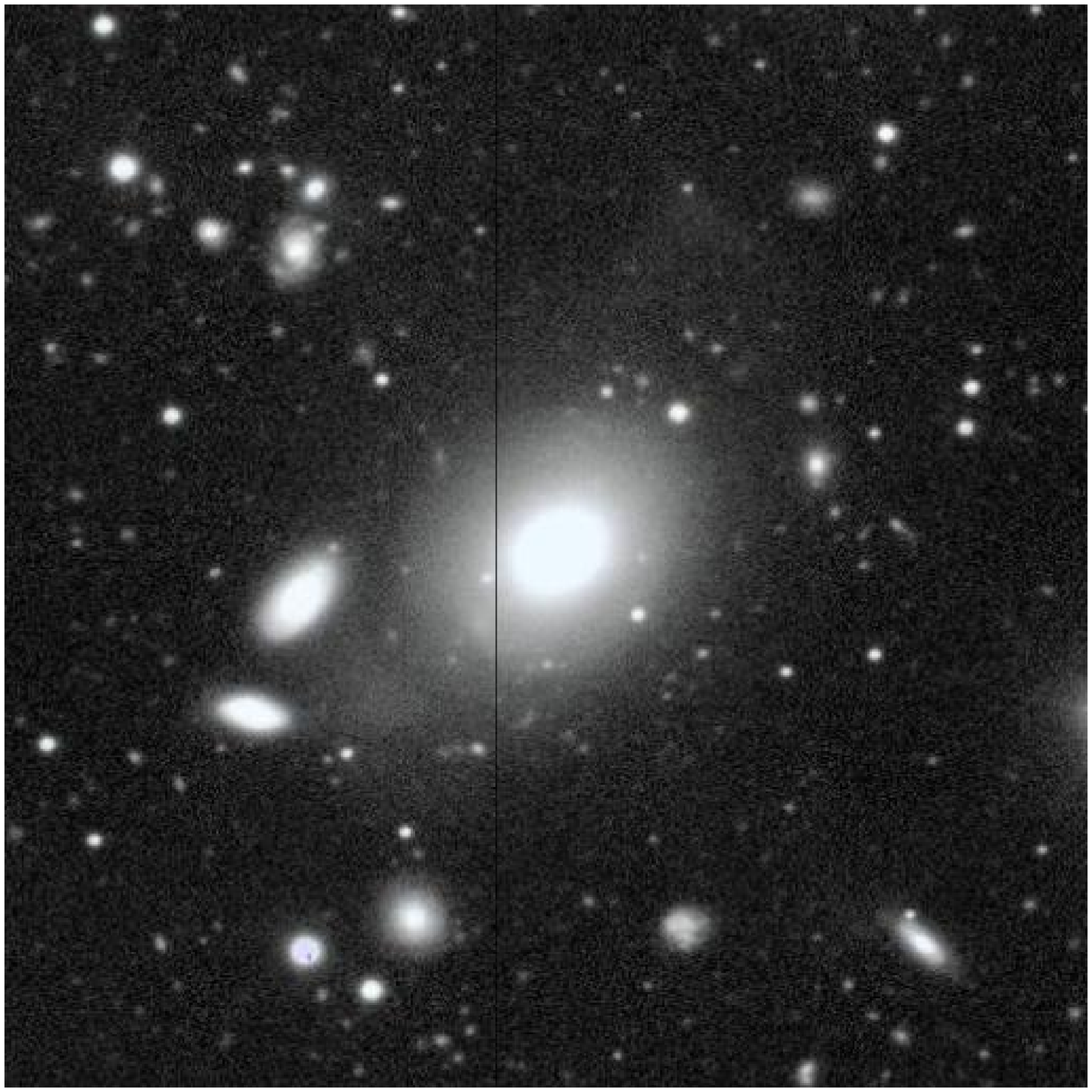}\\
\includegraphics[width=3in]{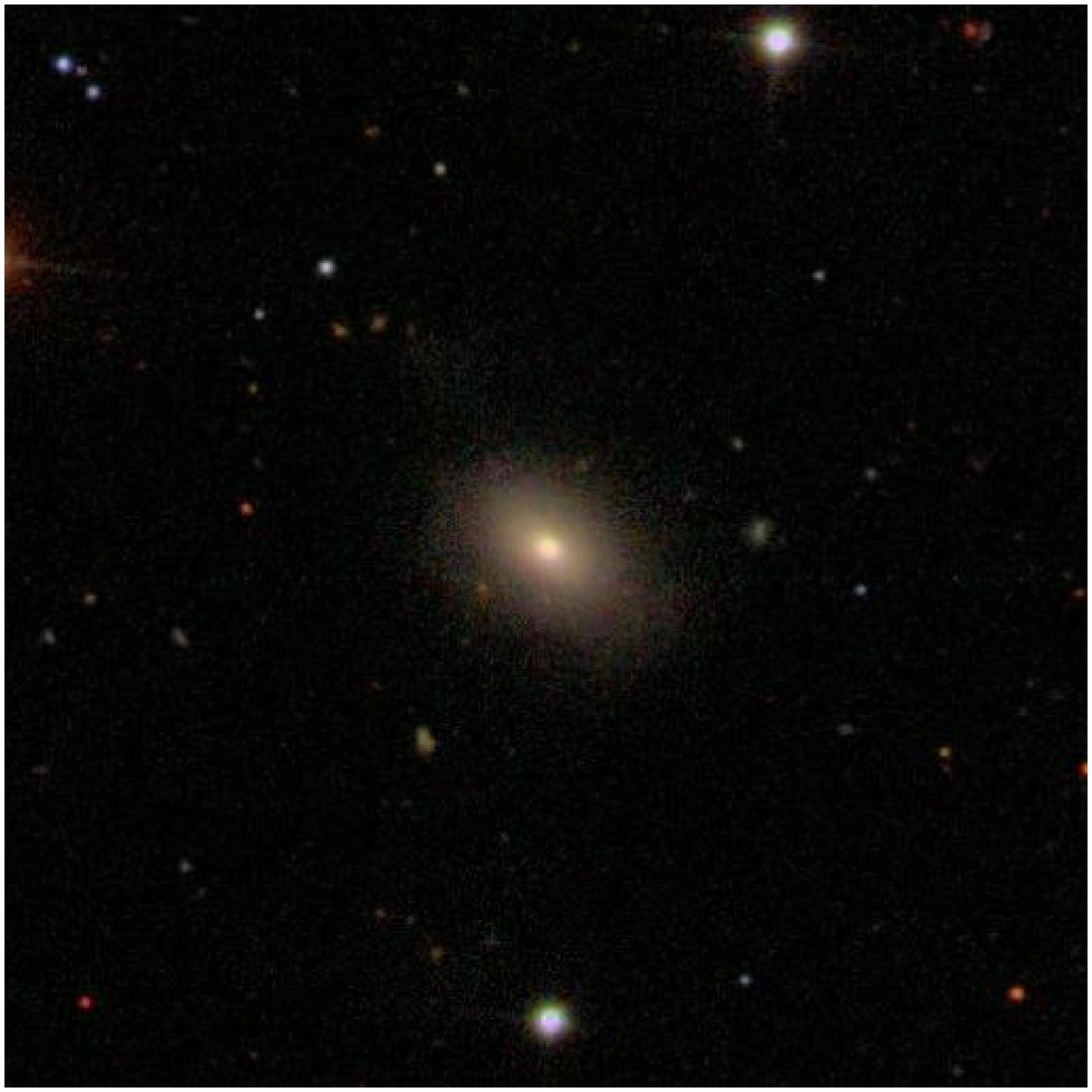}&\includegraphics[width=3in]{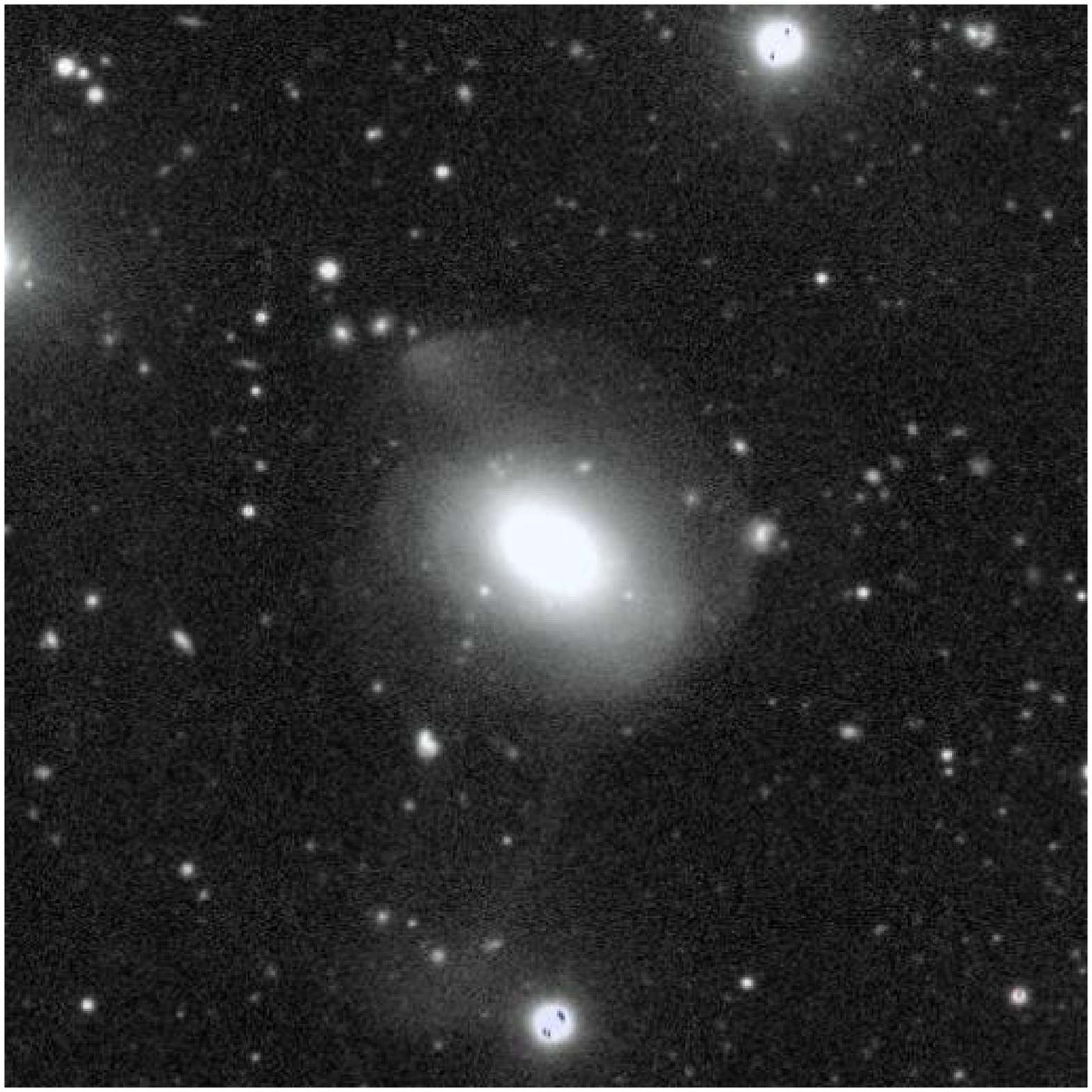}\\
\end{array}$
\caption{More examples of early-types with morphological
disturbances (shells and tidal debris). Note that the features are
only visible in the Stripe82 images (right-hand column) and are
invisible in the standard-depth images (left-hand column).}
\label{fig:tid_etg2}
\end{center}
\end{minipage}
\end{figure*}


\begin{figure*}
\begin{minipage}{172mm}
\begin{center}
$\begin{array}{cc}
\includegraphics[width=3in]{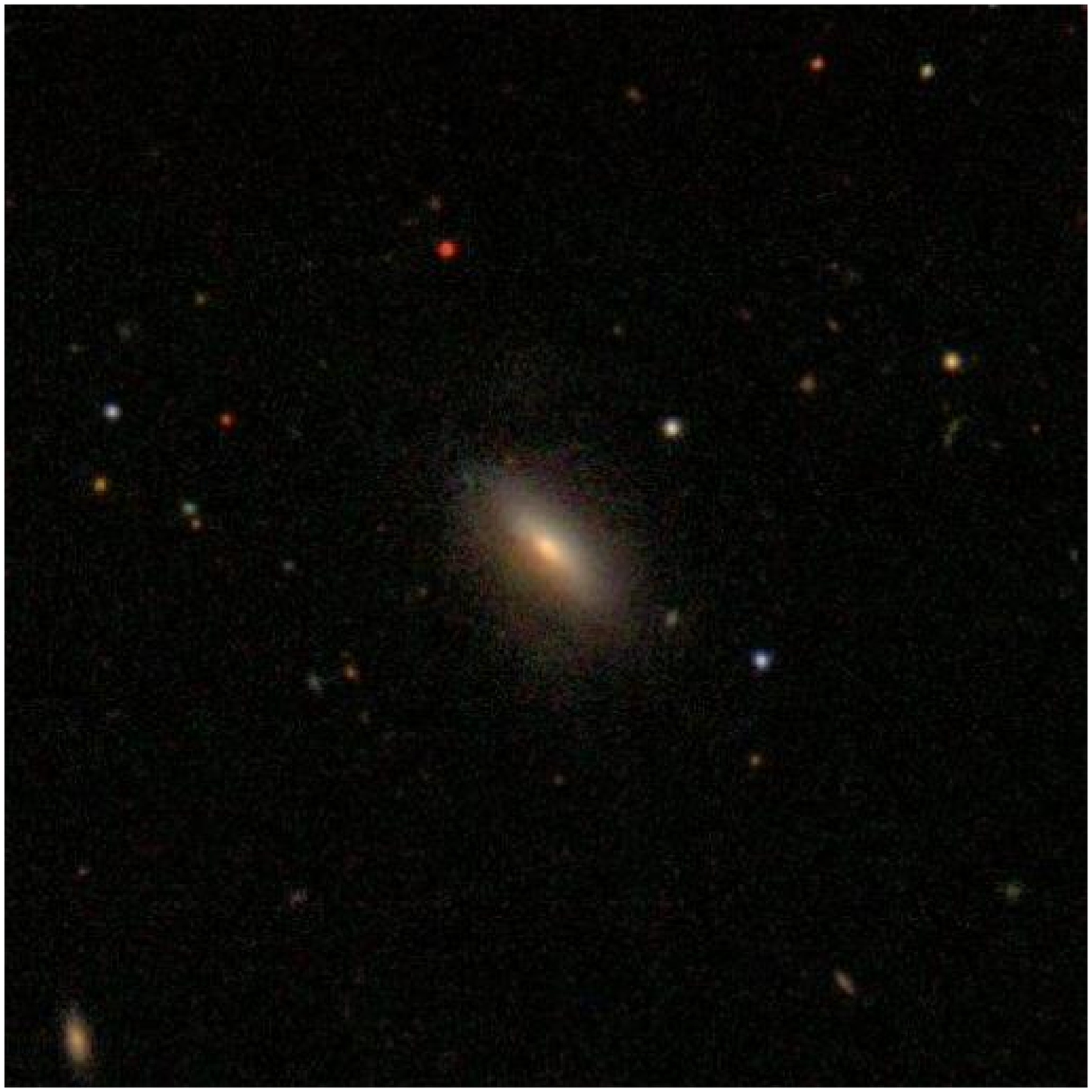}&\includegraphics[width=3in]{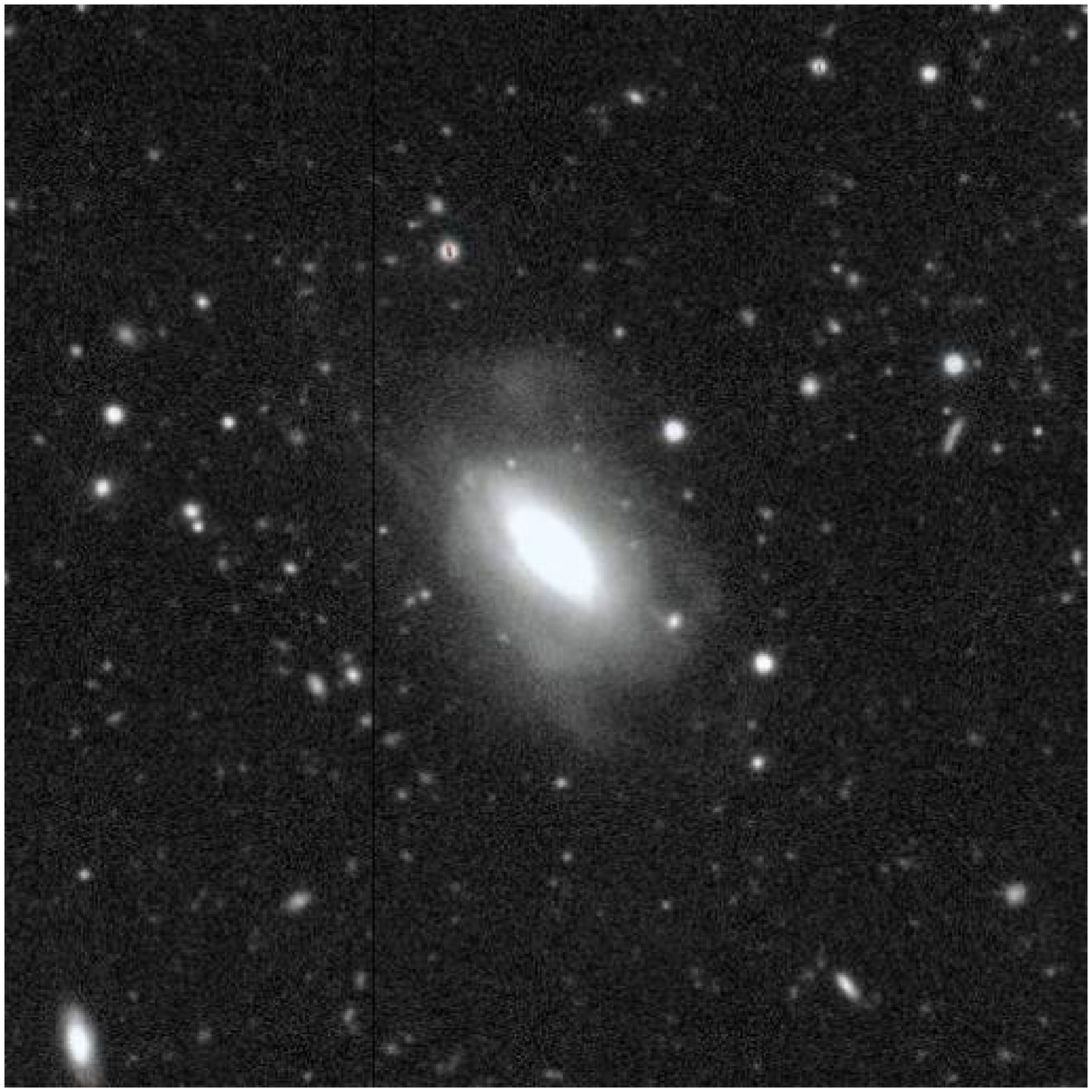}\\
\includegraphics[width=3in]{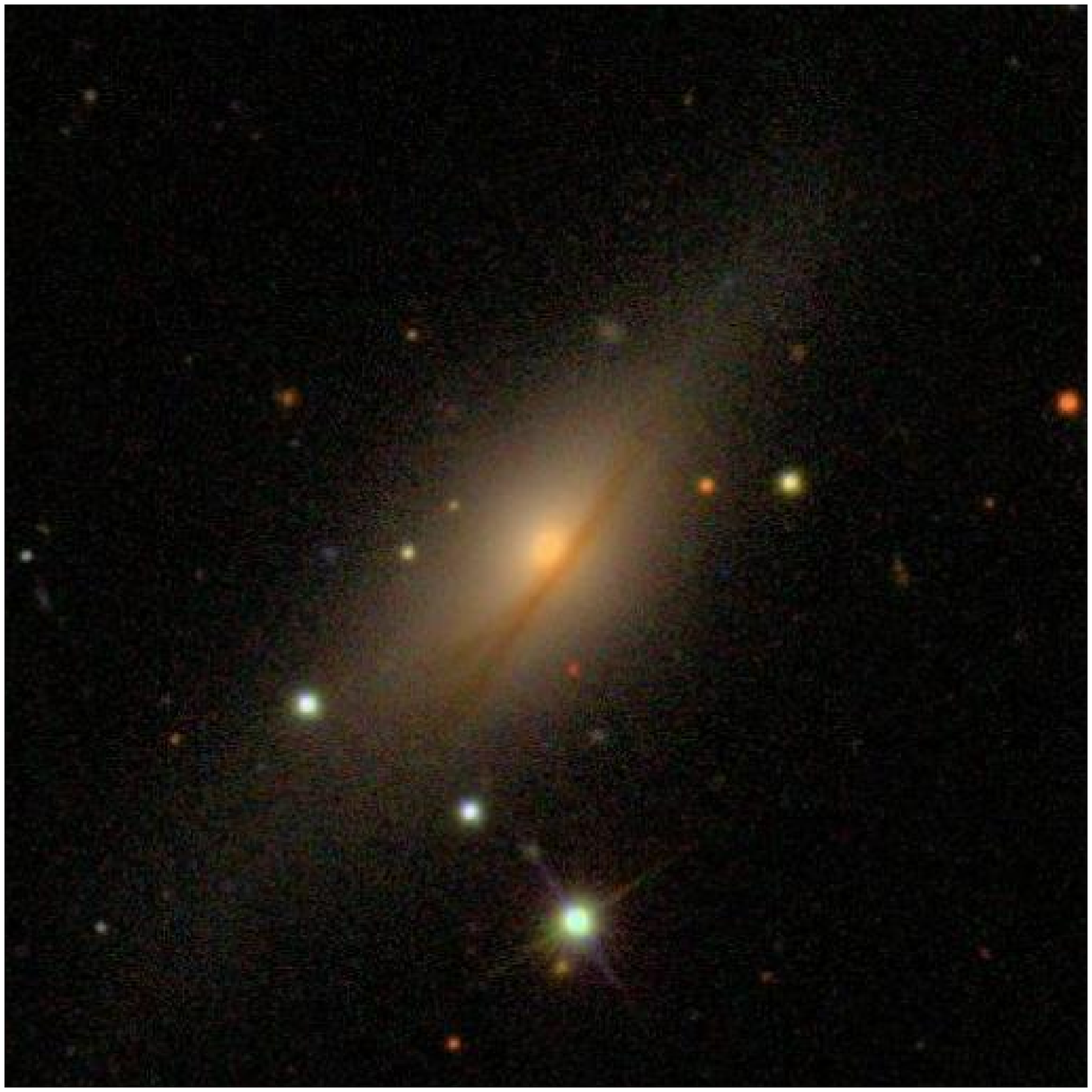}&\includegraphics[width=3in]{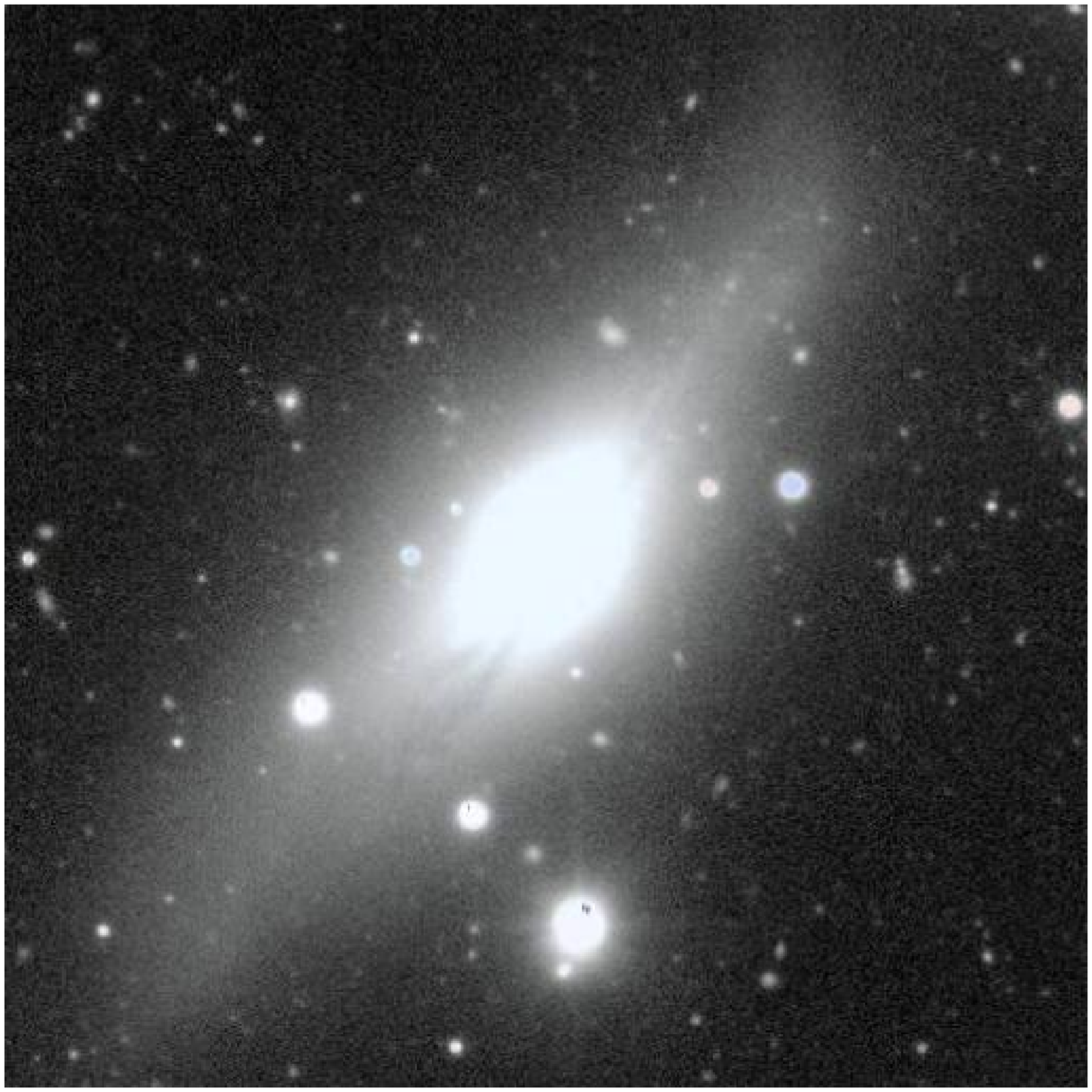}\\
\includegraphics[width=3in]{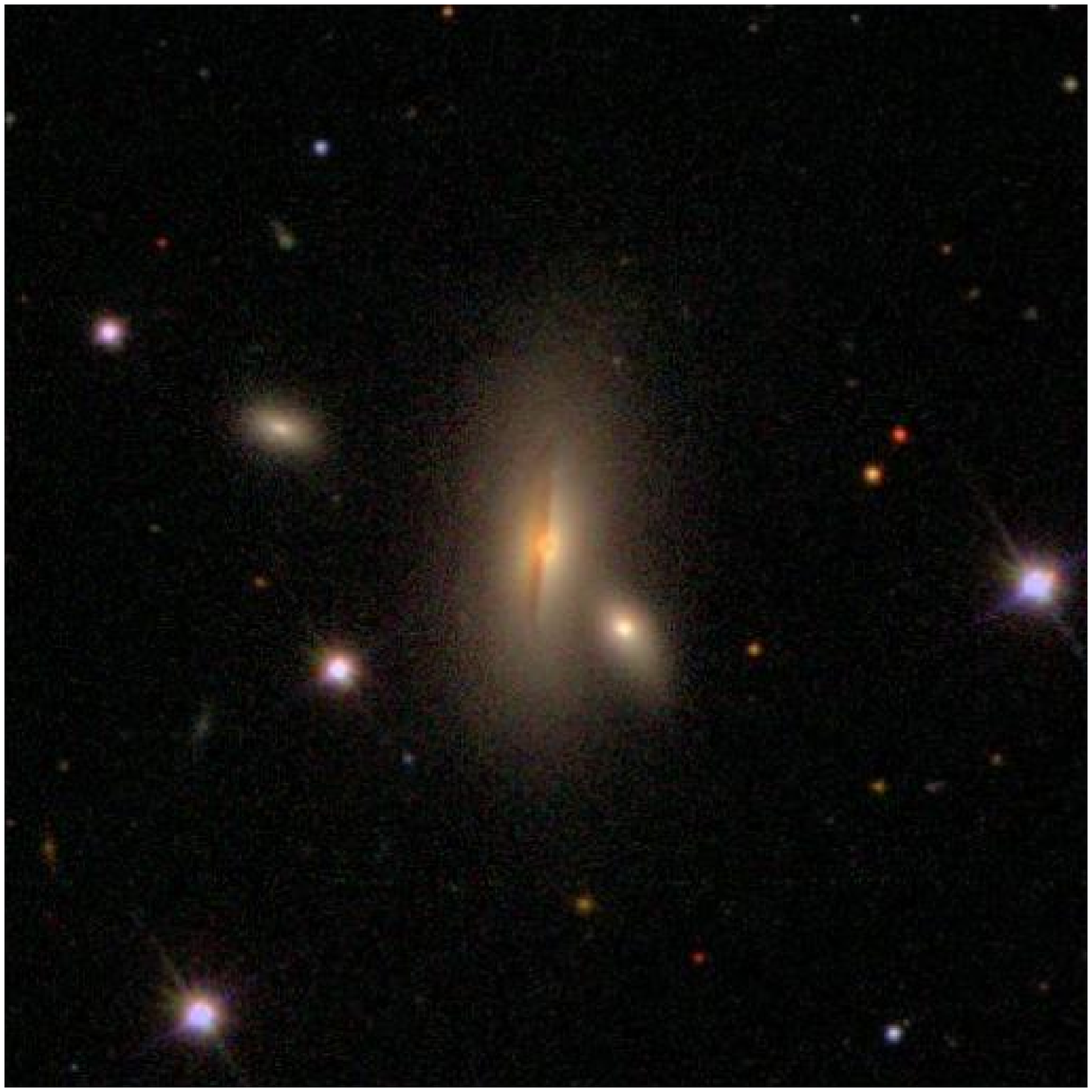}&\includegraphics[width=3in]{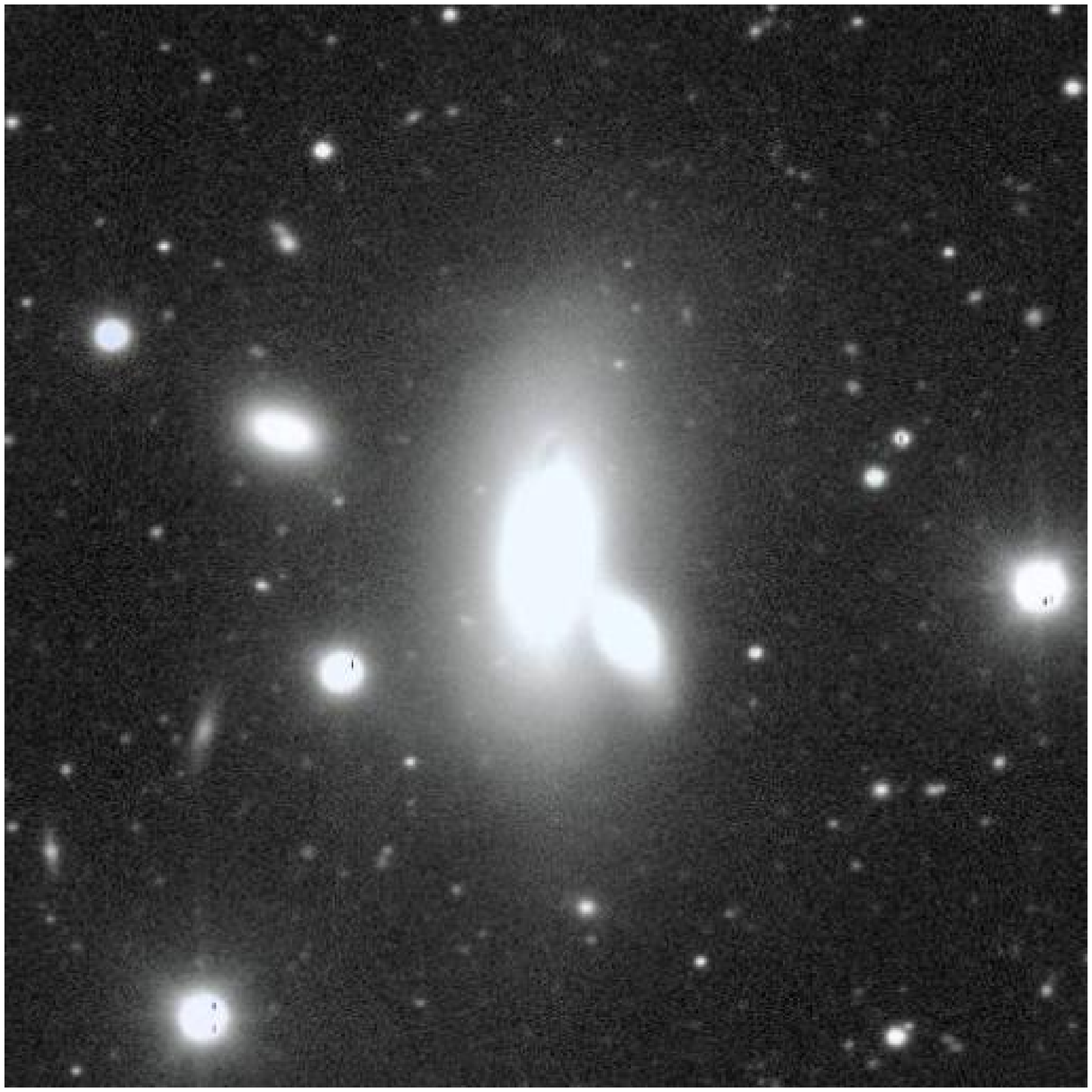}\\
\end{array}$
\caption{Examples of early-types with dust lanes, visible in the
multi-colour SDSS imaging. Typically these objects also show tidal
features in the deep Stripe82 imaging.} \label{fig:dust_etg}
\end{center}
\end{minipage}
\end{figure*}


\begin{figure*}
\begin{minipage}{172mm}
\begin{center}
$\begin{array}{cc}
\includegraphics[width=3in]{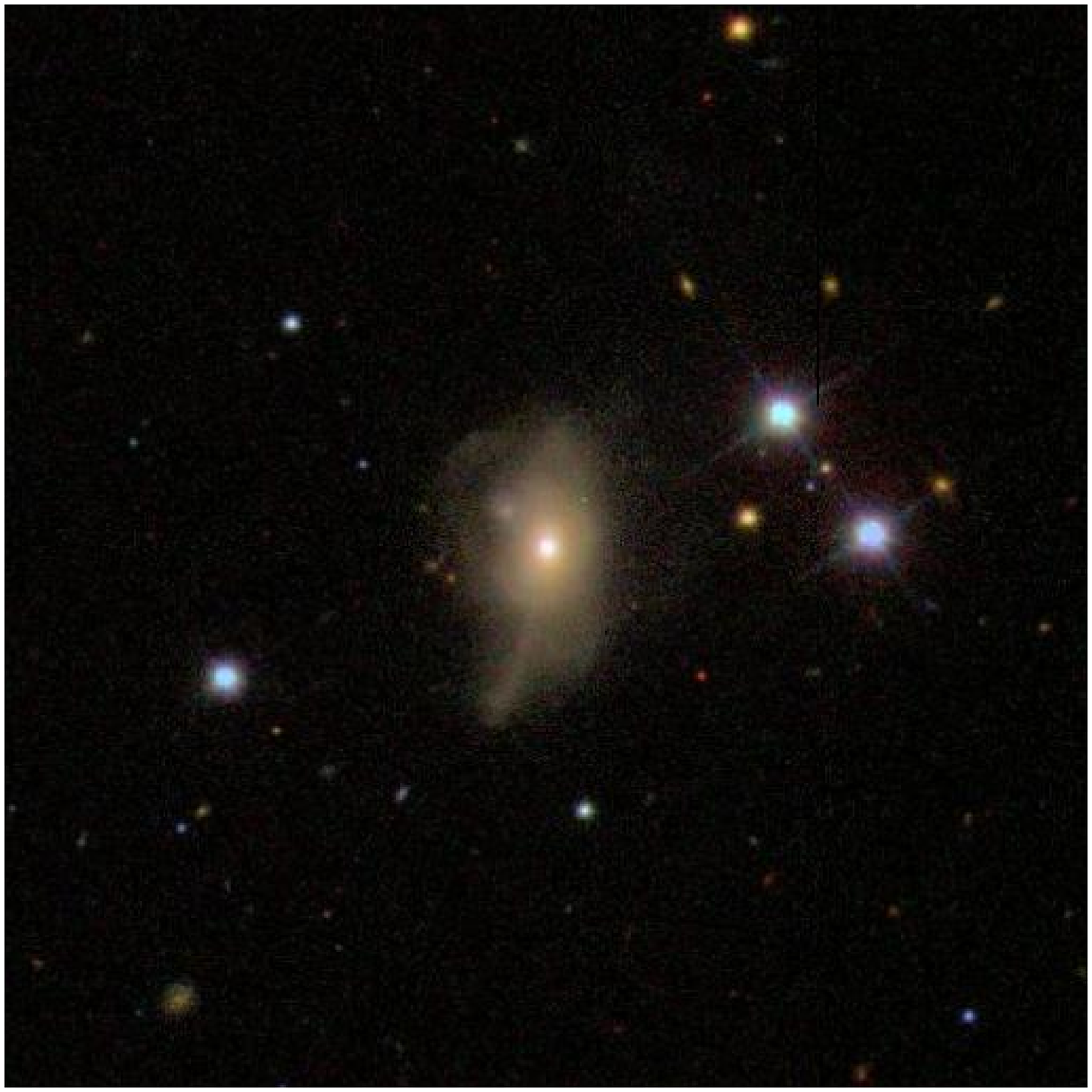}&\includegraphics[width=3in]{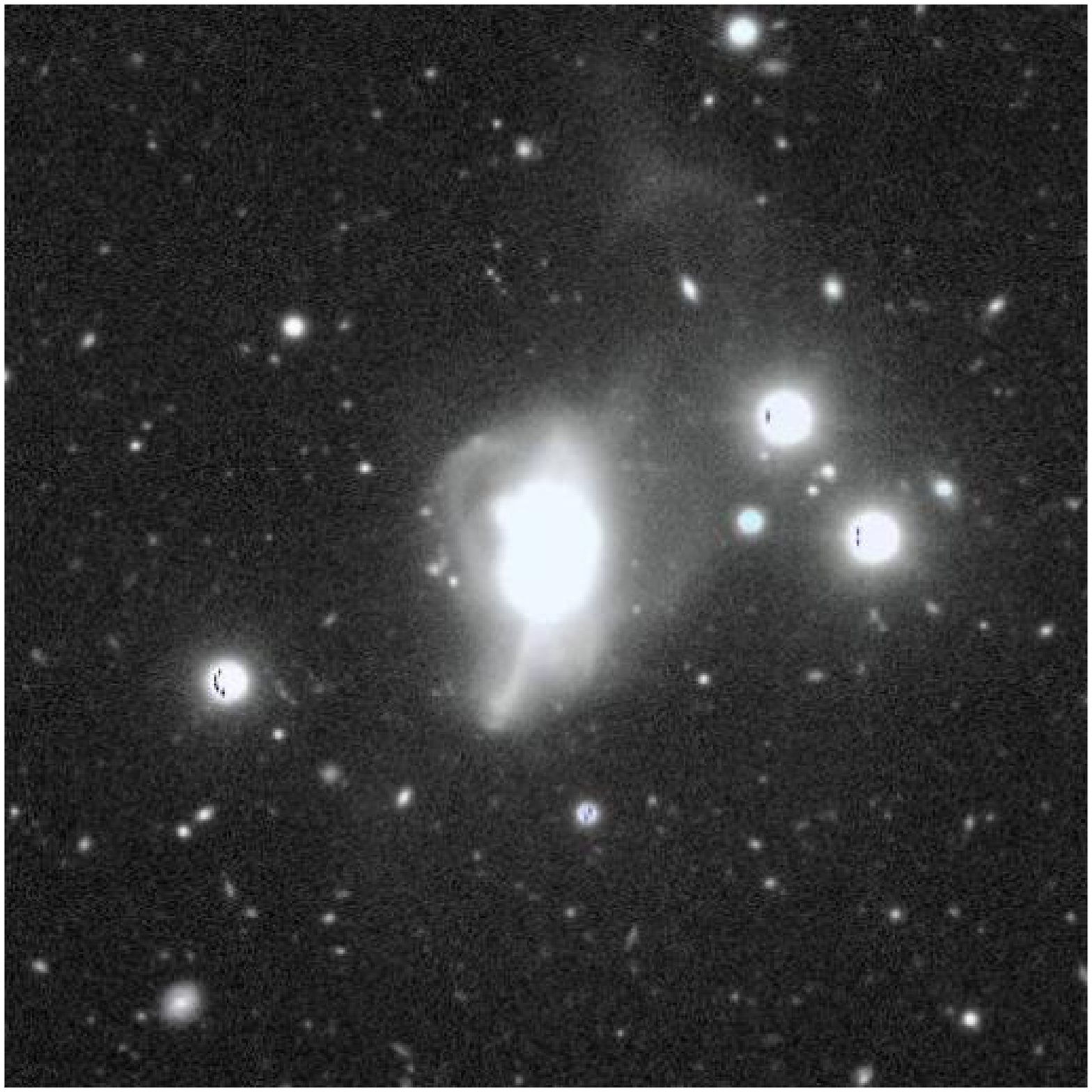}\\
\includegraphics[width=3in]{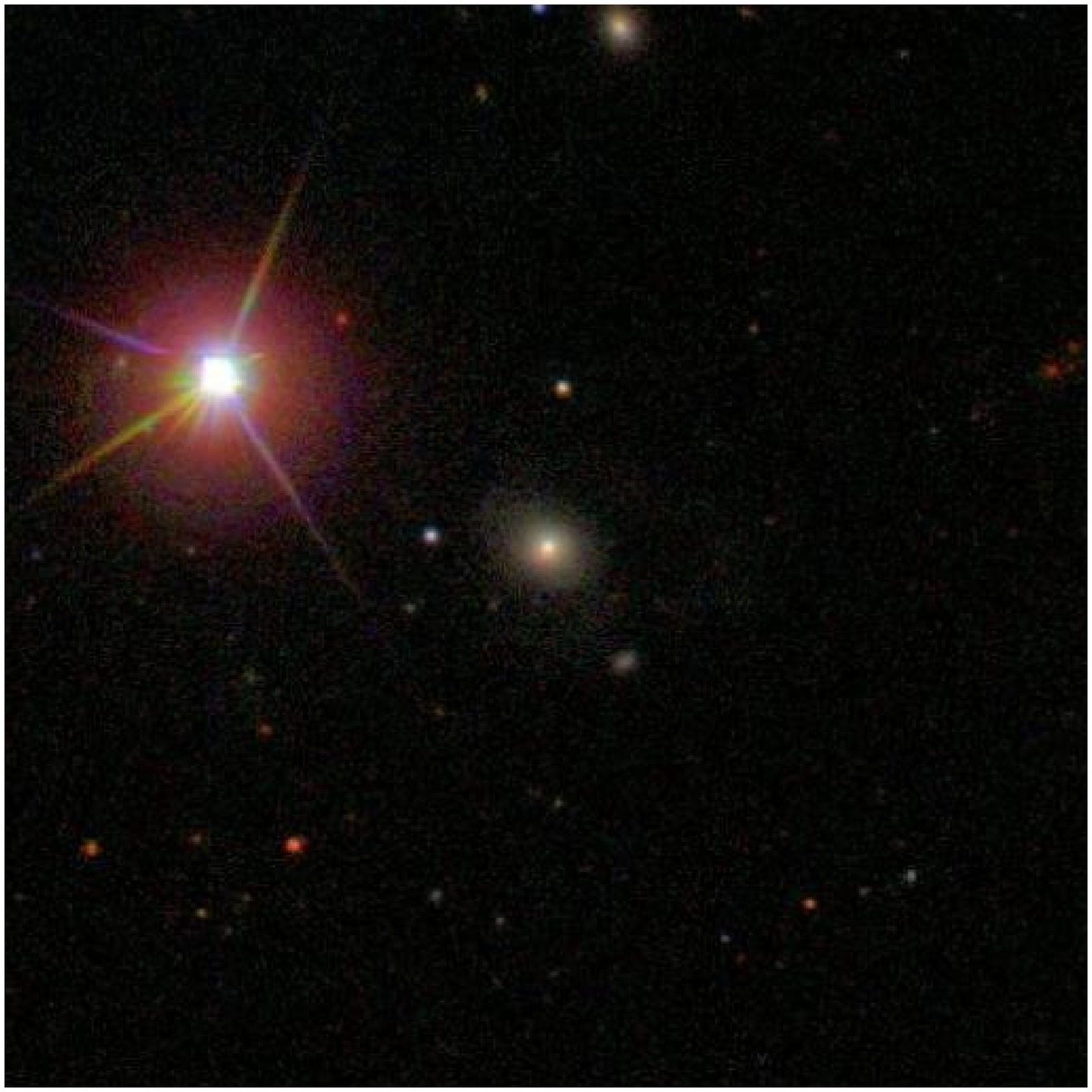}&\includegraphics[width=3in]{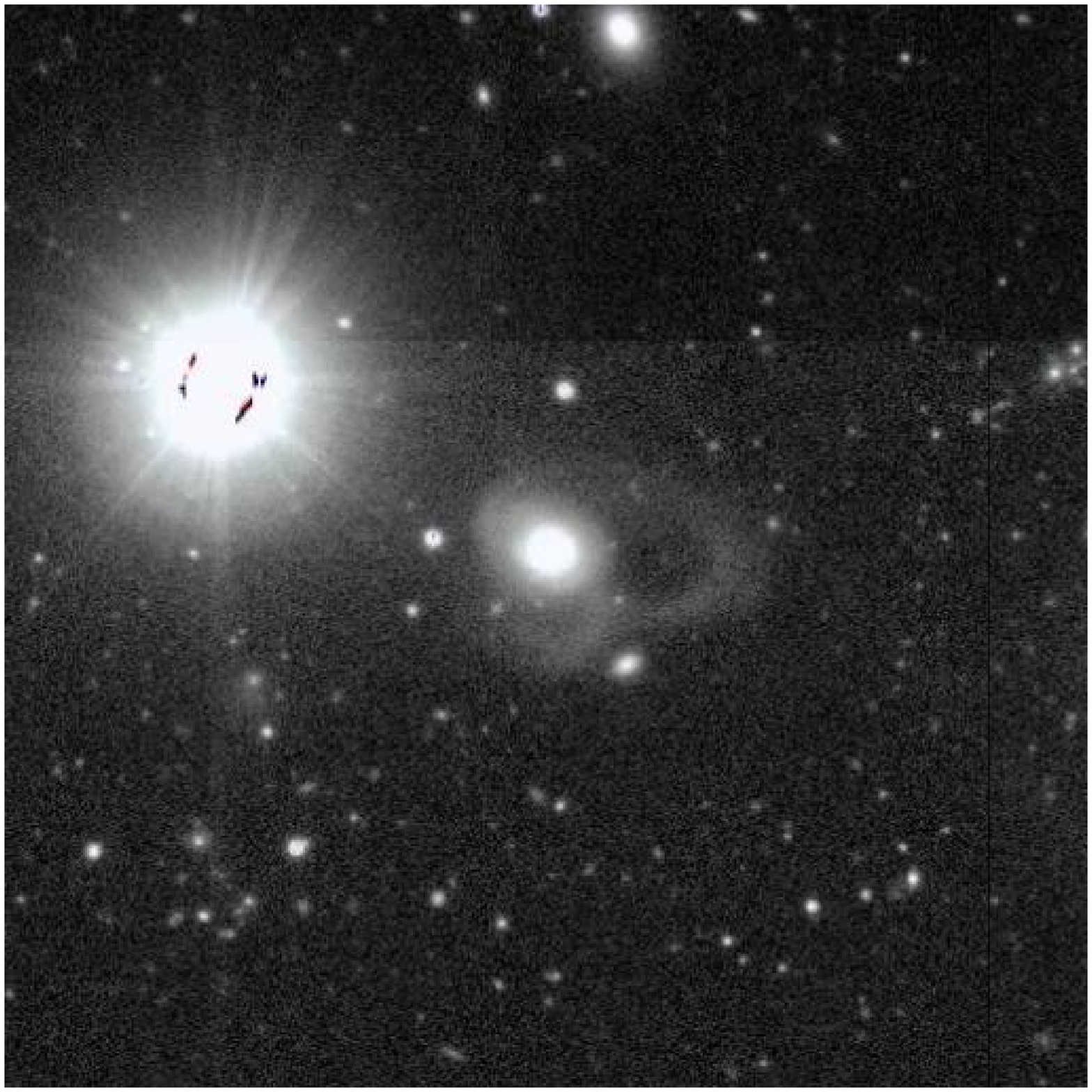}\\
\includegraphics[width=3in]{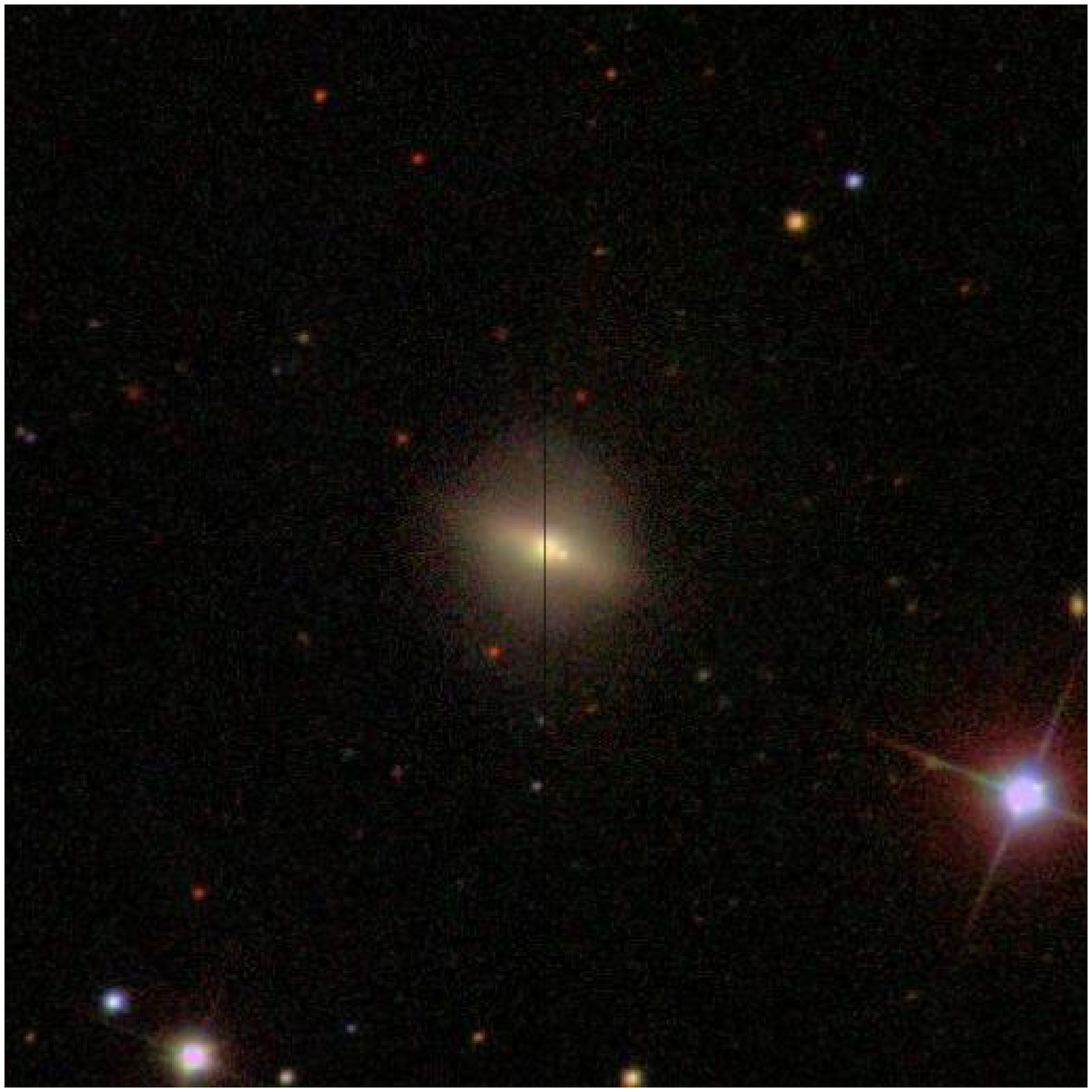}&\includegraphics[width=3in]{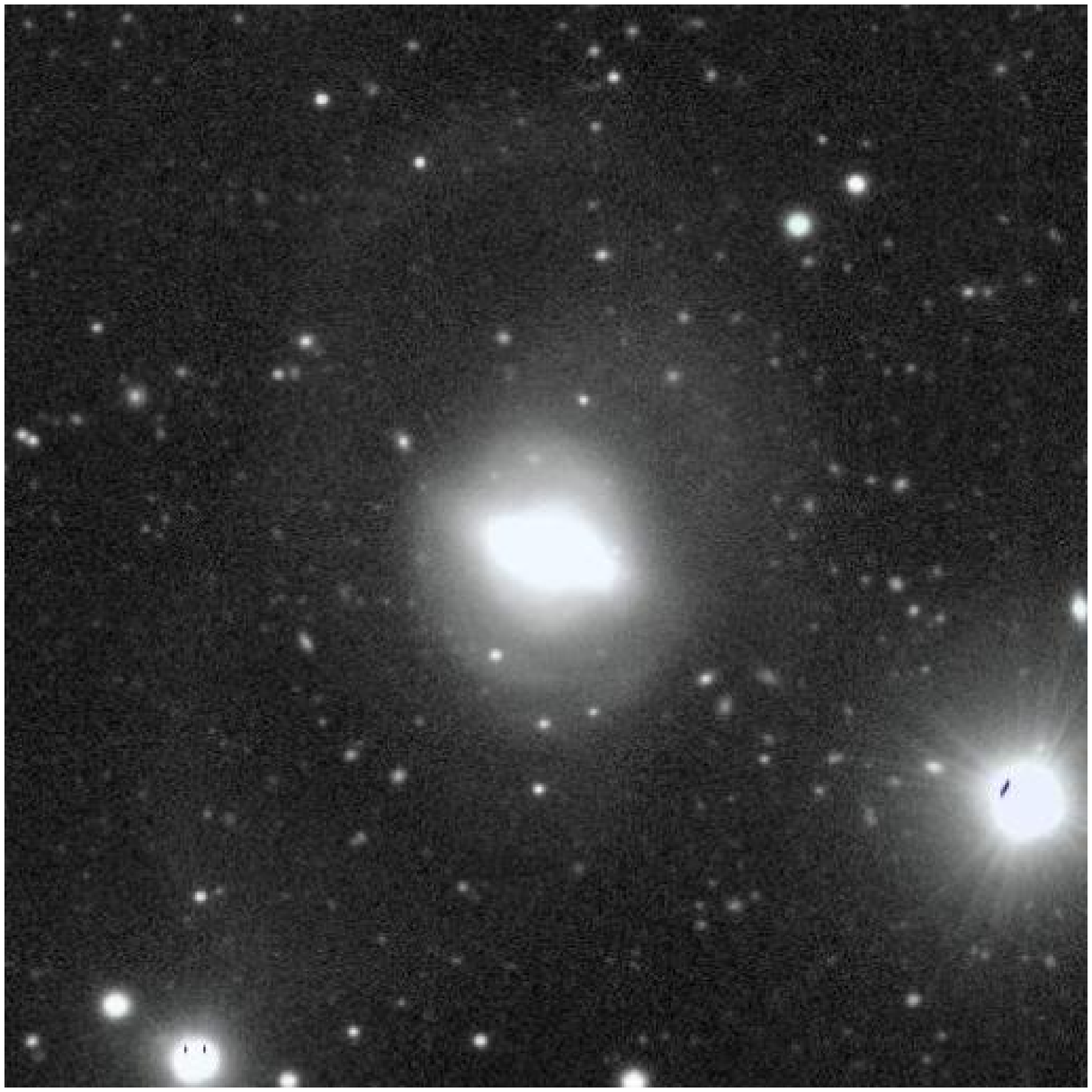}\\
\end{array}$
\caption{Examples of systems that appear to be the progenitors of
early-type galaxies, in the early stages of relaxation after a
recent merger or interaction.} \label{fig:prog_etg}
\end{center}
\end{minipage}
\end{figure*}


\section{Galaxy sample and visual inspection}
We begin by describing the process of visual inspection carried
out on the sample of galaxies explored in this study.
{\color{black} To maximise the accuracy of the morphological
classifications, we restrict ourselves to bright galaxies
($M_r<-20.5$) in the nearby Universe ($z<0.05$).The total number
of galaxies in the parent sample is 902.}

Each galaxy is morphologically classified by simultaneous
inspection of its multi-colour standard-depth image and its deeper
monochromatic Stripe82 counterpart. Note that the images used in
this study are those provided by the public SDSS DR7 release. The
objects are carefully assigned to three main morphological
classes: early-type galaxies (ETGs), late-type galaxies (LTGs) and
`Sa-like' systems which are bulge-dominated galaxies with faint
spiral features that are visible in the Stripe82 imaging but
invisible in the standard-depth images. The ETG population is
further sub-divided into the following categories:\\

\noindent\textbf{(1) Relaxed ETGs}\\
Early-type galaxies that do not exhibit any signs of morphological
disturbances at the depth of the Stripe82 images.\\

\noindent\textbf{(2) ETGs with tidal features}\\
Early-type galaxies that exhibit shells, fans and tidal tails
(typically in the Stripe82 images). {\color{black}Note that this
category includes a small number of systems that are in the early
stages of relaxation after a merger and could be early-type
progenitors.}\\

\noindent\textbf{(3) ETGs with dust features}\\
Galaxies that show the presence of dust patches and dust lanes
(typically in the multi-colour standard-depth imaging).\\

Note that, in what follows, we refer to categories (2) and (3)
collectively as `peculiar ETGs'. The classification process is
repeated until the numbers in the ETG, LTG and Sa categories begin
to converge and the numbers in these categories between
classification runs are within 1\% of each other. In our case,
such convergence was achieved after four such runs. Note also that
between runs 3 and 4 the sample of peculiar ETGs and Sa-like
systems did not change.

We find that $\sim34$\% of the Stripe82 galaxy population are
classified as ETGs. At the depth of the Stripe82 and DR7 images,
$\sim18$\% of ETGs show tidal features, while $\sim$7\% show
evidence for dust lanes and patches.

In Figure \ref{fig:etg_ltg} we present examples of galaxies
classified as relaxed ETGs (top three rows). For comparison we
also show objects classified as LTGs (bottom three rows). The
standard-depth multi-colour image is shown in the left-hand column
and its deeper Stripe82 counterpart is shown on the right. Figures
\ref{fig:tid_etg1} and \ref{fig:tid_etg2} present typical examples
of peculiar ETGs that exhibit tidal features such as shells.
Notice that the features are invisible in the standard DR7 images.
Figure \ref{fig:dust_etg} presents examples of galaxies which
exhibit dust lanes. The presence of dust features is often
accompanied by morphological disturbances, indicating that the
dust may have been deposited by a recent interaction event.

{\color{black}In Figure \ref{fig:prog_etg} we present a subset of
the ETGs with tidal features that appear to be in the early stages
of relaxation into an early-type remnant. Note that, while
galaxies in this morphological subclass have clear bulges, the
`early-stage' nature of the merger remnant can often make it
difficult to pin down the morphology of the final remnant
robustly. For example, the galaxy in the bottom row of Figure
\ref{fig:prog_etg}, in addition to a strong bulge component, may
also be in the process of building a disk. Note that the
bulge-dominated nature of this object is strongly supported by
pipeline measures of morphology such as the SDSS
{\color{black}\tt{fracdev}} parameter, which measures the
contribution of a bulge profile to the overall light profile of
the galaxy; in principle {\color{black}\tt{fracdev}=0} indicates a
pure disk, while {\color{black}\tt{fracdev}=1} indicates a pure
bulge, although prior experience indicates that high fracdev
values do not exclude the possibility of a (visually detected)
disk-like component. The galaxy in question has
{\color{black}\tt{fracdev}=1} in all SDSS filters.

Recent work (e.g. Springel \& Hernquist 2005; Robertson et al.
2006; Hopkins et al. 2009) has shown that disks can be rebuilt in
(major) mergers where the initial gas fractions are high
($\gg0.1$). Feedback from star formation and supernovae act to
pressurise the interstellar medium, redistributing gas spatially
and to larger radii where it does not feel the internal torques
induced by the merger. As a result this gas does not lose angular
momentum and contribute to fuelling the starburst, allowing it to
then reform a disk. It is possible that the galaxy in the bottom
row of Figure \ref{fig:prog_etg} is an example of such a system
that is the product of a gas-rich major merger.}


{\color{black}\subsection{Comparison to previous studies of
morphologically disturbed early-types} It is useful to compare our
results to previous surveys of morphologically disturbed ETGs in
the local Universe. The detection of tidal features is clearly
dependent on the depth of the imaging in question or, more
specifically, on the surface brightness (SB) limits reached by the
survey. In the 1980s and 1990s two deep studies of early-type
galaxies were performed by \citet{Malin1983a} and
\citet{Schweizer1992}. Both studies reached surface brightness
(SB) limits of $\mu_{B} \sim 26.5$ mag arcsec$^{-2}$, which
translates to $\mu_{r} \sim 26$ mag arcsec$^{-2}$ for a purely old
stellar population. The Malin \& Carter study detected tidal
features, shells and ripples in roughly 17\% of their elliptical
galaxy sample (which was comprised of 137 objects). The SB limit
of the standard exposure SDSS $r$-band images is $\mu_r \sim 24$
mag arcsec$^{-2}$ (see the SDSS project
book\footnote{http://www.astro.princeton.edu/PBOOK/strategy/strategy.htm})
which yields an estimate of $\mu_r \sim 26$ mag arcsec$^{-2}$ for
the deeper SDSS Stripe82 images used in this study. Given the
similar SB limits, it is not surprising that the fraction of ETGs
labelled as morphological disturbed in this paper are similar to
the fractions reported by \citet{Malin1983a}.

The deepest survey of ETGs in the local Universe has been
performed by \citet{VD2005}, who studied the red sequence
population observed by the MUSYC (Gawiser et al. 2006) and NOAO
Deep Wide-Field (Jannuzi et al. 2004) surveys down to SB limits of
$\mu \sim 29$ mag arcsec$^{-2}$ (the point source limits were
$\sim$27 mags). The striking result from this study was that
$\sim70$\% of the ETGs on the optical red sequence showed signs of
disturbed morphologies. In other words, it seems likely that
morphological disturbances may be ubiquitous in the ETG population
and their successful detection is driven simply by the depth of
the imaging available. The peculiar ETG population catalogued here
is therefore biased towards disturbed ETGs with more prominent
morphological disturbances.}


\section{Emission line diagnostics: star formation and AGN activity}
\begin{figure}
\begin{center}
\includegraphics[width=3.5in]{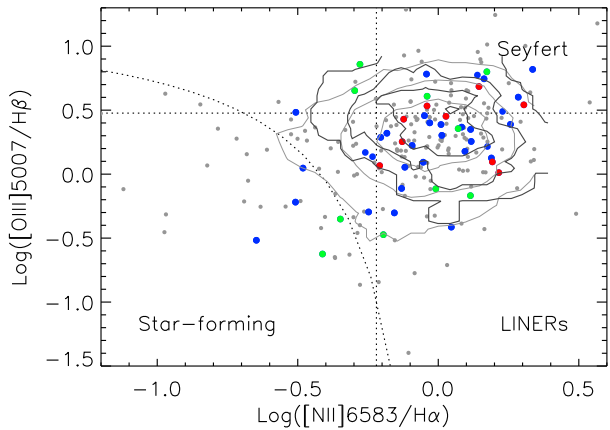}
\caption{A `BPT' plot for the Stripe82 ETG sample. Note that
galaxies only appear on this plot if the S/N in all emission lines
is $>3$. Grey points indicate ETGs that are relaxed. Peculiar ETGs
with tidal distortions are shown using blue points, those with
dust features are shown using red points and those which show both
dust features and tidal distortions are indicated using green
points. The dark contours indicate the distribution of peculiar
ETGs, while the grey contours indicate the distribution of relaxed
ETGs. The distribution of AGN types is similar between the relaxed
and peculiar ETG populations and dominated by LINER-like
activity.} \label{fig:bpt_plot}
\end{center}
\end{figure}

We begin our spectro-photometric analysis by exploring the
emission line diagnostics of our galaxy sample. Optical emission
line ratios are commonly used to determine the presence of (Type
II) AGN and separate the AGN population into its various types
using a `BPT' diagram (e.g. Baldwin et al. 1981; Kauffmann et al.
2003a). Figure \ref{fig:bpt_plot} shows the BPT diagram for the
Stripe82 galaxies. Note that galaxies only appear in this plot if
the S/N in all emission lines is $>3$. Grey points indicate ETGs
that are relaxed. ETGs with tidal distortions are shown using blue
points, ETGs with dust features are shown using red points and
ETGs which show both dust features and tidal distortions are
indicated using green points. The dark contours indicate the
distribution of peculiar ETGs while the grey contours indicate the
distribution of relaxed ETGs. Objects that lie below the curved
line are classified as `Star-forming'. Seyferts and LINERs occupy
separate sections on the right-hand side of the plot while
`Composites' (which host both star formation and AGN activity)
occupy the intermediate regions of the diagram\footnote{Note that
Composites are also sometimes called `Transition Region'
objects.}. Table 1 summaries the AGN properties of the Stripe82
sample.

{\color{black}It is worth noting that the fraction of ETGs that
are actively star-forming is typically very small. Only 1.3\% of
relaxed ETGs fall on the star-forming locus, while the equivalent
value for the peculiar ETG population is 4.5\%. Our results are
consistent with Schawinski et al. (2007), who find a similar value
for the fraction of star-forming ETGs (4.3\%). Table 1 indicates
that, not unexpectedly, the star-forming fraction in the peculiar
ETG population is a few factors higher than that in their relaxed
counterparts, presumably due to star formation associated with the
interaction events that produce the observed morphological
disturbances. The small fraction of actively star-forming ETGs
suggests that the peak star formation rates in these events is
low, consistent with what might be expected from gas-poor major
mergers or, more plausibly, minor mergers, that are predicted to
be the principal driver of star formation in ETGs at late epochs
(Kaviraj et al. 2009; Kaviraj et al. 2010). We revisit this issue
in Section 4, where the UV colour is used as a more detailed probe
of the recent star formation (RSF) in the ETG sample.}

{\color{black}We find that the fraction of objects classified as
`AGN' (i.e. Seyferts + LINERs) doubles in the peculiar ETGs
compared to that in their relaxed counterparts. However, it is
worth noting that the overwhelming bulk of the emission in both
relaxed and peculiar ETGs is LINER-like.} This type of activity
can be driven by low-intensity emission from AGN, stellar
photoionisation or shocks propagating through the gas. Recent work
by \citet{Stasinska2008} has shown that the presence of diffuse
LINER-like emission from stellar photoionisation can be
indistinguishable from AGN activity. The source of the emission is
better probed using integral-field-spectroscopy (IFS) data which
can be used to explore the spatial extent of the emission and
discriminate between central emission (indicative of an AGN) or
spatially extended (diffuse) emission from the stellar population
in the galaxy. Using IFS data {\color{black}Sarzi et al. (2010)}
have explored the nature of the ionised gas emission in the SAURON
sample of galaxies, originally studied by Sarzi et al. (2006).
Their results indicate that only in a small fraction of objects
that have radio and X-ray cores is the nebular emission consistent
with a central AGN. In particular, the source of the emission-line
activity in galaxies that carry signatures of RSF is likely to be
OB stars or a PAGB phase associated with the young stellar
component. In light of these results, it seems plausible that the
increased LINER activity observed in peculiar ETGs is primarily
driven by stellar photoionisation associated with the RSF that is
induced by recent interactions that give rise to these objects.
The enhanced star formation in peculiar ETGs is indicated both by
the higher fraction of peculiar ETGs that are classified as
`Star-forming' in the BPT diagnostics above and by an excess of
UV-blue colours in the population, as discussed in Section 4
below.

\begin{table}
\begin{center}
\caption{AGN properties of the Stripe82 early-type galaxy sample
in the magnitude range $M(r)<-20.5$. The AGNs in both the relaxed
and peculiar ETG populations are shown (columns 1 and 2) as well
as the change in the fractions of AGN types between peculiar and
relaxed ETGs (column 3).}
\begin{tabular}{cccc}

                      & Relaxed ETGs & Peculiar ETGs & f(pec)/f(rel)\\\hline\hline
    Star-forming      & 1.3\%        & 4.5\%         & 3.5\\
    Composites        & 1.3\%        & 1.5\%         & 1.1\\
    Seyferts          & 1.6\%        & 3.0\%         & 1.9\\
    LINERs            & 9.2\%        & 16.4\%        & 1.8\\

\end{tabular}
\end{center}
\label{tab:AGN_properties}
\end{table}


{\color{black}\section{Colours and recent star formation}} We now
turn to the photometric properties of the Stripe82 galaxy
population. We explore both the optical colours of our sample
(using photometry from the SDSS) and their UV colours (using
photometry from the recent GALEX space telescope, see Martin et
al. 2005). Recent work has shown that the UV spectrum is very
sensitive to small amounts of RSF. This makes the UV especially
useful in identifying ETGs that may have had small amounts of RSF
contributing a few percent of their stellar mass over the last
$\sim1$ Gyr (Kaviraj et al. 2007b; Kaviraj et al. 2008) and makes
it particularly pertinent to the study of the peculiar ETG
population in our sample.

\begin{figure}
$\begin{array}{cc}
\includegraphics[width=3.5in]{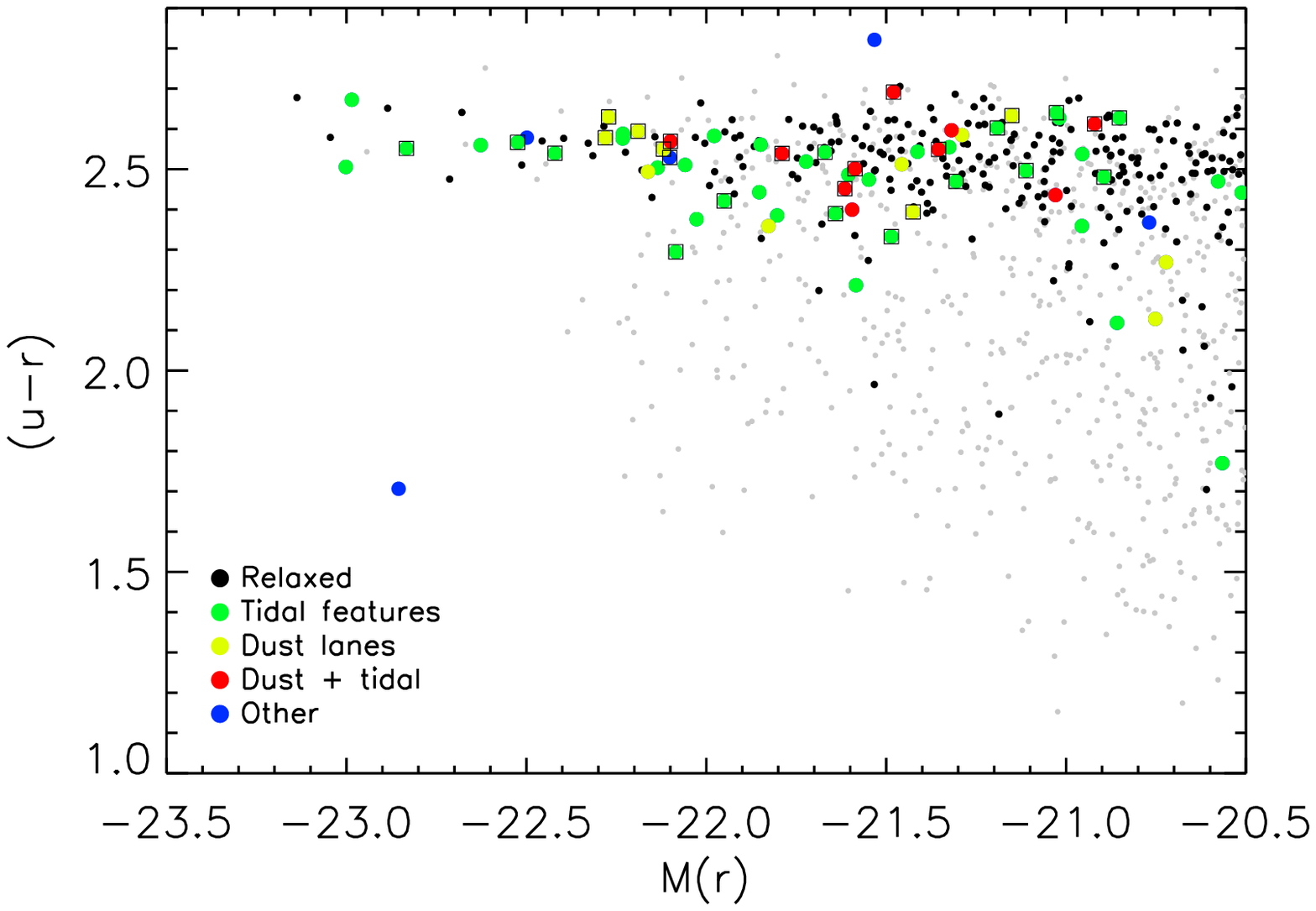}\\
\includegraphics[width=3.5in]{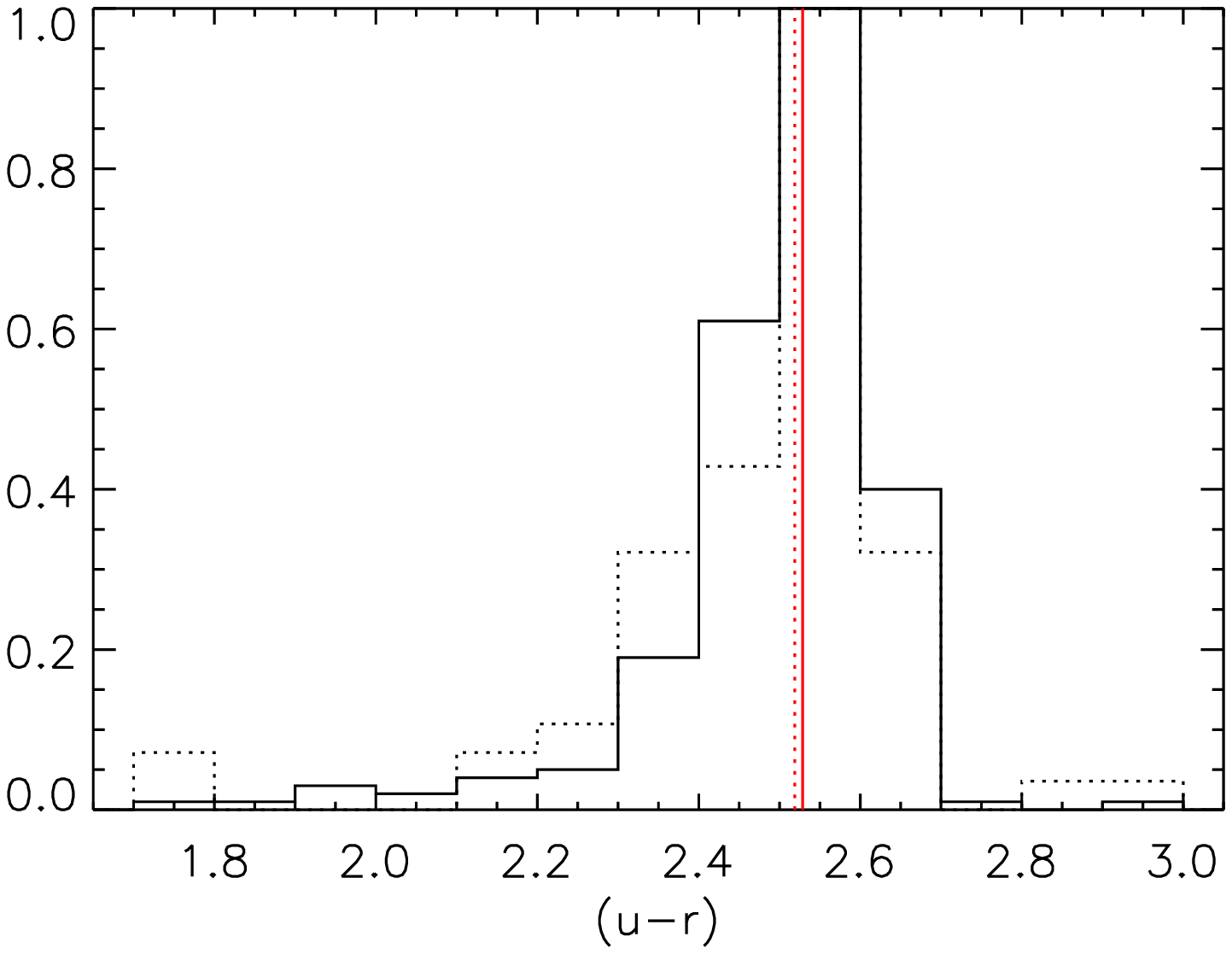}
\end{array}$
\caption{TOP: The optical $(u-r)$ colour-magnitude relation of the
Stripe82 galaxy sample. K-corrections from the SDSS pipeline have
been applied to the photometry. The relaxed ETGs are shown using
small black circles while the various classes of peculiar ETGs are
indicated using colours. Galaxies with Seyfert or LINER-like
emission are indicated using boxes. Late-type galaxies are
indicated using small grey circles. BOTTOM: The $(u-r)$ histogram
for the Stripe82 ETGs. {\color{black}Median} values of the
histograms are indicated using the red lines. The solid line
represents relaxed ETGs while the dotted line indicates peculiar
ETGs. The relaxed and peculiar ETG populations have very similar
distributions in the $(u-r)$ colour, indicating that any star
formation accompanying the interaction events is weak and does not
perturb the optical colours.} \label{fig:ur_cmr}
\end{figure}

In Figure \ref{fig:ur_cmr} we present the optical $(u-r)$
colour-magnitude relation (CMR; top) of the Stripe82 galaxy sample
and the $(u-r)$ colour histograms (bottom) for the relaxed and
peculiar ETG populations. The relaxed ETGs are shown using small
black circles while the various classes of peculiar ETGs are
indicated using colours. Galaxies with Seyfert or LINER-like
emission are indicated using boxes. Late-type galaxies are
indicated in grey. Median values of the histograms are indicated
using the red lines. We find that the relaxed and peculiar ETG
populations have very similar optical colour distributions with
the peculiar ETGs bluer, on average, by less than 0.05 mags.
{\color{black}We use a Mann-Whitney (MW; Walpole \& Myers 1985)
U-test to quantitatively compare the colour distributions of the
relaxed and peculiar ETGs and test the hypothesis that they have
the same median. For the $(u-r)$ colour, the MW test yields a
nearly-normal test statistic Z $\sim$ -0.5 and a probability
$p\sim 0.31$ of achieving a value greater than Z. Thus, at the 5\%
significance level we cannot reject the hypothesis that the two
distributions have the same median.}

\begin{figure}
$\begin{array}{cc}
\includegraphics[width=3.5in]{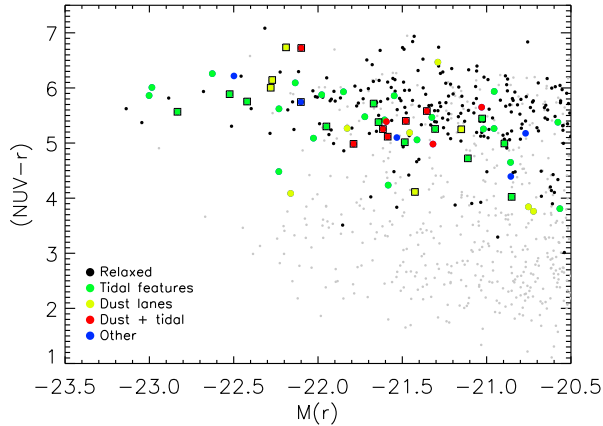}\\
\includegraphics[width=3.5in]{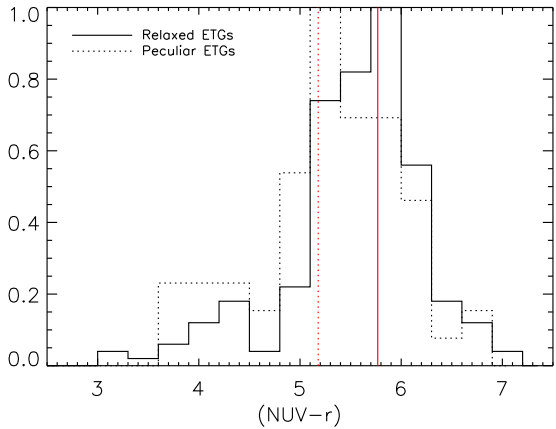}
\end{array}$
\caption{Same as Figure \ref{fig:ur_cmr} but for the $(NUV-r)$
colour.} \label{fig:nuv_cmr}
\end{figure}

Figure \ref{fig:nuv_cmr} shows the UV CMR for the Stripe82 galaxy
population (top) and $(NUV-r)$ colour histograms for the ETG
population, split into its relaxed and peculiar subsets (bottom).
Note that the GALEX $NUV$ filter is centred around 2300\AA.
{\color{black} In contrast to the optical analysis above, several
differences become visible between the UV colours of the two ETG
sub-populations. The top panel of this figure indicates that, in
the magnitude range $M(r)<-22$, the $(NUV-r)$ colour distribution
of the relaxed ETGs is virtually indistinguishable from that of
their peculiar counterparts. However, in the range $-22<M(r)<-21$
the peculiar ETGs become distinctly bluer in the $(NUV-r)$ colour,
which drives the overall offset in the two histograms in the
bottom panel of this figure. This difference in behaviour can be
quantified by noting that the slope of the $(NUV-r)$ CMR for the
relaxed ETGs is -0.27 while that for the peculiar ETGs is -0.61
i.e. twice the value for their relaxed counterparts. The
difference in the median $(NUV-r)$ of the two ETG subclasses,
driven by the blueward trend of the less luminous galaxies, is
$\sim0.6$ mags.

An MW test in the $(NUV-r)$ colour distributions of the relaxed
and peculiar ETG populations yields a p-value of 0.02, indicating
that, unlike in the case for $(u-r)$, the medians of these two
distributions are different at the 5\% significance level. To
further confirm this result we estimate the standard deviation in
the medians using bootstrapping and compare these to the
difference in the medians themselves. The standard deviation of
the ($NUV-r$) median for relaxed ETGs is 0.05 mags, while the
corresponding value for their peculiar counterparts is 0.09 mags.
The sum of these standard deviations (0.14 mags) is a few factors
smaller than the difference in the medians themselves (0.6 mags),
which supports the results of the MW test and indicates that the
$(NUV-r)$ colour distributions of the relaxed and peculiar ETGs
are indeed different.}


\subsection{Tracing recent star formation using the UV colour}
The UV offers an alternative route to exploring the RSF in
early-type galaxies. While emission lines such as H$\alpha$ trace
`instantaneous' star formation activity, the UV spectrum is an
integral of flux from the hot, massive, main sequence stellar
population in galaxies which survive for $\sim$1 Gyr. Furthermore,
the photometric nature of the indicator means that a robust
detection of UV flux can be made in \emph{all} ETGs within $z\sim
0.1$ (including those on the red sequence), while emission lines
are typically useful only if the S/N is satisfactorily large
($\geq 3$), which makes it difficult to study weakly star forming
ETGs using SDSS emission lines {\color{black}such as H${\alpha}$}.

At low redshift care needs to be taken in interpreting the UV flux
from early-type galaxies because there is a potential contribution
from old, horizontal branch stars that leads to the `UV upturn'
phenomenon around 1500\AA in some nearby giant ellipticals
\citep[see e.g.][]{Yi97,Yi99}. However, using the $(NUV-r)$ colour
of one of the strongest UV upturn galaxies as a guide, Kaviraj et
al. (2007b) have argued that objects bluer than $(NUV-r)\sim5.4$
are highly likely to harbour some RSF i.e. their UV colours cannot
be satisfied simply by invoking UV upturn flux.

In Table 2 we use this threshold to estimate the fraction of
relaxed and peculiar ETGs that have experienced some RSF over the
last Gyr. We find that, not unexpectedly, in all magnitude ranges,
the fraction of peculiar ETGs that have experienced RSF is a
factor of $\sim$1.5 higher than that in their relaxed
counterparts. The RSF fraction declines (in all categories) as we
restrict ourselves to more luminous galaxies as a result of the
downsizing phenomenon \citep[e.g.][]{Cowie1996}.

\begin{table}
\begin{center}
\caption{The fraction of ETGs, in various magnitude ranges, that
have $(NUV-r)<5.4$. Galaxies bluer than this colour are highly
likely to have experienced some recent star formation within the
last Gyr. Note that, regardless of the magnitude range considered,
the fraction of peculiar ETGs with RSF is a factor of $\sim$1.5
larger than its counterpart in the relaxed ETG population.}
\begin{tabular}{cccc}

                   & Relaxed ETGs & Peculiar ETGs & All ETGs \\\hline\hline
    $M(r)<-20.5$   & 34\%         & 54\%          & 38\%\\
    $M(r)<-21.0$   & 29\%         & 46\%          & 34\%\\
    $M(r)<-21.5$   & 26\%         & 41\%          & 31\%

\end{tabular}
\end{center}
\label{tab:RSF_properties}
\end{table}

\begin{figure}
$\begin{array}{cc}
\includegraphics[width=3.5in]{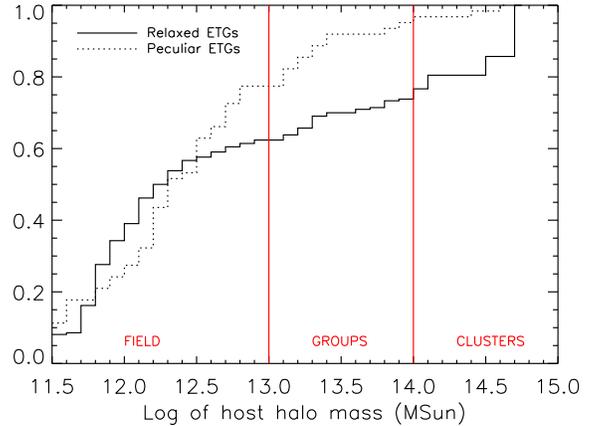}\\
\includegraphics[width=3.5in]{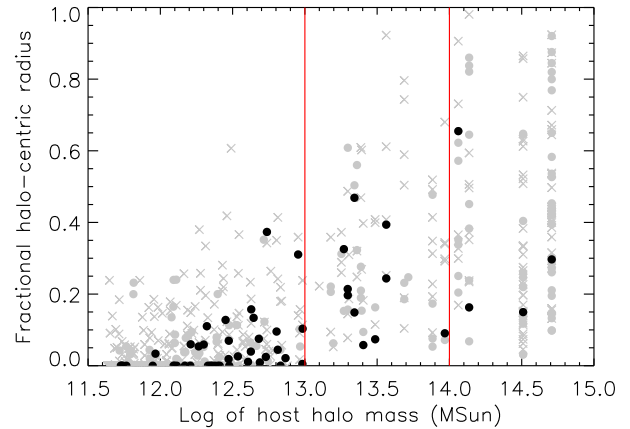}
\end{array}$
\caption{TOP: Cumulative histogram of host halo masses for the
relaxed ETGs (solid lines) and peculiar ETGs (dotted lines). The
`traditional' definitions of local environment are indicated in
red. BOTTOM: The fractional halo-centric radii plotted against the
host halo mass. Grey circles indicate relaxed ETGs, black circles
indicate peculiar ETGs and crosses indicate late-type galaxies.}
\label{fig:environment}
\end{figure}

The bluer colours of the peculiar ETGs in the UV and higher
fraction of objects with RSF in these galaxies is consistent with
the higher fraction of `Star-forming' and LINER-like objects in
the emission-line BPT diagnostics. If the source of the LINER
emission is stellar photoionisation then, taken together, these
results strongly suggest that {\color{black}\emph{the interaction
events that produce the morphological peculiarities are also
responsible for the higher levels of star formation and nebular
activity that is found in the peculiar ETG population}. It is
worth noting that these results are consistent with those of
\citet{Schweizer1990} and \citet{Schweizer1992}, who detected
enhanced H$\beta$ absorption and slightly bluer $UBV$ colours in
ETGs with more pronounced fine structure, and interpreted this as
evidence for lower mean ages. While the robustness of the analysis
in this paper clearly benefits from the added sensitivity of the
UV to RSF, it is reassuring that our results are qualitatively
consistent with those in the literature.

In light of the emission-line analysis and the UV/optical colours,
the properties of the peculiar ETGs offer useful insights into the
`dryness' of typical mergers that take place in the ETG population
at low redshift. The lack of impact of the RSF on the optical
colours and the virtually negligible fraction of `star-forming'
ETGs in the BPT diagnostics indicates that the intensity of star
formation in these events is low and that the encounters are not
particularly gas-rich in comparison to the stellar mass of the
remnant. This is supported by the UV colours of the peculiar ETGs
which, although offset from those of the relaxed ETGs, are largely
contained in the green valley and not the UV blue cloud. It is
likely, therefore, that the mergers that drive the star formation
(and evolution) of the ETG population at low redshift produce
low-level star formation events, consistent with the small mass
fractions of young stars (a few percent) that have been measured
in previous work (e.g. Kaviraj et al. 2008). The processes at work
are likely to involve either dry major mergers or minor mergers
where a large spheroid accretes a small gas-rich satellite which
then fuels a small burst of star formation (Kaviraj et al. 2009,
2010).}


\section{Local environment}
Previous work has strongly suggested that peculiar ETGs
preferentially inhabit regions of lower density \citep[see
e.g.][]{Malin1983a}. To explore the environments of our ETGs we
cross-match our Stripe82 galaxy sample with the SDSS group
catalogue constructed by \citet{Yang2007}, who use a halo-based
group finder to separate the SDSS into over 300,000 groups in a
broad dynamical range, spanning rich clusters to isolated
galaxies. We use this catalogue to identify both the masses of the
parent, dark-matter (DM) haloes that host the galaxies in our
sample and the halo-centric radii of these galaxies within their
respective haloes.

Note that the traditional definitions of environment (`Field',
`Groups' and `Clusters') are related to the underlying dark matter
haloes that host the structures that these definitions refer to.
`Cluster-sized' haloes typically have masses greater than
$10^{14}$M$_{\odot}$. `Group-sized' haloes have masses between
$10^{13}$M$_{\odot}$ and $10^{14}$M$_{\odot}$, while smaller
haloes constitute what is commonly termed the `field'. Figure
\ref{fig:environment} summarises the local environment of the
Stripe82 early-type galaxies in our sample. In the top panel we
plot a cumulative histogram of the halo masses and in the bottom
panel we plot the fractional halo-centric radii against the host
halo mass. The traditional definitions of local environment are
also indicated in the plot.

We find that peculiar ETGs (shown dotted) indeed have a tendency
to inhabit lower density regions. Only 5\% of the peculiar ETGs
inhabit cluster-sized haloes, typically avoiding the core (inner
15\%) of the cluster. Almost 75\% of peculiar ETGs inhabit the
field, while the remaining 20\% inhabit the intermediate-density
environment of group-sized haloes. In contrast, more than 20\% of
relaxed ETGs reside in cluster-sized haloes, consistent with the
established result that early-types typically dominate the galaxy
census in nearby clusters \citep[see e.g.][]{Dressler80}.

The absence of peculiar ETGs in the densest regions of the
Universe is plausibly driven by several factors. Peculiar
velocities in clusters are typically high ($\sim1000$ kms$^{-1}$),
making mergers less likely since interacting systems do not bind
gravitationally. While fly-bys, which are presumably more common
in cluster cores, may also give rise to shell structures
\citep{Thomson1990}, this requires a (hypothetical) cold thick
disk i.e. a dynamically cold population of stars in nearly
circular orbits to be present in the parent spheroid. The lack of
shell early-types in cluster cores might indicate that most ETGs
do not possess a cold, thick disk population, although the size of
our sample is too small to draw a robust conclusion. A final
possibility is that shell structures do form but, given frequent
interactions with other galaxies in the dense cluster cores, they
are disrupted within a cluster crossing time, which is around a
Gyr for a typical core extent of 1 Mpc and a peculiar velocity of
$\sim1000$ kms$^{-1}$.


\section{Summary and conclusions}
We have explored the spectro-photometric properties of the
early-type galaxy (ETG) population, drawn from the SDSS Stripe82
($-50^{\circ} < \alpha < 59^{\circ}, -1.25^{\circ} < \delta <
1.25^{\circ}$). Standard-depth ($\sim51$s exposure) multi-colour
($gri$) galaxy images from the SDSS DR7 have been combined with
the significantly ($\sim$2 mags) deeper
{\color{black}monochromatic} images from the public SDSS Stripe82
to extract, through careful visual inspection, a robust sample of
nearby ($z<0.05$), luminous ($M_r<-20.5$) ETGs. {\color{black}We
have identified, for the first time, a sample of peculiar ETGs
from the SDSS which carry faint morphological signatures of recent
interactions, such as tidal tails and dust lanes. $\sim34$\% of
the Stripe82 galaxy sample are classified as ETGs, and at the
depth of the DR7 and Stripe82 images, $\sim18$\% of the ETG
population show tidal features, while $\sim$7\% show evidence for
extended dust features. Given the depth of the Stripe82 images
used in this paper, the fraction of ETGs that exhibit tidal
features compares well with what might be expected from previous
work in the literature that has employed images of a similar
surface brightness.}

{\color{black}A `BPT' analysis of optical emission line ratios
indicates that the fraction of ETGs that are classified as
`star-forming' is very small - only 1.3\% of relaxed ETGs fall on
the star-forming locus, while the equivalent value for the
peculiar ETG population is 4.5\%. It is worth noting that the
number of peculiar ETGs in this category is three times the
corresponding value in their relaxed counterparts. In a similar
vein, the fraction of peculiar ETGs that are classified as either
Seyferts or LINERs (19.4\%) is twice the equivalent value in their
relaxed counterparts (10.1\%). The dominant type of emission-line
activity in all types of ETGs (relaxed or peculiar) is found to be
LINER-like.

A photometric analysis indicates that the distributions of optical
($u-r$) colours of the relaxed and peculiar ETG populations are
very similar, with a Mann-Whitney test indicating that the median
($u-r$) values of these two ETG subclasses are statistically
indistinguishable. However, significant differences are found in
their UV colours, which are considerably more sensitive to
low-level recent star formation (RSF). The nominal difference in
the median ($NUV-r$) of the two ETG subclasses is $\sim0.6$ mags.
A Mann-Whitney test indicates, that unlike the ($u-r$) colours,
the $(NUV-r)$ distributions of the relaxed and peculiar ETG
populations do indeed have different medians at the 5\%
significance level.

The bluer colours of the peculiar ETGs in the UV and higher
fraction of objects with RSF in these galaxies is consistent with
the higher fraction of `star-forming' and LINER-like objects in
the emission-line BPT diagnostics, if the source of the LINER
emission is, as recent work suggests, driven by stellar
photoionisation associated with the RSF. Taken together, these
results indicate that the merger events that produce the
morphological peculiarities also enhance the star formation and
nebular activity in the peculiar ETG population. However, the
negligible impact of this star formation on the \emph{optical}
colours suggests that the intensity of star formation in these
events is low and that the encounters are not particularly
gas-rich. This is supported by the UV colours of the peculiar ETGs
which, although offset from those of the relaxed ETGs, are largely
contained in the green valley and not the UV blue cloud. Our
results point to a scenario in which the interaction events and
their associated low-level star formation are driven by either
(largely) dry major mergers or minor mergers where a gas-rich
satellite is accreted by a massive spheroid, in agreement with the
recent literature on ETGs.}

A study of the local environments of the relaxed and peculiar ETG
populations indicates that, in agreement with previous studies,
the peculiar ETG population preferentially inhabits regions of low
density (outskirts of clusters, groups and the field). In
particular, peculiar ETGs are not found in the cores (the inner
$\sim15$\%) of rich clusters. This implies either that the
peculiar velocities in cluster centres are too high for mergers
(that produce the shells) to take place or that shell structures
do form in these regions but are rapidly disrupted through
frequent interactions within a cluster crossing time.

While visual inspection of survey images has become increasingly
common in modern galaxy evolution studies, the work presented in
this paper suggests that the depth of wide-area surveys (which are
necessarily optimised for areal coverage) is often insufficient to
reveal the signatures of a galaxy's morphological history.
Interactions are a key driver of galaxy evolution at all epochs.
While the spectro-photometric evolution of galaxies can be
efficiently mapped using contemporary data, the underlying
morphological transformations that may drive that evolution could
remain hidden due to insufficient exposure times. Since a complete
picture of galaxy evolution ideally requires information on both
aspects, deeper imaging than what is available currently is
required in future surveys to enable us to construct an accurate
picture of the evolution of the galaxy population over the
lifetime of the Universe.


\section*{Acknowledgements}
I am grateful to the referee for a number of important suggestions
that added depth to the original manuscript. I acknowledge a
Research Fellowship from the Royal Commission for the Exhibition
of 1851, an Imperial College Junior Research Fellowship from
Imperial College London, a Senior Research Fellowship from
Worcester College, Oxford and support from the BIPAC institute at
Oxford. Marc Sarzi, Chris Lintott, Kevin Schawinski, Scott Trager
and Sukyoung Yi are thanked for useful discussions.

GALEX (Galaxy Evolution Explorer) is a NASA Small Explorer,
launched in April 2003, developed in cooperation with the Centre
National d'Etudes Spatiales of France and the Korean Ministry of
Science and Technology.

Funding for the SDSS and SDSS-II has been provided by the Alfred
P. Sloan Foundation, the Participating Institutions, the National
Science Foundation, the U.S. Department of Energy, the National
Aeronautics and Space Administration, the Japanese Monbukagakusho,
the Max Planck Society, and the Higher Education Funding Council
for England. The SDSS Web Site is http://www.sdss.org/.

The SDSS is managed by the Astrophysical Research Consortium for
the Participating Institutions. The Participating Institutions are
the American Museum of Natural History, Astrophysical Institute
Potsdam, University of Basel, University of Cambridge, Case
Western Reserve University, University of Chicago, Drexel
University, Fermilab, the Institute for Advanced Study, the Japan
Participation Group, Johns Hopkins University, the Joint Institute
for Nuclear Astrophysics, the Kavli Institute for Particle
Astrophysics and Cosmology, the Korean Scientist Group, the
Chinese Academy of Sciences (LAMOST), Los Alamos National
Laboratory, the Max-Planck-Institute for Astronomy (MPIA), the
Max-Planck-Institute for Astrophysics (MPA), New Mexico State
University, Ohio State University, University of Pittsburgh,
University of Portsmouth, Princeton University, the United States
Naval Observatory, and the University of Washington.


\nocite{SDSSDR6} \nocite{Bernardi2003a} \nocite{Bernardi2003b}
\nocite{Bernardi2003c} \nocite{Schawinski2007b}
\nocite{Eisenstein2001} \nocite{Park2005} \nocite{Fukugita2007}
\nocite{Kaviraj2007a} \nocite{Kaviraj2007b} \nocite{Kaviraj2008a}
\nocite{Kaviraj2008b} \nocite{Lintott2008} \nocite{Abazajian2009}
\nocite{Baldwin1981} \nocite{Kauffmann2003a}
\nocite{Kauffmann2003b} \nocite{Frieman2008}
\nocite{Abazajian2009} \nocite{sauron5} \nocite{Martin2005}
\nocite{Yi2005} \nocite{Kaviraj2010} \nocite{Walpole1985}
\nocite{Springel2005d} \nocite{Robertson2006} \nocite{Sarzi2010}
\nocite{Gawiser2006} \nocite{Jannuzi2004} \nocite{Darg1}
\nocite{Darg2} \nocite{Hopkins2009}


\bibliographystyle{mn2e}
\bibliography{references}

\begin{thebibliography}{}

\bibitem[\protect\citeauthoryear{{Abazajian} \& {et al.}}{{Abazajian} \& {et
  al.}}{2009}]{Abazajian2009}
{Abazajian} K.~N.,  {et al.} 2009, ApJS, 182, 543

\bibitem[\protect\citeauthoryear{{Adelman-McCarthy} \& {et
  al.}}{{Adelman-McCarthy} \& {et al.}}{2008}]{SDSSDR6}
{Adelman-McCarthy} J.~K.,  {et al.} 2008, ApJS, 175, 297

\bibitem[\protect\citeauthoryear{{Baldwin}, {Phillips} \&
  {Terlevich}}{{Baldwin} et~al.}{1981}]{Baldwin1981}
{Baldwin} J.~A.,  {Phillips} M.~M.,    {Terlevich} R.,  1981, PASP, 93, 5

\bibitem[\protect\citeauthoryear{{Bell}, {Naab}, {McIntosh}, {Somerville},
  {Caldwell}, {Barden}, {Wolf}, {Rix}, {Beckwith}, {Borch}, {H{\"a}ussler},
  {Heymans}, {Jahnke}, {Jogee}, {Koposov}, {Meisenheimer}, {Peng}, {Sanchez} \&
  {Wisotzki}}{{Bell} et~al.}{2006}]{Bell2006}
{Bell} E.~F.,  {Naab} T.,  {McIntosh} D.~H.,  {Somerville} R.~S.,  {Caldwell}
  J.~A.~R.,  {Barden} M.,  {Wolf} C.,  {Rix} H.-W.,  {Beckwith} S.~V.,  {Borch}
  A.,  {H{\"a}ussler} B.,  {Heymans} C.,  {Jahnke} K.,  {Jogee} S.,  {Koposov}
  S.,  {Meisenheimer} K.,  {Peng} C.~Y.,  {Sanchez} S.~F.,    {Wisotzki} L.,
  2006, ApJ, 640, 241

\bibitem[\protect\citeauthoryear{{Bernardi} \& {et al.}}{{Bernardi} \& {et
  al.}}{2003a}]{Bernardi2003a}
{Bernardi} M.,  {et al.} 2003a, AJ, 125, 1817

\bibitem[\protect\citeauthoryear{{Bernardi} \& {et al.}}{{Bernardi} \& {et
  al.}}{2003b}]{Bernardi2003b}
{Bernardi} M.,  {et al.} 2003b, AJ, 125, 1882

\bibitem[\protect\citeauthoryear{{Bernardi} \& {et al.}}{{Bernardi} \& {et
  al.}}{2003c}]{Bernardi2003c}
{Bernardi} M.,  {et al.} 2003c, AJ, 125, 1849

\bibitem[\protect\citeauthoryear{{Bower}, {Lucey} \& {Ellis}}{{Bower}
  et~al.}{1992}]{BLE92}
{Bower} R.~G.,  {Lucey} J.~R.,    {Ellis} R.,  1992, MNRAS, 254, 589

\bibitem[\protect\citeauthoryear{{Bundy}, {Fukugita}, {Ellis}, {Targett},
  {Belli} \& {Kodama}}{{Bundy} et~al.}{2009}]{Bundy2009}
{Bundy} K.,  {Fukugita} M.,  {Ellis} R.~S.,  {Targett} T.~A.,  {Belli} S.,
  {Kodama} T.,  2009, ApJ, 697, 1369

\bibitem[\protect\citeauthoryear{{Cole}, {Lacey}, {Baugh} \& {Frenk}}{{Cole}
  et~al.}{2000}]{Cole2000}
{Cole} S.,  {Lacey} C.~G.,  {Baugh} C.~M.,    {Frenk} C.~S.,  2000, MNRAS, 319,
  168

\bibitem[\protect\citeauthoryear{{Conselice}, {Bershady}, {Dickinson} \&
  {Papovich}}{{Conselice} et~al.}{2003}]{Conselice2003}
{Conselice} C.~J.,  {Bershady} M.~A.,  {Dickinson} M.,    {Papovich} C.,  2003,
  AJ, 126, 1183

\bibitem[\protect\citeauthoryear{{Cowie}, {Songaila}, {Hu} \& {Cohen}}{{Cowie}
  et~al.}{1996}]{Cowie1996}
{Cowie} L.~L.,  {Songaila} A.,  {Hu} E.~M.,    {Cohen} J.~G.,  1996, AJ, 112,
  839

\bibitem[\protect\citeauthoryear{{Darg}, {Kaviraj}, {Lintott}, {Schawinski},
  {Sarzi}, {Bamford}, {Silk}, {Andreescu}, {Murray}, {Nichol}, {Raddick},
  {Slosar}, {Szalay}, {Thomas} \& {Vandenberg}}{{Darg} et~al.}{010b}]{Darg2}
{Darg} D.~W.,  {Kaviraj} S.,  {Lintott} C.~J.,  {Schawinski} K.,  {Sarzi} M.,
  {Bamford} S.,  {Silk} J.,  {Andreescu} D.,  {Murray} P.,  {Nichol} R.~C.,
  {Raddick} M.~J.,  {Slosar} A.,  {Szalay} A.~S.,  {Thomas} D.,    {Vandenberg}
  J.,  2010b, MNRAS, 401, 1552

\bibitem[\protect\citeauthoryear{{Darg}, {Kaviraj}, {Lintott}, {Schawinski},
  {Sarzi}, {Bamford}, {Silk}, {Proctor}, {Andreescu}, {Murray}, {Nichol},
  {Raddick}, {Slosar}, {Szalay}, {Thomas} \& {Vandenberg}}{{Darg}
  et~al.}{010a}]{Darg1}
{Darg} D.~W.,  {Kaviraj} S.,  {Lintott} C.~J.,  {Schawinski} K.,  {Sarzi} M.,
  {Bamford} S.,  {Silk} J.,  {Proctor} R.,  {Andreescu} D.,  {Murray} P.,
  {Nichol} R.~C.,  {Raddick} M.~J.,  {Slosar} A.,  {Szalay} A.~S.,  {Thomas}
  D.,    {Vandenberg} J.,  2010a, MNRAS, 401, 1043

\bibitem[\protect\citeauthoryear{{De Lucia}, {Poggianti},
  {Arag{\'o}n-Salamanca}, {White}, {Zaritsky}, {Clowe}, {Halliday}, {Jablonka},
  {von der Linden}, {Milvang-Jensen}, {Pell{\'o}}, {Rudnick}, {Saglia} \&
  {Simard}}{{De Lucia} et~al.}{2007}]{deLucia2007}
{De Lucia} G.,  {Poggianti} B.~M.,  {Arag{\'o}n-Salamanca} A.,  {White}
  S.~D.~M.,  {Zaritsky} D.,  {Clowe} D.,  {Halliday} C.,  {Jablonka} P.,  {von
  der Linden} A.,  {Milvang-Jensen} B.,  {Pell{\'o}} R.,  {Rudnick} G.,
  {Saglia} R.~P.,    {Simard} L.,  2007, MNRAS, 374, 809

\bibitem[\protect\citeauthoryear{{De Lucia}, {Springel}, {White}, {Croton} \&
  {Kauffmann}}{{De Lucia} et~al.}{2006}]{deLucia2006}
{De Lucia} G.,  {Springel} V.,  {White} S.~D.~M.,  {Croton} D.,    {Kauffmann}
  G.,  2006, MNRAS, 366, 499

\bibitem[\protect\citeauthoryear{{Dressler}}{{Dressler}}{1980}]{Dressler80}
{Dressler} A.,  1980, ApJ, 236, 351

\bibitem[\protect\citeauthoryear{{Eisenstein} \& {et al.}}{{Eisenstein} \& {et
  al.}}{2001}]{Eisenstein2001}
{Eisenstein} D.~J.,  {et al.} 2001, AJ, 122, 2267

\bibitem[\protect\citeauthoryear{{Ellis}, {Smail}, {Dressler}, {Couche},
  {Oemler}, {Butcher} \& {Sharples}}{{Ellis} et~al.}{1997}]{Ellis1997}
{Ellis} R.~S.,  {Smail} I.,  {Dressler} A.,  {Couche} W.~J.,  {Oemler} A.~J.,
  {Butcher} H.,    {Sharples} R.~M.,  1997, ApJ, 483, 582

\bibitem[\protect\citeauthoryear{{Frieman} \& {et al.}}{{Frieman} \& {et
  al.}}{2008}]{Frieman2008}
{Frieman} J.~A.,  {et al.} 2008, AJ, 135, 338

\bibitem[\protect\citeauthoryear{{Fukugita}, {Nakamura}, {Okamura}, {Yasuda},
  {Barentine}, {Brinkmann}, {Gunn}, {Harvanek}, {Ichikawa}, {Lupton},
  {Schneider}, {Strauss} \& {York}}{{Fukugita} et~al.}{2007}]{Fukugita2007}
{Fukugita} M.,  {Nakamura} O.,  {Okamura} S.,  {Yasuda} N.,  {Barentine} J.~C.,
   {Brinkmann} J.,  {Gunn} J.~E.,  {Harvanek} M.,  {Ichikawa} T.,  {Lupton}
  R.~H.,  {Schneider} D.~P.,  {Strauss} M.~A.,    {York} D.~G.,  2007, AJ, 134,
  579

\bibitem[\protect\citeauthoryear{{Gawiser} \& {MUSYC collaboration}}{{Gawiser}
  \& {MUSYC collaboration}}{2006}]{Gawiser2006}
{Gawiser} E.,  {MUSYC collaboration} 2006, ApJS, 162, 1

\bibitem[\protect\citeauthoryear{{Gladders}, {Lopez-Cruz}, {Yee} \&
  {Kodama}}{{Gladders} et~al.}{1998}]{Gladders98}
{Gladders} M.~D.,  {Lopez-Cruz} O.,  {Yee} H.~K.~C.,    {Kodama} T.,  1998,
  ApJ, 501, 571

\bibitem[\protect\citeauthoryear{{Graham}}{{Graham}}{1979}]{Graham1979}
{Graham} J.~A.,  1979, ApJ, 232, 60

\bibitem[\protect\citeauthoryear{{Hatton}, {Devriendt}, {Ninin}, {Bouchet},
  {Guiderdoni} \& {Vibert}}{{Hatton} et~al.}{2003}]{Hatton2003}
{Hatton} S.,  {Devriendt} J.~E.~G.,  {Ninin} S.,  {Bouchet} F.~R.,
  {Guiderdoni} B.,    {Vibert} D.,  2003, MNRAS, 343, 75

\bibitem[\protect\citeauthoryear{{Hawarden}, {Longmore}, {Tritton}, {Elson} \&
  {Corwin} Jr.}{{Hawarden} et~al.}{1981}]{Hawarden1981}
{Hawarden} T.~G.,  {Longmore} A.~J.,  {Tritton} S.~B.,  {Elson} R.~A.~W.,
  {Corwin} Jr. H.~G.,  1981, MNRAS, 196, 747

\bibitem[\protect\citeauthoryear{{Hernquist} \& {Spergel}}{{Hernquist} \&
  {Spergel}}{1992}]{Hernquist1992}
{Hernquist} L.,  {Spergel} D.~N.,  1992, ApJ, 399, L117

\bibitem[\protect\citeauthoryear{{Hopkins}, {Cox}, {Younger} \&
  {Hernquist}}{{Hopkins} et~al.}{2009}]{Hopkins2009}
{Hopkins} P.~F.,  {Cox} T.~J.,  {Younger} J.~D.,    {Hernquist} L.,  2009, ApJ,
  691, 1168

\bibitem[\protect\citeauthoryear{{Jannuzi}, {Dey}, {Brown}, {Ford}, {Hogan},
  {Miller}, {Ryan}, {Tiede}, {Valdes} \& {NDWFS}}{{Jannuzi}
  et~al.}{2004}]{Jannuzi2004}
{Jannuzi} B.~T.,  {Dey} A.,  {Brown} M.~J.~I.,  {Ford} A.,  {Hogan} E.,
  {Miller} M.,  {Ryan} E.,  {Tiede} G.~P.,  {Valdes} F.,    {NDWFS} 2004
  Vol.~36 of Bulletin of the American Astronomical Society.
pp 1478--+

\bibitem[\protect\citeauthoryear{{Jorgensen}, {Franx} \&
  {Kjaergaard}}{{Jorgensen} et~al.}{1996}]{Jorg1996}
{Jorgensen} I.,  {Franx} M.,    {Kjaergaard} P.,  1996, MNRAS, 280, 167

\bibitem[\protect\citeauthoryear{{Kauffmann} \& {et al.}}{{Kauffmann} \& {et
  al.}}{2003a}]{Kauffmann2003a}
{Kauffmann} G.,  {et al.} 2003a, MNRAS, 346, 1055

\bibitem[\protect\citeauthoryear{{Kauffmann} \& {et al.}}{{Kauffmann} \& {et
  al.}}{2003b}]{Kauffmann2003b}
{Kauffmann} G.,  {et al.} 2003b, MNRAS, 341, 33

\bibitem[\protect\citeauthoryear{{Kaviraj}}{{Kaviraj}}{2008}]{Kaviraj2008a}
{Kaviraj} S.,  2008, Modern Physics Letters A, 23, 153

\bibitem[\protect\citeauthoryear{{Kaviraj}, {Devriendt}, {Ferreras} \&
  {Yi}}{{Kaviraj} et~al.}{2005}]{Kaviraj2005a}
{Kaviraj} S.,  {Devriendt} J.~E.~G.,  {Ferreras} I.,    {Yi} S.~K.,  2005,
  MNRAS, 360, 60

\bibitem[\protect\citeauthoryear{{Kaviraj} \& {GALEX collaboration}}{{Kaviraj}
  \& {GALEX collaboration}}{007b}]{Kaviraj2007b}
{Kaviraj} S.,  {GALEX collaboration} 2007b, ApJS, 173, 619

\bibitem[\protect\citeauthoryear{{Kaviraj}, {Khochfar}, {Schawinski}, {Yi},
  {Gawiser}, {Silk}, {Virani}, {Cardamone}, {van Dokkum} \& {Urry}}{{Kaviraj}
  et~al.}{2008}]{Kaviraj2008b}
{Kaviraj} S.,  {Khochfar} S.,  {Schawinski} K.,  {Yi} S.~K.,  {Gawiser} E.,
  {Silk} J.,  {Virani} S.~N.,  {Cardamone} C.~N.,  {van Dokkum} P.~G.,
  {Urry} C.~M.,  2008, MNRAS, 388, 67

\bibitem[\protect\citeauthoryear{{Kaviraj}, {Peirani}, {Khochfar}, {Silk} \&
  {Kay}}{{Kaviraj} et~al.}{2009}]{Kaviraj2009}
{Kaviraj} S.,  {Peirani} S.,  {Khochfar} S.,  {Silk} J.,    {Kay} S.,  2009,
  MNRAS, 394, 1713

\bibitem[\protect\citeauthoryear{{Kaviraj}, {Rey}, {Rich}, {Yoon} \&
  {Yi}}{{Kaviraj} et~al.}{007a}]{Kaviraj2007a}
{Kaviraj} S.,  {Rey} S.-C.,  {Rich} R.~M.,  {Yoon} S.-J.,    {Yi} S.~K.,
  2007a, MNRAS, 381, L74

\bibitem[\protect\citeauthoryear{{Kaviraj}, {Tan}, {Ellis} \& {Silk}}{{Kaviraj}
  et~al.}{2010}]{Kaviraj2010}
{Kaviraj} S.,  {Tan} K.,  {Ellis} R.~S.,    {Silk} J.,  2010, ArXiv:1001.2141

\bibitem[\protect\citeauthoryear{{Le F{\`e}vre}, {Abraham}, {Lilly}, {Ellis},
  {Brinchmann}, {Schade}, {Tresse}, {Colless}, {Crampton}, {Glazebrook},
  {Hammer} \& {Broadhurst}}{{Le F{\`e}vre} et~al.}{2000}]{Lefevre2000}
{Le F{\`e}vre} O.,  {Abraham} R.,  {Lilly} S.~J.,  {Ellis} R.~S.,  {Brinchmann}
  J.,  {Schade} D.,  {Tresse} L.,  {Colless} M.,  {Crampton} D.,  {Glazebrook}
  K.,  {Hammer} F.,    {Broadhurst} T.,  2000, MNRAS, 311, 565

\bibitem[\protect\citeauthoryear{{Lin}, {Koo}, {Willmer}, {Patton},
  {Conselice}, {Yan}, {Coil}, {Cooper}, {Davis}, {Faber}, {Gerke},
  {Guhathakurta} \& {Newman}}{{Lin} et~al.}{2004}]{Lin2004}
{Lin} L.,  {Koo} D.~C.,  {Willmer} C.~N.~A.,  {Patton} D.~R.,  {Conselice}
  C.~J.,  {Yan} R.,  {Coil} A.~L.,  {Cooper} M.~C.,  {Davis} M.,  {Faber}
  S.~M.,  {Gerke} B.~F.,  {Guhathakurta} P.,    {Newman} J.~A.,  2004, ApJ,
  617, L9

\bibitem[\protect\citeauthoryear{{Lintott}, {Schawinski}, {Slosar}, {Land},
  {Bamford}, {Thomas}, {Raddick}, {Nichol}, {Szalay}, {Andreescu}, {Murray} \&
  {Vandenberg}}{{Lintott} et~al.}{2008}]{Lintott2008}
{Lintott} C.~J.,  {Schawinski} K.,  {Slosar} A.,  {Land} K.,  {Bamford} S.,
  {Thomas} D.,  {Raddick} M.~J.,  {Nichol} R.~C.,  {Szalay} A.,  {Andreescu}
  D.,  {Murray} P.,    {Vandenberg} J.,  2008, MNRAS, 389, 1179

\bibitem[\protect\citeauthoryear{{Malin} \& {Carter}}{{Malin} \&
  {Carter}}{1980}]{Malin1980}
{Malin} D.~F.,  {Carter} D.,  1980, Nature, 285, 643

\bibitem[\protect\citeauthoryear{{Malin} \& {Carter}}{{Malin} \&
  {Carter}}{1983}]{Malin1983a}
{Malin} D.~F.,  {Carter} D.,  1983, ApJ, 274, 534

\bibitem[\protect\citeauthoryear{{Malin}, {Quinn} \& {Graham}}{{Malin}
  et~al.}{1983}]{Malin1983b}
{Malin} D.~F.,  {Quinn} P.~J.,    {Graham} J.~A.,  1983, ApJ, 272, L5

\bibitem[\protect\citeauthoryear{{Martin} \& {GALEX collaboration}}{{Martin} \&
  {GALEX collaboration}}{2005}]{Martin2005}
{Martin} D.~C.,  {GALEX collaboration} 2005, ApJ, 619, L1

\bibitem[\protect\citeauthoryear{{McIntosh}, {Guo}, {Hertzberg}, {Katz}, {Mo},
  {van den Bosch} \& {Yang}}{{McIntosh} et~al.}{2008}]{McIntosh2008}
{McIntosh} D.~H.,  {Guo} Y.,  {Hertzberg} J.,  {Katz} N.,  {Mo} H.~J.,  {van
  den Bosch} F.~C.,    {Yang} X.,  2008, MNRAS, 388, 1537

\bibitem[\protect\citeauthoryear{{Park} \& {Choi}}{{Park} \&
  {Choi}}{2005}]{Park2005}
{Park} C.,  {Choi} Y.-Y.,  2005, ApJ, 635, L29

\bibitem[\protect\citeauthoryear{{Patton}, {Carlberg}, {Marzke}, {Pritchet},
  {da Costa} \& {Pellegrini}}{{Patton} et~al.}{2000}]{Patton2000}
{Patton} D.~R.,  {Carlberg} R.~G.,  {Marzke} R.~O.,  {Pritchet} C.~J.,  {da
  Costa} L.~N.,    {Pellegrini} P.~S.,  2000, ApJ, 536, 153

\bibitem[\protect\citeauthoryear{{Patton}, {Pritchet}, {Carlberg}, {Marzke},
  {Yee}, {Hall}, {Lin}, {Morris}, {Sawicki}, {Shepherd} \& {Wirth}}{{Patton}
  et~al.}{2002}]{Patton2002}
{Patton} D.~R.,  {Pritchet} C.~J.,  {Carlberg} R.~G.,  {Marzke} R.~O.,  {Yee}
  H.~K.~C.,  {Hall} P.~B.,  {Lin} H.,  {Morris} S.~L.,  {Sawicki} M.,
  {Shepherd} C.~W.,    {Wirth} G.~D.,  2002, ApJ, 565, 208

\bibitem[\protect\citeauthoryear{{Prieur}}{{Prieur}}{1988}]{Prieur1988}
{Prieur} J.-L.,  1988, ApJ, 326, 596

\bibitem[\protect\citeauthoryear{{Quinn}}{{Quinn}}{1984}]{Quinn1984}
{Quinn} P.~J.,  1984, ApJ, 279, 596

\bibitem[\protect\citeauthoryear{{Robertson}, {Bullock}, {Cox}, {Di Matteo},
  {Hernquist}, {Springel} \& {Yoshida}}{{Robertson}
  et~al.}{2006}]{Robertson2006}
{Robertson} B.,  {Bullock} J.~S.,  {Cox} T.~J.,  {Di Matteo} T.,  {Hernquist}
  L.,  {Springel} V.,    {Yoshida} N.,  2006, ApJ, 645, 986

\bibitem[\protect\citeauthoryear{{Saglia}, {Colless}, {Baggley},
  {Bertschinger}, {Burstein}, {Davies}, {McMahan} \& {Wegner}}{{Saglia}
  et~al.}{1997}]{Saglia1997}
{Saglia} R.~P.,  {Colless} M.,  {Baggley} G.,  {Bertschinger} E.,  {Burstein}
  D.,  {Davies} R.~L.,  {McMahan} R.~K.,    {Wegner} G.,  1997, in {Arnaboldi}
  M.,  {Da Costa} G.~S.,   {Saha} P.,  eds, ASP Conf. Ser. 116: The Nature of
  Elliptical Galaxies; 2nd Stromlo Symposium {The EFAR Fundamental Plane}.
pp 180--+

\bibitem[\protect\citeauthoryear{{Sarzi} \& {et al.}}{{Sarzi} \& {et
  al.}}{2006}]{sauron5}
{Sarzi} M.,  {et al.} 2006, MNRAS, 366, 1151

\bibitem[\protect\citeauthoryear{{Sarzi} \& {et al.}}{{Sarzi} \& {et
  al.}}{2010}]{Sarzi2010}
{Sarzi} M.,  {et al.} 2010, MNRAS, 402, 2187

\bibitem[\protect\citeauthoryear{{Schawinski}, {Thomas}, {Sarzi}, {Maraston},
  {Kaviraj}, {Joo}, {Yi} \& {Silk}}{{Schawinski}
  et~al.}{007b}]{Schawinski2007b}
{Schawinski} K.,  {Thomas} D.,  {Sarzi} M.,  {Maraston} C.,  {Kaviraj} S.,
  {Joo} S.-J.,  {Yi} S.~K.,    {Silk} J.,  2007b, MNRAS, 382, 1415

\bibitem[\protect\citeauthoryear{{Schweizer} \& {Seitzer}}{{Schweizer} \&
  {Seitzer}}{1992}]{Schweizer1992}
{Schweizer} F.,  {Seitzer} P.,  1992, AJ, 104, 1039

\bibitem[\protect\citeauthoryear{{Schweizer}, {Seitzer}, {Faber}, {Burstein},
  {Dalle Ore} \& {Gonzalez}}{{Schweizer} et~al.}{1990}]{Schweizer1990}
{Schweizer} F.,  {Seitzer} P.,  {Faber} S.~M.,  {Burstein} D.,  {Dalle Ore}
  C.~M.,    {Gonzalez} J.~J.,  1990, ApJ, 364, L33

\bibitem[\protect\citeauthoryear{{Sikkema}, {Carter}, {Peletier}, {Balcells},
  {Del Burgo} \& {Valentijn}}{{Sikkema} et~al.}{2007}]{Sikkema2007}
{Sikkema} G.,  {Carter} D.,  {Peletier} R.~F.,  {Balcells} M.,  {Del Burgo} C.,
     {Valentijn} E.~A.,  2007, A\&A, 467, 1011

\bibitem[\protect\citeauthoryear{{Springel} \& {Hernquist}}{{Springel} \&
  {Hernquist}}{2005}]{Springel2005d}
{Springel} V.,  {Hernquist} L.,  2005, ApJ, 622, L9

\bibitem[\protect\citeauthoryear{{Stanford}, {Eisenhardt} \&
  {Dickinson}}{{Stanford} et~al.}{1998}]{Stanford98}
{Stanford} S.~A.,  {Eisenhardt} P.~R.~M.,    {Dickinson} M.,  1998, ApJ, 492,
  461

\bibitem[\protect\citeauthoryear{{Stasi{\'n}ska}, {Vale Asari}, {Cid
  Fernandes}, {Gomes}, {Schlickmann}, {Mateus}, {Schoenell} \& {Sodr{\'e}}
  Jr.}{{Stasi{\'n}ska} et~al.}{2008}]{Stasinska2008}
{Stasi{\'n}ska} G.,  {Vale Asari} N.,  {Cid Fernandes} R.,  {Gomes} J.~M.,
  {Schlickmann} M.,  {Mateus} A.,  {Schoenell} W.,    {Sodr{\'e}} Jr. L.,
  2008, MNRAS, 391, L29

\bibitem[\protect\citeauthoryear{{Thomas}, {Greggio} \& {Bender}}{{Thomas}
  et~al.}{1999}]{Thomas1999}
{Thomas} D.,  {Greggio} L.,    {Bender} R.,  1999, MNRAS, 302, 537

\bibitem[\protect\citeauthoryear{{Thomson} \& {Wright}}{{Thomson} \&
  {Wright}}{1990}]{Thomson1990}
{Thomson} R.~C.,  {Wright} A.~E.,  1990, MNRAS, 247, 122

\bibitem[\protect\citeauthoryear{{van Dokkum}}{{van Dokkum}}{2005}]{VD2005}
{van Dokkum} P.~G.,  2005, AJ, 130, 2647

\bibitem[\protect\citeauthoryear{{van Dokkum} \& {Franx}}{{van Dokkum} \&
  {Franx}}{1996}]{VD1996}
{van Dokkum} P.~G.,  {Franx} M.,  1996, MNRAS, 281, 985

\bibitem[\protect\citeauthoryear{{van Dokkum}, {Franx}, {Fabricant},
  {Illingworth} \& {Kelson}}{{van Dokkum} et~al.}{2000}]{VD2000}
{van Dokkum} P.~G.,  {Franx} M.,  {Fabricant} D.,  {Illingworth} G.~D.,
  {Kelson} D.~D.,  2000, ApJ, 541, 95

\bibitem[\protect\citeauthoryear{{Walpole} \& {Myers}}{{Walpole} \&
  {Myers}}{1985}]{Walpole1985}
{Walpole} R.~E.,  {Myers} R.~H.,  1985, Probability and statistics for
  engineers and scientists, 3rd, edn.
Macmillan

\bibitem[\protect\citeauthoryear{{Yang}, {Mo}, {van den Bosch}, {Pasquali},
  {Li} \& {Barden}}{{Yang} et~al.}{2007}]{Yang2007}
{Yang} X.,  {Mo} H.~J.,  {van den Bosch} F.~C.,  {Pasquali} A.,  {Li} C.,
  {Barden} M.,  2007, ApJ, 671, 153

\bibitem[\protect\citeauthoryear{{Yi}, {Demarque} \& {Oemler}}{{Yi}
  et~al.}{1997}]{Yi97}
{Yi} S.,  {Demarque} P.,    {Oemler} A.~J.,  1997, ApJ, 486, 201

\bibitem[\protect\citeauthoryear{{Yi}, {Lee}, {Woo}, {Park}, {Demarque} \&
  {Oemler}}{{Yi} et~al.}{1999}]{Yi99}
{Yi} S.,  {Lee} Y.-W.,  {Woo} J.-H.,  {Park} J.-H.,  {Demarque} P.,    {Oemler}
  A.~J.,  1999, ApJ, 513, 128

\bibitem[\protect\citeauthoryear{{Yi}, {Yoon}, {Kaviraj}, {Deharveng} \& {the
  GALEX Science Team}}{{Yi} et~al.}{2005}]{Yi2005}
{Yi} S.~K.,  {Yoon} S.-J.,  {Kaviraj} S.,  {Deharveng} J.-M.,    {the GALEX
  Science Team} 2005, ApJ, 619, L111

\end{thebibliography}


\end{document}